\documentclass[aps,prd,superscriptaddress,preprint,notitlepage]{revtex4-1}
\usepackage[utf8]{inputenc}
\usepackage{booktabs}
\usepackage{tabularx}
\usepackage{xstring}
\usepackage{ifthen}
\usepackage{orcidlink}

\newboolean{uprightparticles}
\setboolean{uprightparticles}{false} 

\newcommand\summaryname{Abstract}
\newenvironment{Abstract}%
    {\small\begin{center}%
    \bfseries{\summaryname} \end{center}}

\usepackage{amsmath,amssymb,amsfonts,bm,slashed}
\usepackage{array,multirow,xspace}
\usepackage{graphicx}
\usepackage{lineno}
\usepackage{xcolor}

\definecolor{nicered}{rgb}{0.7,0.1,0.1}
\definecolor{nicegreen}{rgb}{0.1,0.5,0.1}
\definecolor{niceblue}{rgb}{0.1,0.1,0.5}

\graphicspath{ {./Figures/} }

\usepackage{hyperref}
\hypersetup{bookmarks=false,colorlinks,citecolor= nicegreen,linkcolor= niceblue,urlcolor= niceblue}

\usepackage{xspace} 
\usepackage{upgreek}
\usepackage{xcolor}

\def\lhcb   {\mbox{LHCb}\xspace}
\def\atlas  {\mbox{ATLAS}\xspace}
\def\cms    {\mbox{CMS}\xspace}
\def\codexb {\mbox{CODEX-b}\xspace}
\def\codexbeta {\mbox{CODEX-\ensuremath{\beta}}\xspace}

\ifthenelse{\boolean{uprightparticles}} 
{

 \def\PDelta      {\ensuremath{\Delta}\xspace}
 \def\PXi         {\ensuremath{\Xi}\xspace}
 \def\PLambda     {\ensuremath{\Lambda}\xspace}
 \def\PSigma      {\ensuremath{\Sigma}\xspace}
 \def\POmega      {\ensuremath{\Omega}\xspace}
 \def\PUpsilon    {\ensuremath{\Upsilon}\xspace}

 \def\PB      {\ensuremath{\mathrm{B}}\xspace}
 
 \def\PD      {\ensuremath{\mathrm{D}}\xspace}

 \def\PK      {\ensuremath{\mathrm{K}}\xspace}

 \def\Pb      {\ensuremath{\mathrm{b}}\xspace}

 \def\Pi      {\ensuremath{\mathrm{i}}\xspace}

 \def\Ps      {\ensuremath{\mathrm{s}}\xspace}

 \def\thebaroffset{0.0em}
}
{

 \mathchardef\PDelta="7101
 \mathchardef\PXi="7104
 \mathchardef\PLambda="7103
 \mathchardef\PSigma="7106
 \mathchardef\POmega="710A
 \mathchardef\PUpsilon="7107
 
 \def\PB      {\ensuremath{B}\xspace}
 
 \def\PD      {\ensuremath{D}\xspace}

 \def\PK      {\ensuremath{K}\xspace}

 \def\Pb      {\ensuremath{b}\xspace}

 \def\Pi      {\ensuremath{i}\xspace}

 \def\Ps      {\ensuremath{s}\xspace}

 \def\thebaroffset{0.18em}
}
\newcommand{\offsetoverline}[2][\thebaroffset]{\kern #1\overline{\kern -#1 #2}}%

\makeatletter
\gdef\@ptsize{0} 
\ifcase \@ptsize \relax
  \newcommand{\miniscule}{\@setfontsize\miniscule{4}{5}}
\or
  \newcommand{\miniscule}{\@setfontsize\miniscule{5}{6}}
\or
  \newcommand{\miniscule}{\@setfontsize\miniscule{5}{6}}
\fi
\makeatother

\DeclareRobustCommand{\optbar}[1]{\shortstack{{\miniscule (\rule[.5ex]{1.25em}{.18mm})}
  \\ [-.7ex] $#1$}}

\def\squark    {{\ensuremath{\Ps}}\xspace}

\def\bquark    {{\ensuremath{\Pb}}\xspace}

\def\kaon    {{\ensuremath{\PK}}\xspace}

\def\KorKbar {\kern \thebaroffset\optbar{\kern -\thebaroffset \PK}{}\xspace}

\def\KL      {{\ensuremath{\kaon^0_{\mathrm{L}}}}\xspace}

\def\D       {{\ensuremath{\PD}}\xspace}

\def\DorDbar {\kern \thebaroffset\optbar{\kern -\thebaroffset \PD}\xspace}

\def\Dp      {{\ensuremath{\D^+}}\xspace}
\def\Dm      {{\ensuremath{\D^-}}\xspace}

\def\DpDm    {\ensuremath{\Dp {\kern -0.16em \Dm}}\xspace}

\def\B       {{\ensuremath{\PB}}\xspace}

\def\BorBbar {\kern \thebaroffset\optbar{\kern -\thebaroffset \PB}\xspace}

\def\Bd      {{\ensuremath{\B^0}}\xspace}

\def\BdorBdbar {\kern \thebaroffset\optbar{\kern -\thebaroffset \Bd}\xspace}

\def\Bs      {{\ensuremath{\B^0_\squark}}\xspace}

\def\BsorBsbar {\kern \thebaroffset\optbar{\kern -\thebaroffset \Bs}\xspace}

\def\Y#1S{\ensuremath{\PUpsilon{(#1S)}}\xspace}

\def\LorLbar     {\kern \thebaroffset\optbar{\kern -\thebaroffset \PLambda}\xspace}

\newcommand{\nospaceunit}[1]{\ensuremath{\text{#1}}}
\newcommand{\aunit}[1]{\ensuremath{\text{\,#1}}}
\newcommand{\unit}[1]{\aunit{#1}\xspace}

\newcommand{\tev}{\aunit{Te\kern -0.1em V}\xspace}
\newcommand{\gev}{\aunit{Ge\kern -0.1em V}\xspace}
\newcommand{\mev}{\aunit{Me\kern -0.1em V}\xspace}
\newcommand{\kev}{\aunit{ke\kern -0.1em V}\xspace}
\newcommand{\ev}{\aunit{e\kern -0.1em V}\xspace}

\newcommand{\mevc}{\ensuremath{\aunit{Me\kern -0.1em V\!/}c}\xspace}
\newcommand{\gevc}{\ensuremath{\aunit{Ge\kern -0.1em V\!/}c}\xspace}
\newcommand{\mevcc}{\ensuremath{\aunit{Me\kern -0.1em V\!/}c^2}\xspace}
\newcommand{\gevcc}{\ensuremath{\aunit{Ge\kern -0.1em V\!/}c^2}\xspace}

\def\m    {\aunit{m}\xspace}
\def\ma   {\ensuremath{\aunit{m}^2}\xspace}
\def\cm   {\aunit{cm}\xspace}
\def\cma  {\ensuremath{\aunit{cm}^2}\xspace}
\def\mm   {\aunit{mm}\xspace}

\def\mum  {\ensuremath{\,\upmu\nospaceunit{m}}\xspace}

\def\fb   {\ensuremath{\aunit{fb}}\xspace}
\def\invfb   {\ensuremath{\fb^{-1}}\xspace}

\def\mv   {\ensuremath{\aunit{m}^3}\xspace}
\def\L    {\unit{L}}

\def\ns   {\ensuremath{\aunit{ns}}\xspace}
\def\ps   {\ensuremath{\aunit{ps}}\xspace}

\def\ohm   {\ensuremath{\,\Omega}\xspace}
\def\kohm  {\ensuremath{\,\text{k}\Omega}\xspace}

\def\mV    {\unit{mV}}
\def\muV   {\ensuremath{\,\upmu\text{V}}\xspace}

\def\kg   {\unit{kg}}

\begin{document}


\title{Technical design report for the \codexbeta demonstrator}

\author{Giulio Aielli\,\orcidlink{0000-0002-0573-8114}}
\affiliation{Universit\`a e INFN Sezione di Roma Tor Vergata, Roma, Italy}

\author{Juliette Alimena\,\orcidlink{0000-0001-6030-3191}}
\affiliation{Deutsches Elektronen-Synchrotron (DESY), Hamburg, Germany}

\author{Saul Balcarcel-Salazar\,\orcidlink{0009-0005-1104-3502}}
\affiliation{Laboratory for Nuclear Science, Massachusetts Institute of Technology, Cambridge, MA 02139, USA}

\author{James Beacham\,\orcidlink{0000-0003-3623-3335}}
\affiliation{Department of Physics, Duke University, Durham, NC 27706, USA}

\author{Eli Ben Haim\,\orcidlink{0000-0002-9510-8414}}
\affiliation{LPNHE, Sorbonne Universit\'e, Universit{\'e} Paris Cit{\'e}, CNRS/IN2P3, Paris, France}

\author{Andr{\'a}s Barnab{\'a}s Burucs\,\orcidlink{009-0001-6791-1007}}
\affiliation{ELTE E\"otv\"os Lor\'and University, Budapest, Hungary}

\author{Roberto Cardarelli\,\orcidlink{0000-0003-4541-4189}}
\affiliation{Universit\`a e INFN Sezione di Roma Tor Vergata, Roma, Italy}

\author{Matthew J. Charles\,\orcidlink{0000-0003-4795-498X}}
\affiliation{Universit\'e Pierre et Marie Curie, Paris, France}

\author{Xabier Cid Vidal\,\orcidlink{0000-0002-0468-541X}}
\affiliation{Instituto Galego de F\'isica de Altas Enerx\'ias (IGFAE), Universidade de Santiago de Compostela, Santiago de Compostela, Spain}

\author{Albert De Roeck\,\orcidlink{0000-0002-9228-5271}}
\affiliation{European Organization for Nuclear Research (CERN), Geneva, Switzerland}

\author{Biplab Dey\,\orcidlink{0000-0002-4563-5806}}
\affiliation{ELTE E\"otv\"os Lor\'and University, Budapest, Hungary}

\author{Silviu Dobrescu}
\affiliation{Department of Physics, University of Cincinnati, Cincinnati, OH 45221, USA}

\author{Ozgur Durmus\,\orcidlink{0000-0002-8161-7832}}
\affiliation{ELTE E\"otv\"os Lor\'and University, Budapest, Hungary}

\author{Mohamed Elashri\,\orcidlink{0000-0001-9398-953X}}
\affiliation{Department of Physics, University of Cincinnati, Cincinnati, OH 45221, USA}

\author{Vladimir V.~Gligorov\,\orcidlink{0000-0002-8189-8267}}
\affiliation{LPNHE, Sorbonne Universit\'e, Universit{\'e} Paris Cit{\'e}, CNRS/IN2P3, Paris, France}
\affiliation{European Organization for Nuclear Research (CERN), Geneva, Switzerland}

\author{Rebeca Gonzalez Suarez\,\orcidlink{0000-0002-6126-7230}}
\affiliation{Department of Physics and Astronomy, Uppsala Universitet, Uppsala, Sweden}

\author{Thomas Gorordo\,\orcidlink{0009-0001-0342-6205}}
\affiliation{Department of Physics, University of Oregon, Eugene, OR 97403, USA}

\author{Zarria Gray}
\affiliation{Department of Physics, University of Cincinnati, Cincinnati, OH 45221, USA}

\author{Conor Henderson\,\orcidlink{0000-0002-6986-9404}}
\affiliation{Department of Physics, University of Cincinnati, Cincinnati, OH 45221, USA}

\author{Louis Henry\,\orcidlink{0000-0003-3605-832X}}
\affiliation{Ecole Polytechnique F\'ed\'erale Lausanne, Lausanne, Switzerland}

\author{Philip Ilten\,\orcidlink{0000-0001-5534-1732}}
\affiliation{Department of Physics, University of Cincinnati, Cincinnati, OH 45221, USA}

\author{Daniel Johnson\,\orcidlink{0000-0003-3272-6001}}
\affiliation{University of Birmingham, Birmingham, United Kingdom}

\author{Jacob Kautz\,\orcidlink{0000-0001-8482-5576}}
\affiliation{Department of Physics, University of Cincinnati, Cincinnati, OH 45221, USA}

\author{Simon Knapen\,\orcidlink{0000-0002-6733-9231}}
\affiliation{Physics Division, Lawrence Berkeley National Laboratory, Berkeley, CA 94720, USA}
\affiliation{Department of Physics, University of California, Berkeley, CA 94720, USA}

\author{Bingxuan Liu\,\orcidlink{0000-0002-0721-8331}}
\affiliation{School of Science, Shenzhen Campus of Sun Yat-Sen University, Shenzhen, China}

\author{Yang Liu\,\orcidlink{0000-0003-3615-2332}}
\affiliation{School of Science, Shenzhen Campus of Sun Yat-Sen University, Shenzhen, China}

\author{Saul L\'opez Soli\~no\,\orcidlink{0000-0001-9892-5113}}
\affiliation{Instituto Galego de F\'isica de Altas Enerx\'ias (IGFAE), Universidade de Santiago de Compostela, Santiago de Compostela, Spain}

\author{Pablo Eduardo Men\'endez-Vald\'es P\'erez\,\orcidlink{0009-0003-0406-8141}}
\affiliation{Instituto Galego de F\'isica de Altas Enerx\'ias (IGFAE), Universidade de Santiago de Compostela, Santiago de Compostela, Spain}

\author{Titus Momb\"acher\,\orcidlink{0000-0002-5612-979X}}
\affiliation{European Organization for Nuclear Research (CERN), Geneva, Switzerland}

\author{Benjamin Nachman\,\orcidlink{0000-0003-1024-0932}}
\affiliation{Physics Division, Lawrence Berkeley National Laboratory, Berkeley, CA 94720, USA}

\author{David T.~Northacker}
\affiliation{Department of Physics, University of Cincinnati, Cincinnati, OH 45221, USA}

\author{Gabriel M.~Nowak\,\orcidlink{0000-0003-4864-7164}}
\affiliation{Department of Physics, University of Cincinnati, Cincinnati, OH 45221, USA}

\author{Michele Papucci\,\orcidlink{0000-0003-0810-0017}}
\affiliation{Walter Burke Institute for Theoretical Physics, California Institute of Technology, Pasadena, CA 91125, USA}

\author{Gabriella P\'asztor\,\orcidlink{0000-0003-0707-9762}}
\affiliation{ELTE E\"otv\"os Lor\'and University, Budapest, Hungary}

\author{Eloi Pazos Rial}
\affiliation{Department of Physics, University of Cincinnati, Cincinnati, OH 45221, USA}

\author{Mar\'ia Pereira Mart\'inez\,\orcidlink{0009-0006-8577-9560}}
\affiliation{Instituto Galego de F\'isica de Altas Enerx\'ias (IGFAE), Universidade de Santiago de Compostela, Santiago de Compostela, Spain}

\author{Michael Peters\,\orcidlink{0009-0008-9089-1287}}
\affiliation{Department of Physics, University of Cincinnati, Cincinnati, OH 45221, USA}

\author{Jake Pfaller\,\orcidlink{0009-0009-8578-3078}}
\affiliation{Department of Physics, University of Cincinnati, Cincinnati, OH 45221, USA}

\author{Luca Pizzimento\,\orcidlink{0000-0002-1814-2758}}
\affiliation{University of Hong Kong, Hong Kong}

\author{M\'aximo Pl\'o Casas\'us\,\orcidlink{0000-0002-2289-918X}}
\affiliation{Instituto Galego de F\'isica de Altas Enerx\'ias (IGFAE), Universidade de Santiago de Compostela, Santiago de Compostela, Spain}

\author{Gian Andrea Rassati\,\orcidlink{0000-0002-7782-2155}}
\affiliation{Department of Physics, University of Cincinnati, Cincinnati, OH 45221, USA}

\author{Dean J.~Robinson\,\orcidlink{0000-0003-0057-1703}}
\affiliation{Physics Division, Lawrence Berkeley National Laboratory, Berkeley, CA 94720, USA}
\affiliation{Department of Physics, University of California, Berkeley, CA 94720, USA}

\author{Emilio Xos\'e Rodr\'iguez Fern\'andez\,\orcidlink{0000-0002-3040-065X}}
\affiliation{Instituto Galego de F\'isica de Altas Enerx\'ias (IGFAE), Universidade de Santiago de Compostela, Santiago de Compostela, Spain}

\author{Debashis Sahoo\,\orcidlink{0000-0002-5600-9413}}
\affiliation{ELTE E\"otv\"os Lor\'and University, Budapest, Hungary}

\author{Sinem Simsek\,\orcidlink{0000-0002-9650-3846}}
\affiliation{Istynie University, Istanbul, Turkey}

\author{Michael D.~Sokoloff\,\orcidlink{0000-0002-1468-0479}}
\affiliation{Department of Physics, University of Cincinnati, Cincinnati, OH 45221, USA}

\author{Joeal Subash\,\orcidlink{0000-0001-6431-6010}}
\affiliation{University of Birmingham, Birmingham, United Kingdom}

\author{Aditya Suresh\,\orcidlink{0000-0002-8085-2021}}
\affiliation{Physics Department, Stanford University, Stanford, CA 94305, USA}

\author{Paul Swallow\,\orcidlink{0000-0003-2751-8515}}
\affiliation{University of Birmingham, Birmingham, United Kingdom}

\author{James Swanson\,\orcidlink{0000-0002-5501-3867}}
\affiliation{Department of Physics, University of Cincinnati, Cincinnati, OH 45221, USA}

\author{Abhinaba Upadhyay\,\orcidlink{0009-0000-6052-6889}}
\affiliation{ELTE E\"otv\"os Lor\'and University, Budapest, Hungary}

\author{Riccardo Vari\,\orcidlink{0000-0002-2814-1337}}
\affiliation{INFN Sezione di Roma La Sapienza, Roma, Italy}

\author{Carlos V\'azquez~Sierra\,\orcidlink{0000-0002-5865-0677}}
\affiliation{Instituto Galego de F\'isica de Altas Enerx\'ias (IGFAE), Universidade de Santiago de Compostela, Santiago de Compostela, Spain}
\affiliation{Universidade da Coru\~na, A Coru\~na, Galicia, Spain}

\author{G\'abor Veres\,\orcidlink{0000-0002-5440-4356}}
\affiliation{ELTE E\"otv\"os Lor\'and University, Budapest, Hungary}

\author{Nigel Watson\,\orcidlink{0000-0002-8142-4678}}
\affiliation{University of Birmingham, Birmingham, United Kingdom}

\author{Michael K.~Wilkinson\,\orcidlink{0000-0001-6561-2145}}
\affiliation{Department of Physics, University of Cincinnati, Cincinnati, OH 45221, USA}

\author{Michael Williams\,\orcidlink{0000-0001-8285-3346}}
\affiliation{Laboratory for Nuclear Science, Massachusetts Institute of Technology, Cambridge, MA 02139, USA}

\author{Eleanor Winkler\,\orcidlink{0000-0002-5021-4002}}
\affiliation{Laboratory for Nuclear Science, Massachusetts Institute of Technology, Cambridge, MA 02139, USA}

\collaboration{CODEX-b collaboration}

\maketitle

\vspace*{\fill}

\begin{Abstract}
\begin{changemargin}{1cm}{1cm}
The \codexbeta apparatus is a demonstrator for the proposed future \codexb experiment, a long-lived-particle detector foreseen for operation at IP8 during HL-LHC data-taking. The demonstrator project, intended to collect data in 2025, is described, with a particular focus on the design, construction, and installation of the new apparatus.
\end{changemargin}
\end{Abstract}

\vspace*{\fill}

\vspace*{2.0cm}

\begin{center}
  Published in
  JINST 20 (2025) T07007
\end{center}

contacts: \href{mailto:philten@cern.ch}{philten@cern.ch} and \href{mailto:daniel.johnson@cern.ch}{daniel.johnson@cern.ch}

\clearpage

\tableofcontents
\clearpage

\section{Introduction}
\label{sec:Introduction}
Long-standing open questions in particle physics have motivated ongoing searches for physics Beyond the Standard Model (BSM). However, using conventional detectors, the LHC has not yet observed any direct evidence for New Physics (NP). These searches by the LHC have already begun to shift the theory landscape, ruling out well-known BSM extensions and requiring theoretical assumptions to be broadened. This has led to signature-based searches including decays-in-flight of Long-Lived Particles (LLP), which could emerge from any number of BSM scenarios involving hidden and dark sectors.

While the main LHC detectors can probe parts of the LLP parameter space, a new class of hermetic detectors that are both shielded and displaced transverse to the beam-line, are necessary for further coverage. These transverse detectors can search for LLPs with lighter masses (on the order of \mev to \gev) and long lifetimes (on the order of $c\tau > 10\m$). Additionally, the high parton center-of-mass energy available for transversely-produced states in LHC collisions allows these transverse detectors to access possible high-mass portals such as the Higgs boson or other mediators above the electroweak scale; currently, the limit on the invisible width of the Higgs is 13\%~\cite{Workman:2022ynf}. Consequently, this is a physics opportunity unique to the LHC and is of pressing interest to the physics community. The COmpact Detector for EXotics at \lhcb (\codexb) proposal is such a detector and utilizes the space around the \lhcb detector to build an economical yet competitive shielded transverse detector.
\subsection{Goals}
\label{subsec:Goals}
The \codexbeta detector is a demonstrator for the proposed full-scale \codexb detector. Its primary design goal is 
to validate the key concepts which justify the full design, construction, and operation of \codexb. Specifically:
\begin{enumerate}
    \item Validate the preliminary background estimates from the \codexb proposal and the background measurement campaign in 2018~\cite{Dey:2019vyo}. Demonstrate the ability to reconstruct known SM backgrounds expected to decay inside UX85A (the proposed location for \codexbeta and the nominal location of \codexb) and validate the agreement of the background rates with a full simulation of the LHCb detector and cavern environment. These studies will demonstrate that \codexb can be operated as a zero-background experiment.
    \item Demonstrate the seamless integration of the detector with the LHCb software-based trigger, so that any events of interest in \codexb can be tagged with the corresponding LHCb detector information to aid in their interpretation. This feature is unique to \codexb because LLP detectors linked to \atlas or \cms must, by necessity, implement hardware triggers. Transverse detectors are sensitive to many signatures that correspond to high activity proton-proton interactions with unique topologies, so such integration with a traditional detector is invaluable.
    \item Demonstrate the suitability of Resistive Plate Chambers (RPCs) as a baseline tracking technology for \codexb in terms of spatial granularity, hermeticity, and timing resolution, as well as establishing operational expertise within the \codexb team. 
    \item Demonstrate the suitability of the mechanical support required for these RPCs and its scalability to the full \codexb detector.
\end{enumerate}
The production and installation of \codexbeta will provide critical experience to the \codexb team, as well as ensure any possible integration issues with \lhcb are encountered in a controlled and low-risk/impact environment.

The timeline outlined in this document for the production and installation of \codexbeta supports these goals above. The urgency of this timeline is motivated by producing a viable prototype for a shielded and displaced transverse detector. The high luminosity LHC (HL-LHC) is a once-in-a-lifetime opportunity to search for LLPs, but to utilise the full HL-LHC dataset, a demonstration of the feasibility for near zero-background measurements by such a detector needs to happen in the very near future. Already \codexbeta can perform novel measurements both for specific new physics and SM measurements, such as kaon production. Furthermore, for any transverse detector, feasible large-scale instrumentation such as RPCs will be necessary. \codexbeta will allow the community to understand the performance of such detectors in this context. These background measurements will permit data-driven modeling of background simulations relevant to a wide range of LLP detector proposals, while the trigger integration will demonstrate how transverse detectors can be run in tandem with traditional detectors. To convince the community that IP8 remains the most plausible location for a zero-background transverse detector, a successful demonstration with \codexbeta is necessary.  Such a demonstration can lead to the convergence among proponents of transverse LLP detector proposals towards a detector located at IP8. Regardless of this specific outcome, achieving all these milestones will help validate the goal of building a transverse detector that runs during the HL-LHC.

The support of \codexbeta by \lhcb benefits the entire \lhcb collaboration, which is now recognised for physics well beyond the original design of the \lhcb detector. Nearly one third of the physics case for \lhcb Upgrade II~\cite{LHCb:2018roe} corresponds to this expanded physics portfolio. Some key measurements justify running \lhcb for the full HL-LHC, including dark photon searches, the charm Yukawa coupling of the Higgs, axion searches, and more. These capabilities are well known by the broader community, and are featured in the CERN HL-LHC Yellow Reports~\cite{CidVidal:2018eel,Cepeda:2019klc} and in a number of 2022 Snowmass summaries~\cite{Artuso:2022ouk,Bose:2022obr}. Already this physics programme has attracted significant funding and personnel, including many early career physicists. By supporting \codexbeta, the \lhcb collaboration is able to further contribute to the BSM community with very little cost and significant benefits, including: attracting new collaborators, institutes, funding agencies, and countries; supporting current collaborators; generating good will in the wider high-energy physics community; enhancing \lhcb's reputation for leadership; and most importantly, contributing to important physics goals and results.

The remainder of this document addresses how the proposed \codexbeta demonstrator will achieve the enumerated goals above. In addition to these goals, the \codexbeta demonstrator may be able to set first limits on certain BSM signatures, as discussed further in the next section.
\subsection{Physics Reach}
\label{subsec:PhysicsReach}
Residual neutron scattering inside the decay volume is one of the most important backgrounds 
identified in the \codexb proposal~\cite{Gligorov:2017nwh, Aielli:2019ivi}.
In addition, a second important background arises  from decays of muon-scattering-produced secondaries, such as $K^0_L$ mesons, produced by strange muoproduction in shielding elements.
These backgrounds require a combination of active vetoes and passive shielding to be reducible.
Accurate and fully data-driven estimates of the backgrounds present in the proposed \codexb installation site 
are crucial to the design of these passive and active shielding elements,
allowing them to be optimized to be as small and inexpensive as possible. 

During Run~2, preliminary background measurements were performed in the UX85A hall using a pair of scintillators~\cite{Dey:2019vyo}.
However, these were simply measurements of hit counts---particle trajectories were not reconstructed---and 
were used to approximately cross-check the expectations from numerical simulations of the background~\cite{Aielli:2019ivi}.
The \codexbeta demonstrator detector will, by contrast, allow tracking of particles with similar performance to the full detector
and, in particular, allow for the separation of particle backgrounds produced outside the decay volume from backgrounds 
induced by particle scattering inside the decay volume itself.
This includes allowing for the study of potential soft charged particles that could be bent around the \codexb shield by LHCb's magnet ``focusing''  and produce background within the detector~\cite{Dey:2019vyo}.

Aside from measuring background levels, observing long-lived SM particles decaying inside the detector acceptance
will allow us to calibrate the detector reconstruction and the RPC timing resolution.
The most natural candidates are \KL mesons.
Table~\ref{tab:bkg-tracks} summarizes the expected multitrack production from decay or scattering on air by neutral fluxes entering \codexbeta, as well as the contribution from just \KL mesons entering the detector.
A $\text{few} \times 10^7$ \KL decays to two or more tracks are expected in the \codexbeta volume in $15\invfb$ of data taking in Run~3,
so that \codexbeta will have the opportunity to reconstruct a variety of \KL decays. 
Measurements of the decay vertices and decay product trajectories will allow the boost of the LLPs to be reconstructed independently from time-of-flight information~\cite{Curtin:2017izq}.
Moreover, measurement of the distribution of \KL decay vertices can be compared to expectations from the background simulations of the expected \KL boost distribution 
combined with  the \KL decay probability, allowing calibration and validation of our detector simulation and reconstruction. 
Conversely, one may possibly combine the predicted boost distribution and measured vertex distribution to extract the \KL lifetime itself.

\begin{table*}[tbp]
\renewcommand*{\arraystretch}{0.9}
\newcolumntype{C}{ >{\centering\arraybackslash $} m{3.25cm} <{$}}
\newcolumntype{E}{ >{\centering\arraybackslash $} c <{$}}
\begin{tabular*}{0.6\linewidth}{@{\extracolsep{\fill}}E|CC}
\hline
\text{Tracks} & \text{Total} & \KL~\text{contribution}\\
\hline\hline
1	&		(3.9 \pm 0.1) \times 10^{8}	&	(2.9 \pm 0.1) \times 10^{8}	 \\
2	&		(4.1 \pm 0.1) \times 10^{7}	&	(3.7 \pm 0.1) \times 10^{7}	 \\
3	&		(6 \pm 1) \times 10^{5}	&	(2.9 \pm 0.4) \times 10^{5}	 \\
4+	&		(9 \pm 2) \times 10^{4}	&	(7 \pm 2) \times 10^{4}	 \\
\hline
\end{tabular*}
\caption{Total neutral-initiated (together with charged-initiated after a $10^{-4}$ rejection factor from the front face veto) and \KL-initiated multitrack production on air during Run~3 inside the \codexbeta volume for total integrated luminosity $\mathcal{L} = 15\invfb$, 
requiring $E_{\text{kin}} > 0.4\gev$ per track~\cite{Aielli:2019ivi}.}
\label{tab:bkg-tracks}
\end{table*}

The acceptance of \codexbeta is roughly $1\%$ that of the full \codexb detector~\cite{Aielli:2019ivi}, 
and no shielding beyond the existing UX85 concrete wall will be in place;
its reach for physics beyond the Standard Model (BSM) is therefore limited.
Roughly $10^{13}$ \bquark-hadrons will be produced at IP8 during Run~3, however, 
which will allow \codexbeta to probe some novel regions of parameter space; specifically, those where
the branching ratio of $b$-hadron decays to LLPs is independent of the LLP lifetimes.
Further, as shown in Tab.~\ref{tab:bkg-tracks}, because the multitrack background falls relatively fast with the number of tracks, 
a potential LLP signal region for \codexbeta can be defined as those events with $4$ or more reconstructed tracks
(requiring $E_{\text{kin}}>0.4\gev$ per track from expected minimum tracking requirements).

This type of signal morphology can arise, for example, in models that address the baryogenesis puzzle, 
in which one searches for $B \to X_s \chi$ decays, where $X_s$ is a SM (multi)hadronic state with baryon number $\pm1$,
and $\chi$ is an LLP decaying to an (anti-)baryon plus a number of light mesons.
Though the Run~3 reach for LHCb itself for such a model exceeds that of \codexbeta by an order of magnitude~\cite{Aielli:2019ivi},
\codexbeta may be able to set a first limit well before the completion of a (more complicated) LHCb study.
Further, in such a portal, the full \codexb experiment will outperform LHCb, 
so examining these multitrack signatures would provide an early proof-of-concept for probing these types of models.

\clearpage

\section{Detector Design}
\label{sec:DetectorDesign}
\label{sec:detectordesign}

\begin{figure}[tbp]
    \begin{center}
        \includegraphics[width=0.45\textwidth]{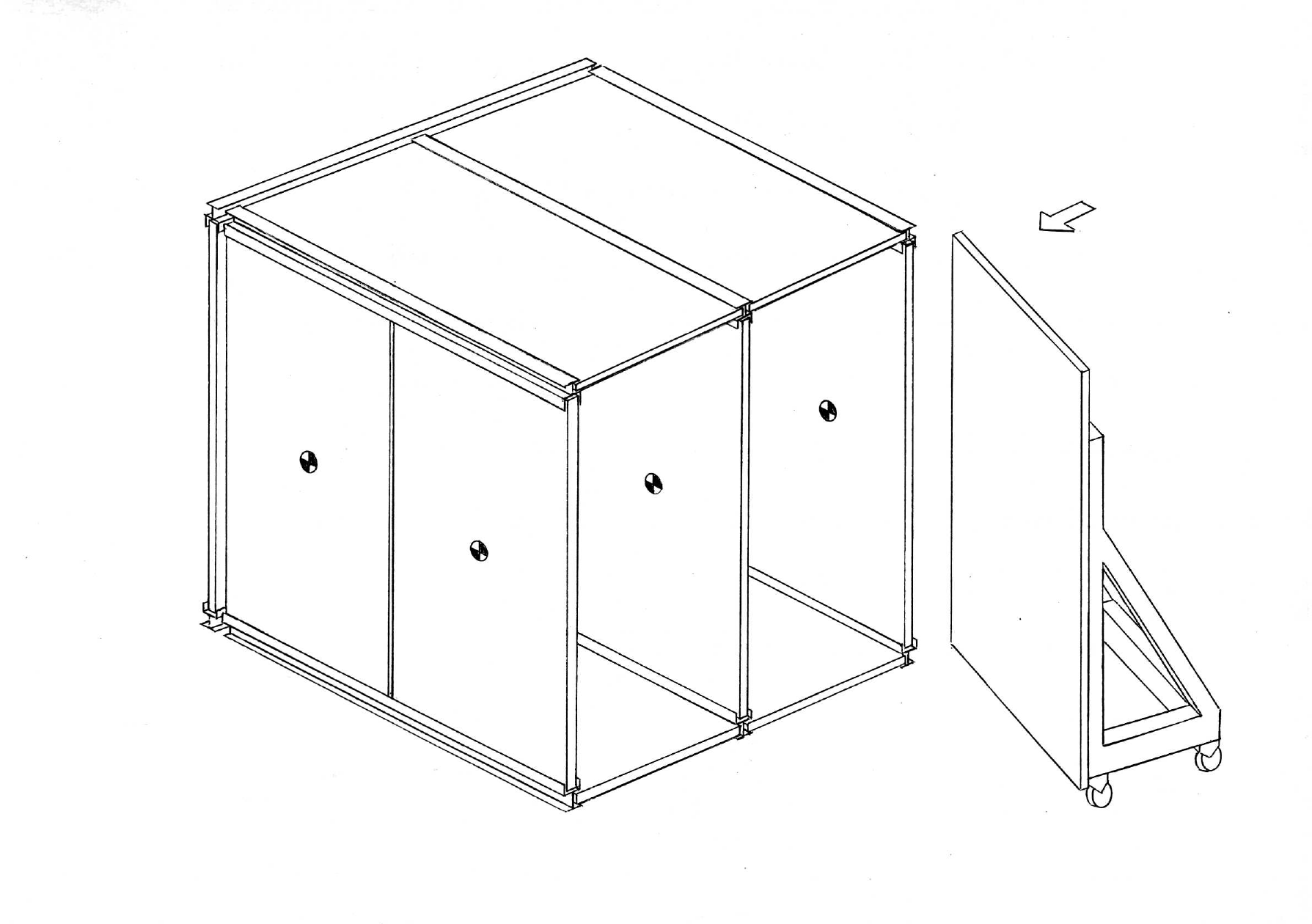}
        \includegraphics[width=0.45\textwidth]{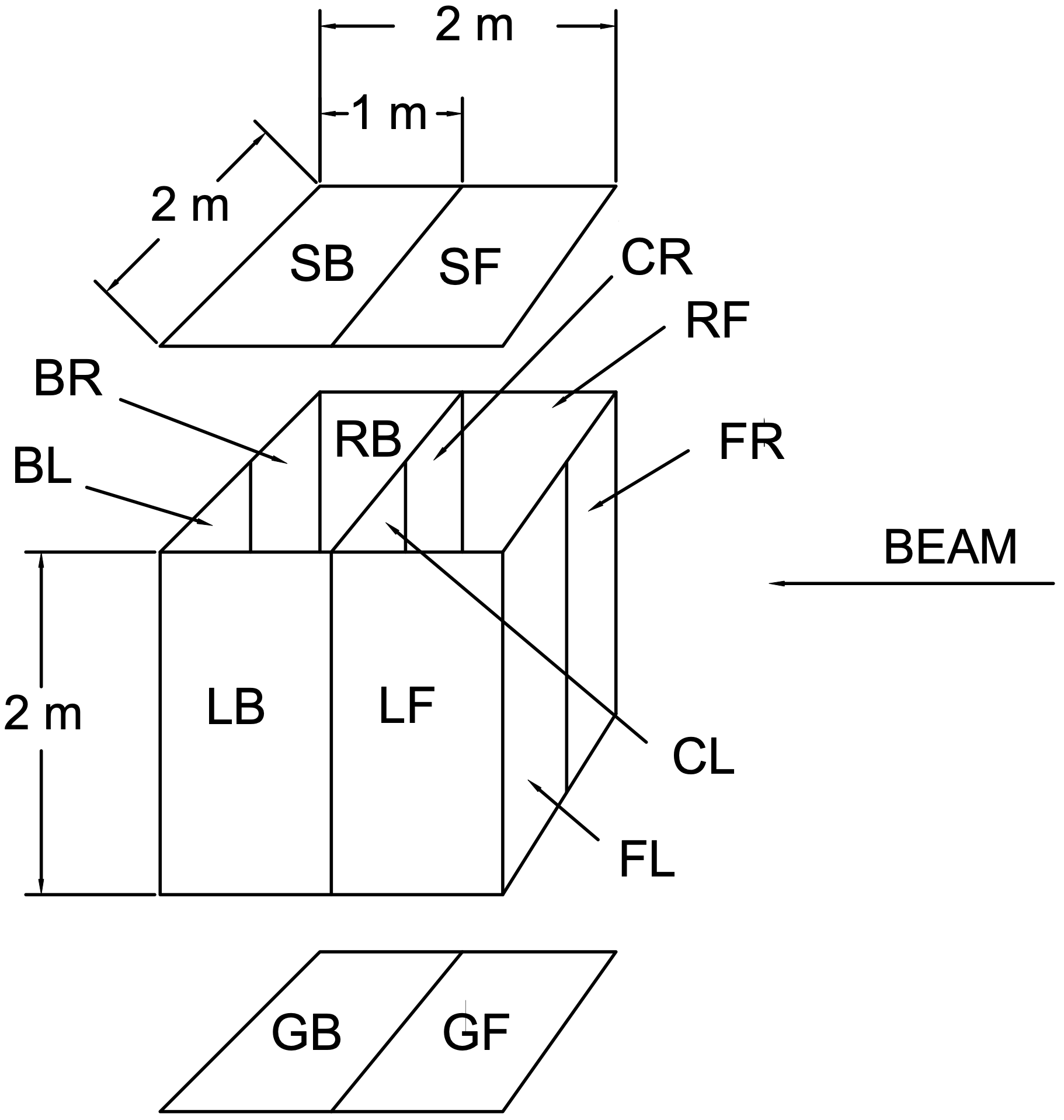}
    \end{center}
    \caption{
        Two sketches of \codexbeta from two perspectives.
        (Left) a diagram showing the mechanical framework for the RPC modules; the arrow indicates the direction of incoming particles; module LB, which will also be on a rolling cart, is not shown.
        (Right) an exploded sketch of \codexbeta with module labels and dimensions; the arrow labeled ``BEAM'' indicates the direction of incoming particles, corresponding to the arrow on the left diagram; note that this diagram shows \codexbeta from a different perspective than the one on the left.
    }
    \label{fig:demosketch}
\end{figure}

As shown in Fig.~\ref{fig:demosketch}, \codexbeta will comprise a 2\m cube with an additional face spanning the interior.
Each face will contain two modules, which will each house a stacked triplet of $2\times 1\ma$ RPCs, requiring a total of $(6+1) \times 2 \times 3 = 42$ RPCs housed in 14 modules.

Each module is described by a two-letter name.
The first letter indicates its face (Front, Center, Back, Left, Right, Ground, or Sky) and the second its position within that face (Front, Back, Left, or Right).
Figure~\ref{fig:demosketch} summarizes the relative positions of each module.

\subsection{Requirements and Operation Objectives}
\label{subsec:RequirementsAndOperationObjectives}
Beyond its raw geometric acceptance, the physics reach of the full \codexb detector is determined mainly by its signal efficiency and the ability to separate this signal from any background.
Considering the broad, inclusive searches considered, the best strategy for background separation is its suppression (both passive suppression and active veto) to a negligible level, so that any detection of a synchronous particle from \lhcb can be investigated closely.
One of the key objectives of \codexbeta will be to provide more information about the distributions and production rates of background particles. These particles are mainly
\begin{itemize}
    \item neutrons, as they are difficult to  track and can produce multitrack LLP-like events from scattering on air;
    \item \KL mesons, for the same reasons as neutrons, because they are muoproduced secondaries, and because they decay with topologies mimicking LLPs;
    \item neutrinos, which are produced in such quantities at LHCb that they need to be considered, including rare neutrino scattering on shielding;
    \item muons, which are difficult to absorb and can also create secondary particles close to \codexb.
\end{itemize}

\subsubsection{Principle of the \codexb shield}

\begin{figure}[tbp]
    \centering
    \includegraphics[width=0.80\textwidth]{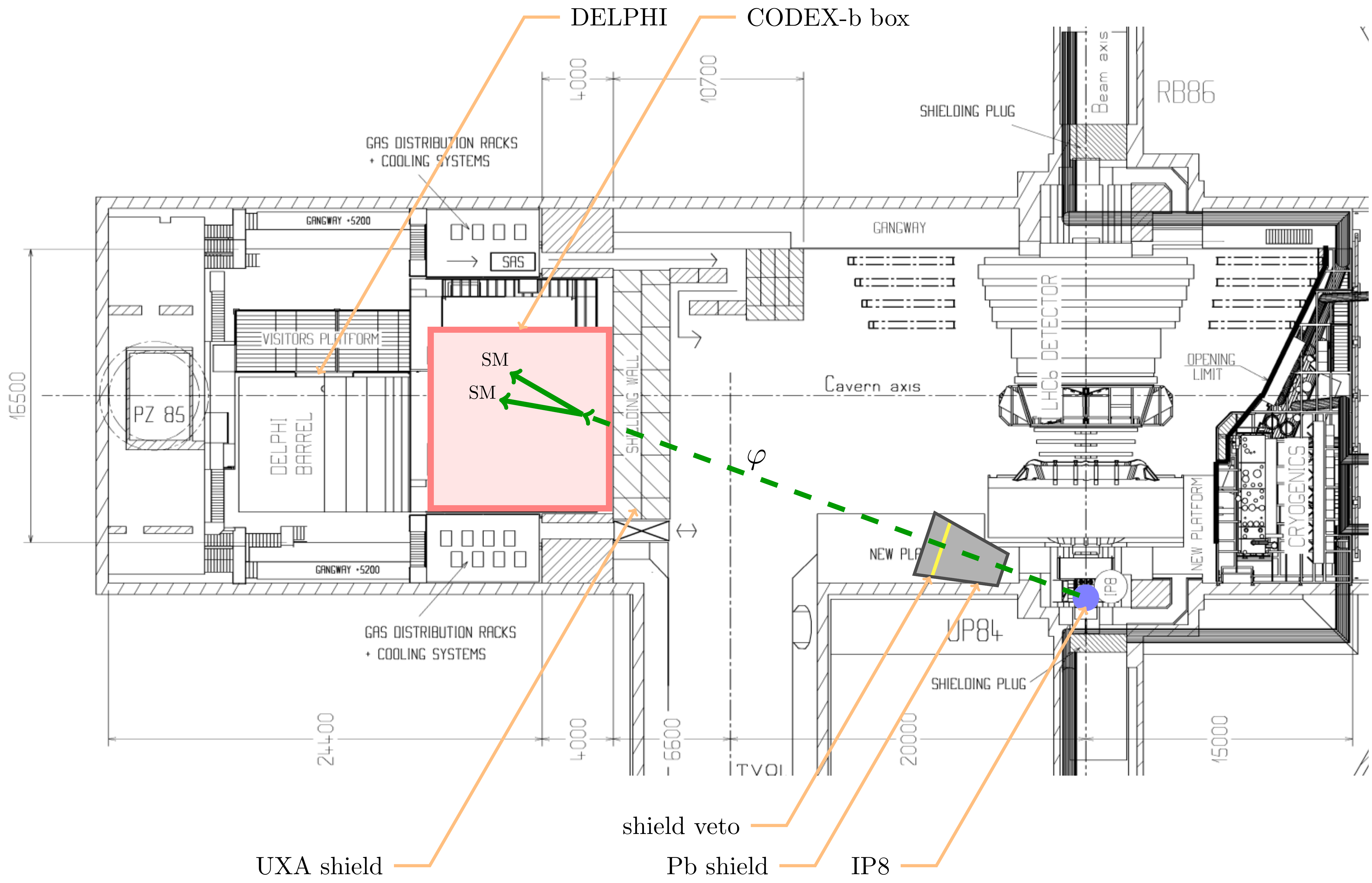}
    \caption{Positions of the LHCb detector and the \codexb volume, adapted from Ref.~\cite{EDMSUX}. For \codexbeta, the lead shield and active veto will not be present, leaving only the $\approx 7\,\lambda$ shielding provided by the UXA shield.}
    \label{fig:LHCbCavern}
\end{figure}

The total expected number of neutrons and \KL mesons in the \codexb acceptance, absent any shielding, is around $10^{14}/300\invfb$.
To reduce this flux to less than 1 expected particle (modulo kinematic cuts), we need to use a $32\,\lambda$-thick hadronic shield.
Of these interaction lengths, 7 are already provided by the concrete wall, meaning a $25\,\lambda$ lead shield would be installed along the axis from the LHCb IP to \codexb.
To reduce the required volume of this shield, it would be installed close to the IP, as shown in Fig.~\ref{fig:LHCbCavern}.

Additionally, muons slowly stopping in the lead absorber can create a \KL meson or a neutron, which then could not be absorbed by the smaller amount of shielding left.
We call this background ``stopped-parent secondaries''.
To address this background, we would introduce a layer of active material inside the shield.
This active material is yet to be determined, but current simulations use a silicon plane with a $2\cma$ granularity~\cite{Aielli:2019ivi}.
It would  then be placed 20 interaction lengths inside the shield, so that most particles would have already stopped and that the veto would only remove those rare muons that decay far enough into the shield that their decay products could reach the detector.

\subsubsection{Role of \codexbeta}

The base design of \codexbeta is shown in Fig.~\ref{fig:demosketch}
and described at the beginning of Sec.~\ref{sec:detectordesign}.
Since \codexb searches are extremely dependent on the precision of the background veto, the main role of \codexbeta will be to validate the flux predictions in the presence of the concrete wall.
These predictions are based on hypotheses of the primary fluxes, the material budget, and the propagation of these particles in said budget.
In the absence of PID and calorimetry, only absolute fluxes will be measured as well as angular dependencies.
It is however possible to consider reconstructing \KL mesons to separate them from other particles.
This represents an opportunity to develop the whole data acquisition chain as well as the reconstruction algorithms.
For a more detailed discussion, see Sec.~\ref{subsec:PhysicsReach}.

\subsection{RPC Design}
\label{subsec:RPCDesign}
\begin{figure}[tbp]
    \centering
    \includegraphics[scale=0.4]{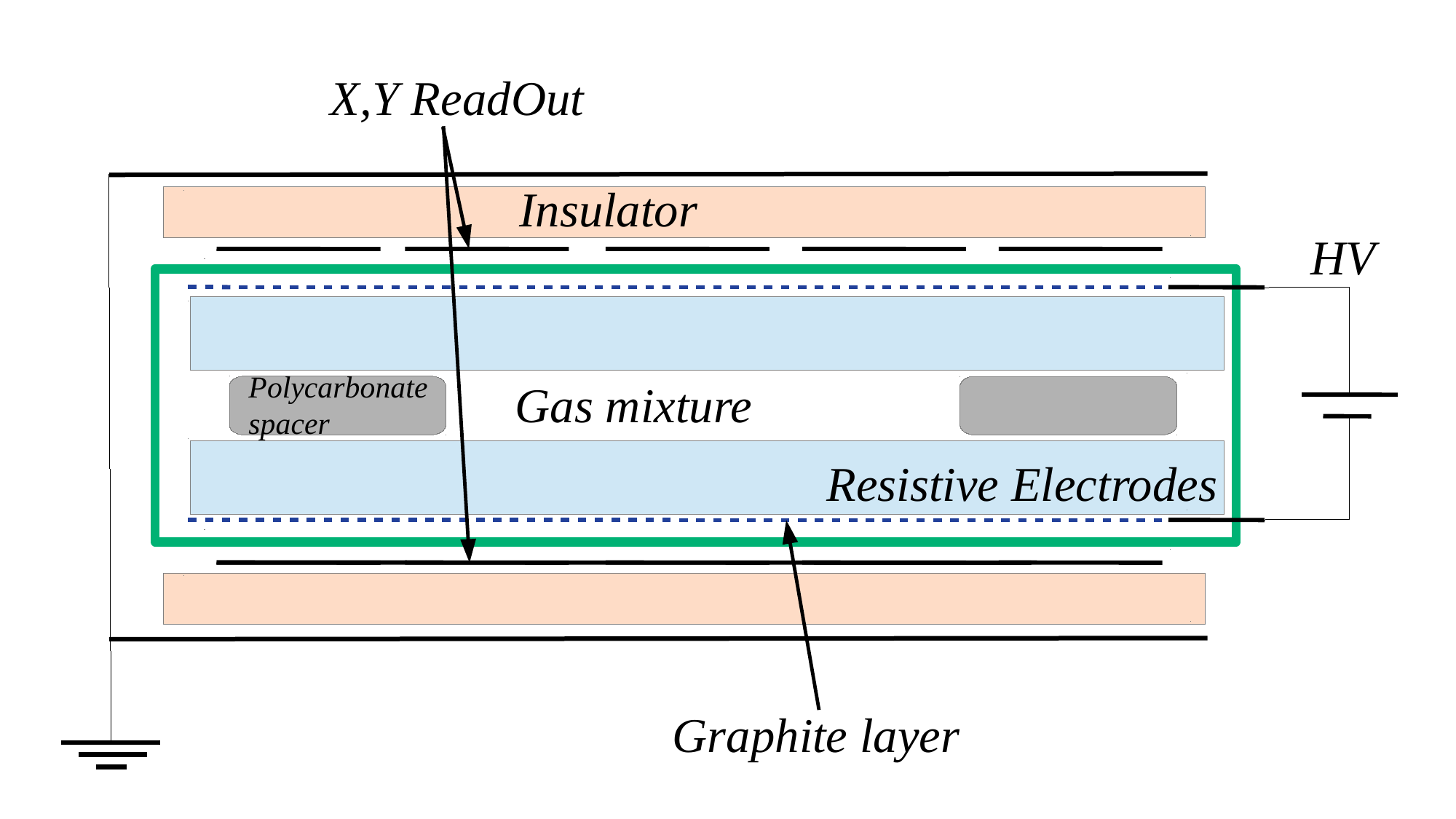} 
    \caption{Schematic of the structure of a singlet RPC, taken from \cite{CERN-LHCC-2017-020}.}
    \label{fig:rpc}
\end{figure}

The CODEX-$\beta$ RPC design is based on the one of the BIS78 project~\cite{Massa:2020hjw}. The entire development performed  in the BIS78 project in terms of gas gap, detector structure, front-end electronics, Faraday cage and detector-electronics integration have been inherited by the CODEX-$\beta$ experiment.  
Each RPC singlet has two panels of orthogonal strips, with strip pitches of $20$--$25\mm$, providing two orthogonal coordinates (see Fig.~\ref{fig:rpc}).
The Faraday cage design of the singlet, suitable for low-threshold operation, has been developed to allow a better shielding of the more sensitive front-end electronics.

The RPC singlet is the base element of the  modular design of \codexbeta and comprises a gas gap sandwiched between two orthogonal readout-strip panels containing the front-end electronics.
Fig.~\ref{fig:rpc} shows a schematic of the singlet RPC design.
A ``module'' is composed of three independent singlets, comprising a triplet, as well as the on-module services and support structures required for installation. The use of triplet RPCs allows the modules to work in self-trigger mode if necessary, avoiding any external reference system for the particle detection. This is not the preferred mode of operation, but is a possible fall back if integration with the \lhcb DAQ system is not possible.

The resistive electrodes consist of two sheets of phenolic  High  Pressure  Laminate (HPL), with resistivity $\rho$ between $10^{10}$ and $10^{11}\ohm\cm$. The electrode thickness is $\approx 1.2\mm$. A matrix of polycarbonate spacers, of $\approx 10\mm$ diameter each, guarantees the uniformity of the gas gap thickness for the entire detector area. The spacer matrix causes an intrinsic inefficiency of the detector which must be taken into account. The spacer separation is $7\cm$, causing a geometric inefficiency of $\approx 1\%$.

The external surface of the electrodes is covered with a paint of graphite with a superficial resistivity $\rho$ of $\approx500\kohm/\square$, which allows a uniform distribution of the high voltage and at the same time allows the inductive pick-up of the charge created within the gas.
The inner surfaces of the electrodes are coated with a few \mum thick layer of polymerized linseed oil to avoid the spike effect, which creates electrical field inhomogeneity and increases the detector noise. This is a fundamental component for the detector to work properly.

Readout strips, separated from the graphite by an insulating material, are used to read the electron-induced signal. The front-end electronics and the readout panels are then integrated with the gas gap within a Faraday cage structure, which suppresses most of the external noise.

The gas mixture which shows the best performance at the moment is $94.7\%$ R-134a (1,1,1,2 tetrafluoroethane), $5\%$ isobutane, and $0.3\%$ SF6 (sulfur hexafluoride).
Tetrafluorethane is characterized by high density and relatively low operating voltage.
Isobutane is a hydrocarbon which acts as streamer quencher, absorbing the UV photons.
The addition into the gas mixture of the sulfur hexafluoride causes a net separation between the avalanche and the streamer mode, allowing a better control on the working regime.
The strong electronegative property of SF6 allows for electron capture during the avalanche formation, leading to a huge suppression of the streamer.
The avalanche mode operation is preferable over the streamer mode due to the huge difference in the average charge per count produced between the two processes and its effect on the detector rate capability and ageing. 

The last point related to the RPC structure which must be stressed is the relationship between gas-gap thickness and time resolution.
The RPC suffers from a contribution to the time resolution given by the statistics of the saturated avalanche development.
The most effective way to reduce the contribution to time resolution given by the saturated avalanche development is to reduce the signal induction time.
A faster signal development reduces its contribution to time resolution.
Experimental observations show that the most direct approach allowing an improvement of the RPC time resolution is the reduction of the gas gap thickness.
The gas-gap width of $1\mm$ permits a time resolution of $\approx400\ps$.
However, a gas-gap thickness reduction comes along with a decrease of the pulse duration.
To exploit such faster signals, very sensitive and high-performing front-end electronics are required.
The preamplifier developed for such a purpose should perform an ideal integration of the pulse, down to a few hundreds picoseconds duration, without deteriorating the rise time of the output signal.

The developed front-end electronics board is composed of eight channels of a new preamplifier coupled with two full-custom ASIC discriminators with four channels each and Low-Voltage Differential Signaling (LVDS) transmitters integrated directly inside the board.
The overall features of the preamplifier and of the ASIC discriminator are reported in Tab.~\ref{tab:FEproperties}.
The preamplifier is described in more detail in Sec.~\ref{subsec:preamplifier},
and the full-custom ASIC discriminator is described in Sec.~\ref{subsec:ASIC}.

\begin{table}[tbp]
\begin{center}
    \caption{Overall properties of preamplifier and full-custom ASIC discriminator.}
    \begin{tabular}{|l|l|}
        \hline
        \multicolumn{2}{|l|}{\textbf{Preamplifier Properties}} \\
        \hline
        Si standard component&\\
        \hline
        Preamplifier sensitivity & 0.2-0.4 mV/fC  \\
        Power Consumption &3-5 V   0.5–1 mA\\
        Bandwidth & 100 MHz\\
        \hline
    \end{tabular}
    \begin{tabular}{|l|l|}
        \hline
        \multicolumn{2}{|l|}{\textbf{Discriminator Properties}} \\
        \hline
        SiGe full custom&\\
        \hline
        Discrimination Threshold &0.5 mV\\
        Power Consumption & 2-3 V   4-5 mA \\
        Bandwidth & 100 MHz\\
        \hline
    \end{tabular}
    \label{tab:FEproperties}  
\end{center}
\end{table}

\subsubsection{Discrete components preamplifier}
\label{subsec:preamplifier}
\begin{figure}[tbp]
\centering
    \includegraphics[width=0.55\textwidth]{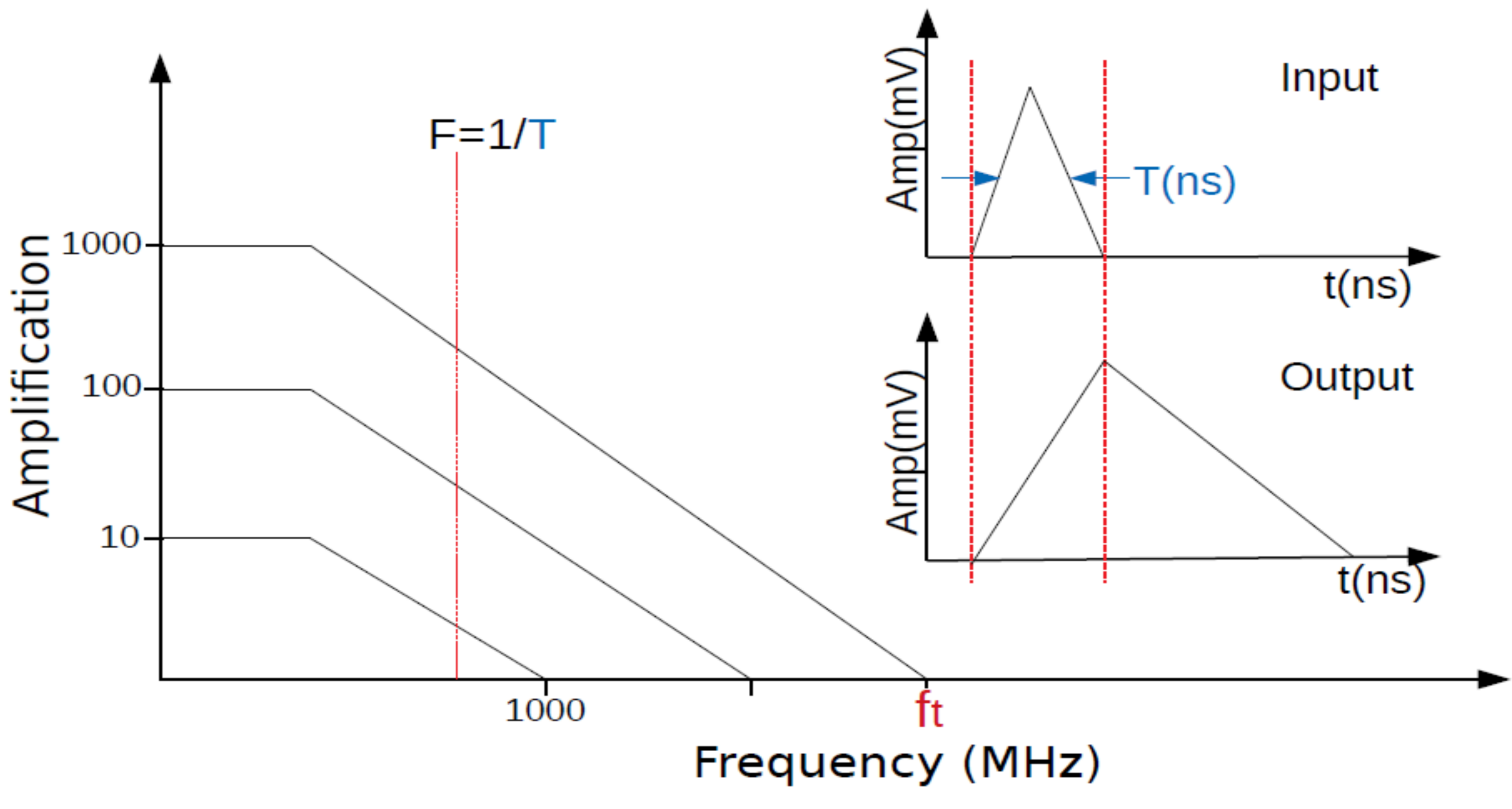} \\
    \includegraphics[width=0.65\textwidth]{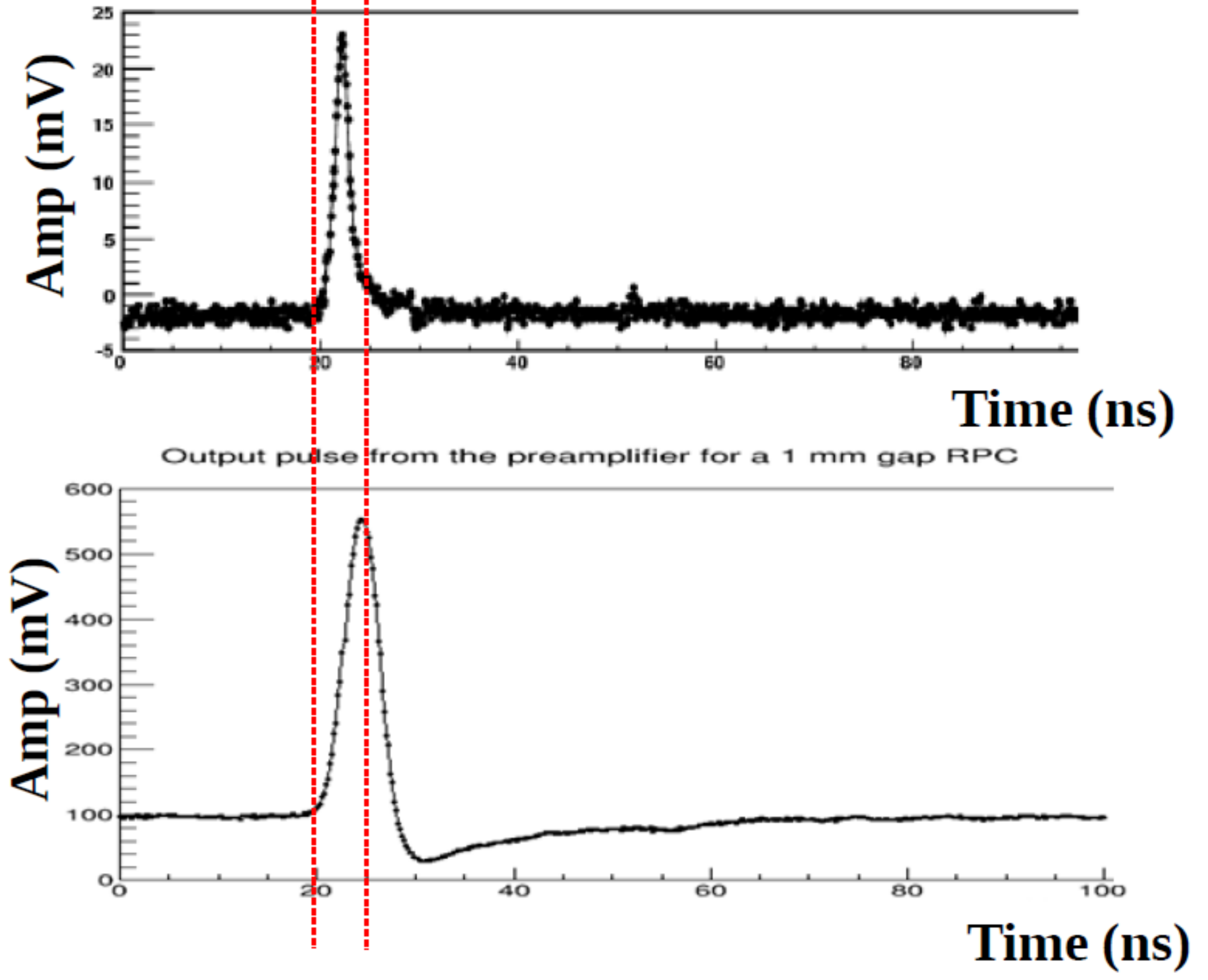}
    \caption{Working principle of the preamplifier, taken from \cite{CERN-LHCC-2017-020}. The plot on top shows the working principle of the fast charge BJT transimpedance amplifier realized for the RPC detector. It shows how the charge integration is achieved by working outside of the band of the circuit. The two plots below are an example of a not-amplified pulse from a 1 mm gap RPC as seen with a current readout (on top) and with the custom charge preamplifier (on bottom)}
    \label{fig:funzamp1}
\end{figure}

The preamplifier developed in the INFN laboratory of Rome Tor Vergata for RPCs is made using Silicon Bipolar Junction Transistor (BJT) technology. The main feature of this new kind of preamplifier is a fast charge integration with the possibility to match the input impedance to a transmission line. The working principle of this amplifier is shown in Fig.~\ref{fig:funzamp1}.

Figure~\ref{fig:funzamp1} shows how the injected signal is integrated. The output of the preamplifier is a signal which has the rise time and the maximum amplitude directly proportional to the duration of the injected signal and its charge. The fall time is an intrinsic parameter of the preamplifier; it is not related with the physical injected signal but depends only on the constants of the preamplifier.
The detailed properties of the silicon BJT preamplifier are shown in Tab.~\ref{tab:propertiesampl}.

\begin{table}[tbp]
\begin{center}
    \caption{Detailed properties of the preamplifier.}
    \begin{tabular}{|l|l|}
        \hline
        \multicolumn{2}{|l|}{\textbf{Amplifier Properties}} \\
        \hline
        Voltage supply&3-5 V\\
        \hline
        Sensitivity & 0.2-0.4 mV/fC  \\
        \hline
        Noise (independent from detector)& 1000 $e^{-}$ RMS\\
        \hline
        Input impedance &100-50 Ohm\\
        \hline
        Bandwidth & 10 - 100 MHz\\
        \hline
        Power Consumption & 5 mW/ch\\
        \hline
        Rise time $\delta (t) $ input& 300-600 ps\\
        \hline
        Radiation hardness & 1 Mrad, $10^{13}$ n cm$^{-2}$\\
        \hline
    \end{tabular}
    \label{tab:propertiesampl}
\end{center}
\end{table}

\subsubsection{Full-custom ASIC discriminator}
\label{subsec:ASIC}
The new full-custom discriminator circuit dedicated to the RPC detector for high rate environments uses silicon-germanium (SiGe) Hetero-junction Bipolar Transistor (HBT) technology.
The principle of SiGe HBT is to introduce a silicon-germanium impurity in the base of the transistor.
The advantage of this device is that the band structure introduces a drift field for electrons into the base of the transistor, thus producing a ballistic effect that reduces the base transit time of the carriers injected in the collector.
The net effect is  to improve the transition frequency and to introduce a directionality in the charge transport, allowing a much lower value of Base-Collector capacitance; hence a much higher charge amplification can be achieved.

\begin{figure}[tbp]
\centering
    \includegraphics[height=0.4\paperheight]{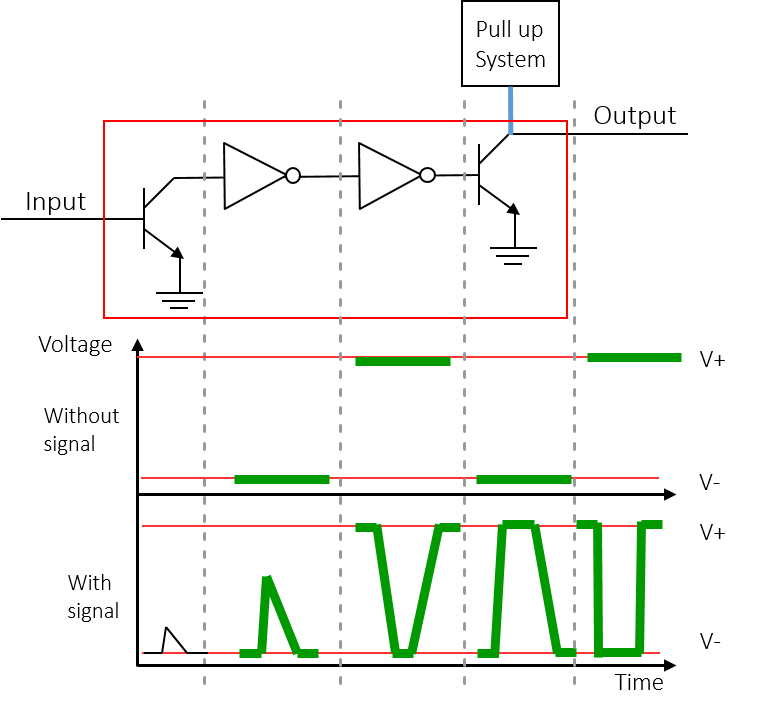}
    \caption{Working principle of the discriminator, taken from \cite{CERN-LHCC-2017-020}. (Top) the conceptual scheme of the limit amplifier discriminator, composed by a first stage of amplification by means of a transimpedance amplifier and three saturation stages, represented by the inverters. (Bottom) the output of each stage, showing how the signal is amplified until saturation leading to a digital signal in output.}
    \label{fig:discriminator}
\end{figure}

The main idea behind this new discriminator is the limit amplifier.
If the signal surpasses the threshold, it will be amplified until saturation, giving as output a square wave (see Fig.~\ref{fig:discriminator}).
Moreover, the Full Width Half Maximum of the discriminator output is proportional to the time that the input signal stays over the threshold, allowing the time over threshold to be measured directly with the discriminator.
The SiGe HBT technology is particularly suitable for this kind of working principle, since the performance of this kind of discriminator is mainly determined by the switching frequency of the transistor and its $\beta$.

\begin{figure}[tbp]
\centering
    \includegraphics[width=0.75\textwidth]{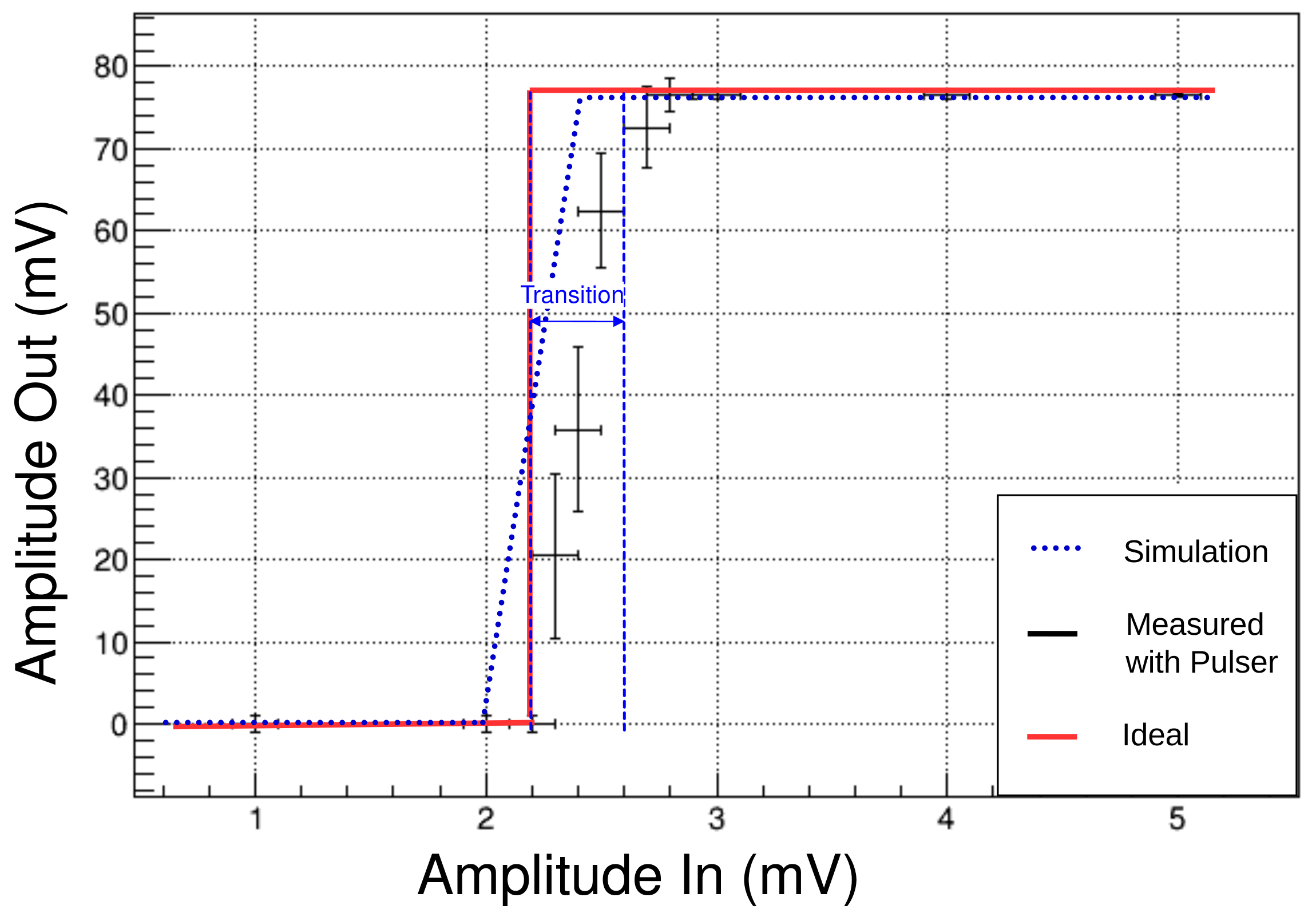}
    \caption{Characteristic function of the discriminator showing the logic transition from 0 to 1 and the length of the transition region, taken from \cite{CERN-LHCC-2017-020}.}
    \label{fig:funzcar}
\end{figure}

\begin{figure}[tbp]
\centering
    \includegraphics[width=0.75\textwidth]{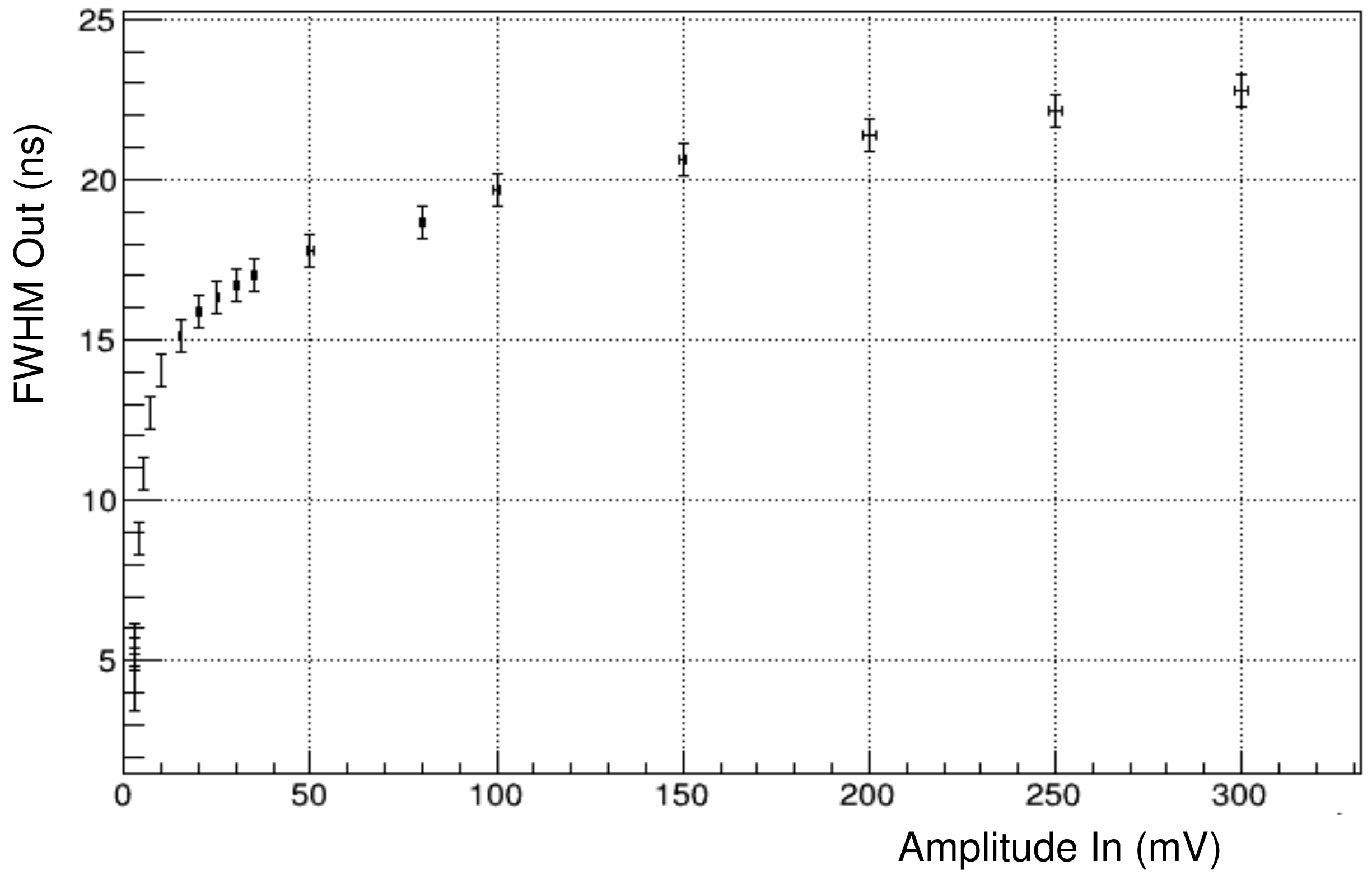}
    \caption{Calibration curve of the Time-Over-Threshold measurement realized by the discriminator, taken from \cite{CERN-LHCC-2017-020}. The amplitude of the input signal is varied checking the correlation with the width of the output signal.}
    \label{fig:disctot}
\end{figure}

\begin{figure}[tbp]
\centering
    \includegraphics[width=0.75\textwidth]{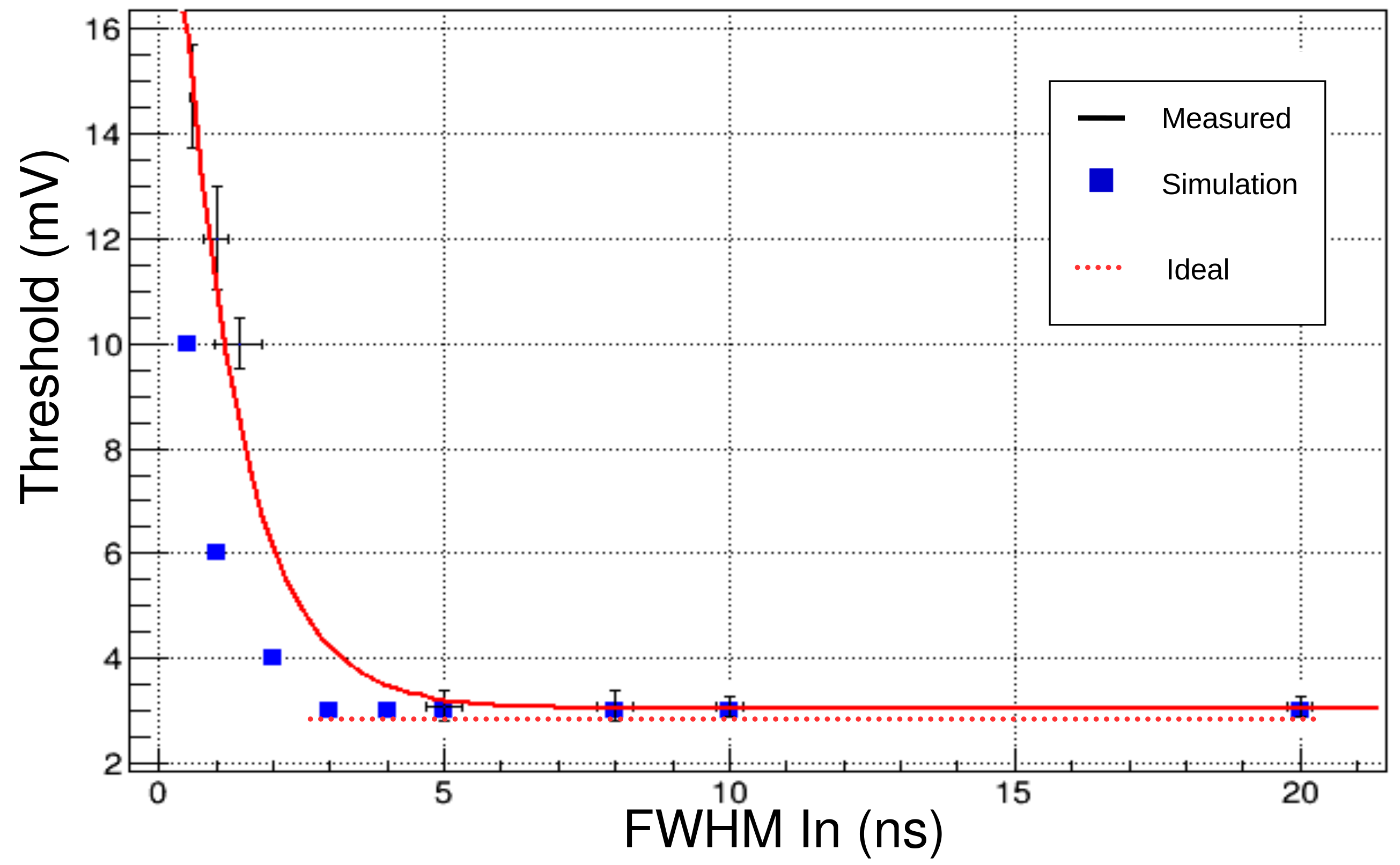}
    \caption{Calibration curve of the minimum injectable signal duration in order to have a linear working regime, taken from \cite{CERN-LHCC-2017-020}.}
    \label{fig:discminsig}
\end{figure}

The main features of this kind of discriminator are:
\begin{itemize}
    \item It provides the optimal characteristic function for the RPC. The characteristic function of the discriminator is reported in Fig.~\ref{fig:funzcar}. It shows a very small logic state transition of around $300\muV$, which becomes practically negligible when this discriminator is operated within the RPC detector and its charge distribution is taken into account. Moreover, the threshold can be easily regulated with just a few\mV.
    \item The time-over-threshold measurement can be performed directly with the discriminator. The calibration curve of this measurement is reported in Fig.~\ref{fig:disctot}.
    \item It has a minimum pulse width of $3\ns$. This is one of the most important parameters for a discriminator since it defines the shortest signal which can be injected inside the discriminator to keep the working regime linear (see Fig.~\ref{fig:discminsig}).
\end{itemize}

\clearpage
\subsection{RPC Layout}
\label{subsec:RPCLayout}
\begin{figure}[tbp]
    \begin{center}
    \includegraphics[width=\textwidth]{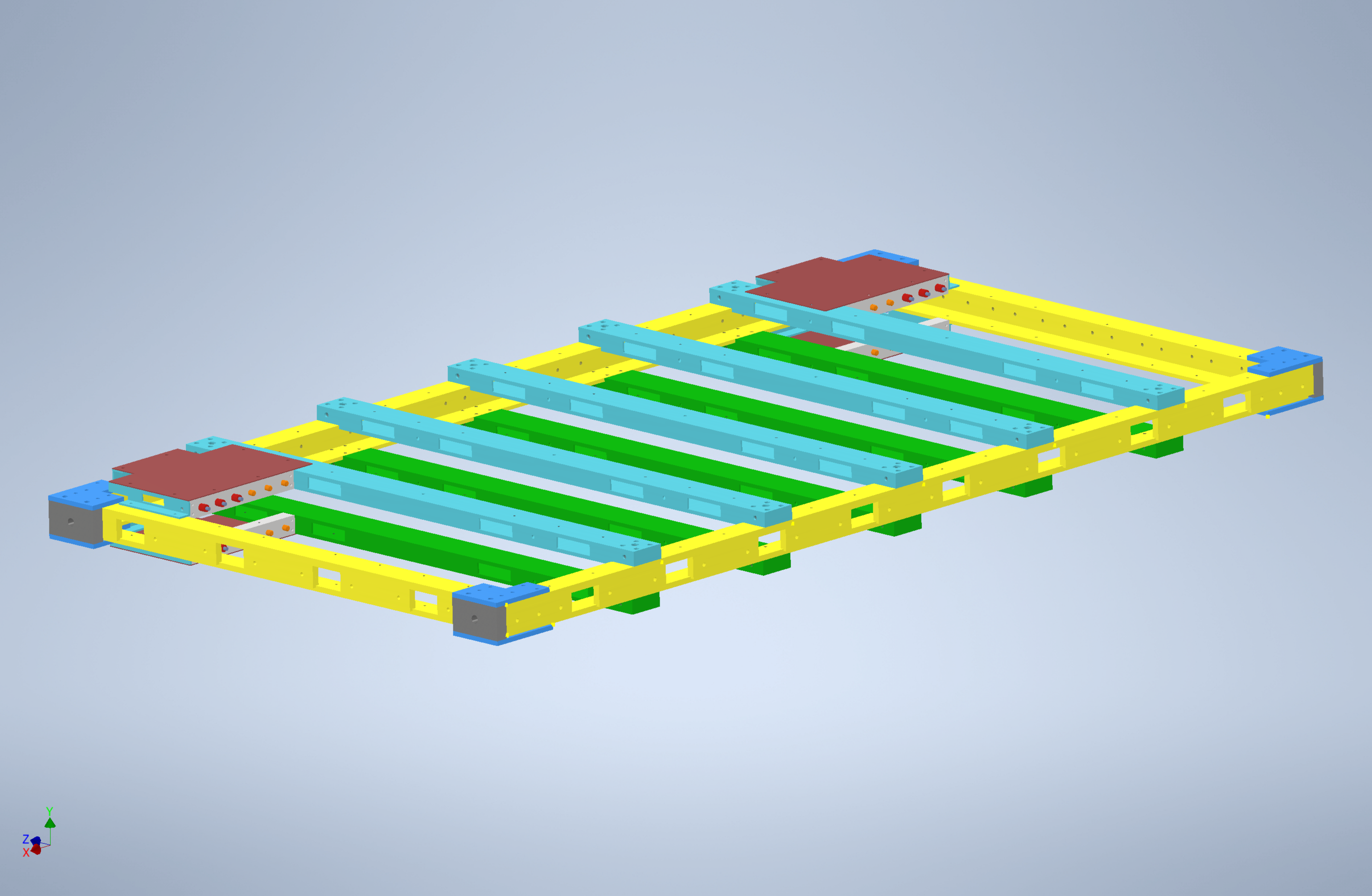}
    \end{center}
    \caption{Diagram of the frame that supports the RPC triplet within each module. Its design is referred to as ``CX1''. It is entirely composed of aluminum (and stainless steel fasteners); the colors shown are only to help visually distinguish certain components.}
    \label{fig:moduleFrame}
\end{figure}

\begin{figure}[tbp]
    \begin{center}
    \includegraphics[width=0.75\textwidth]{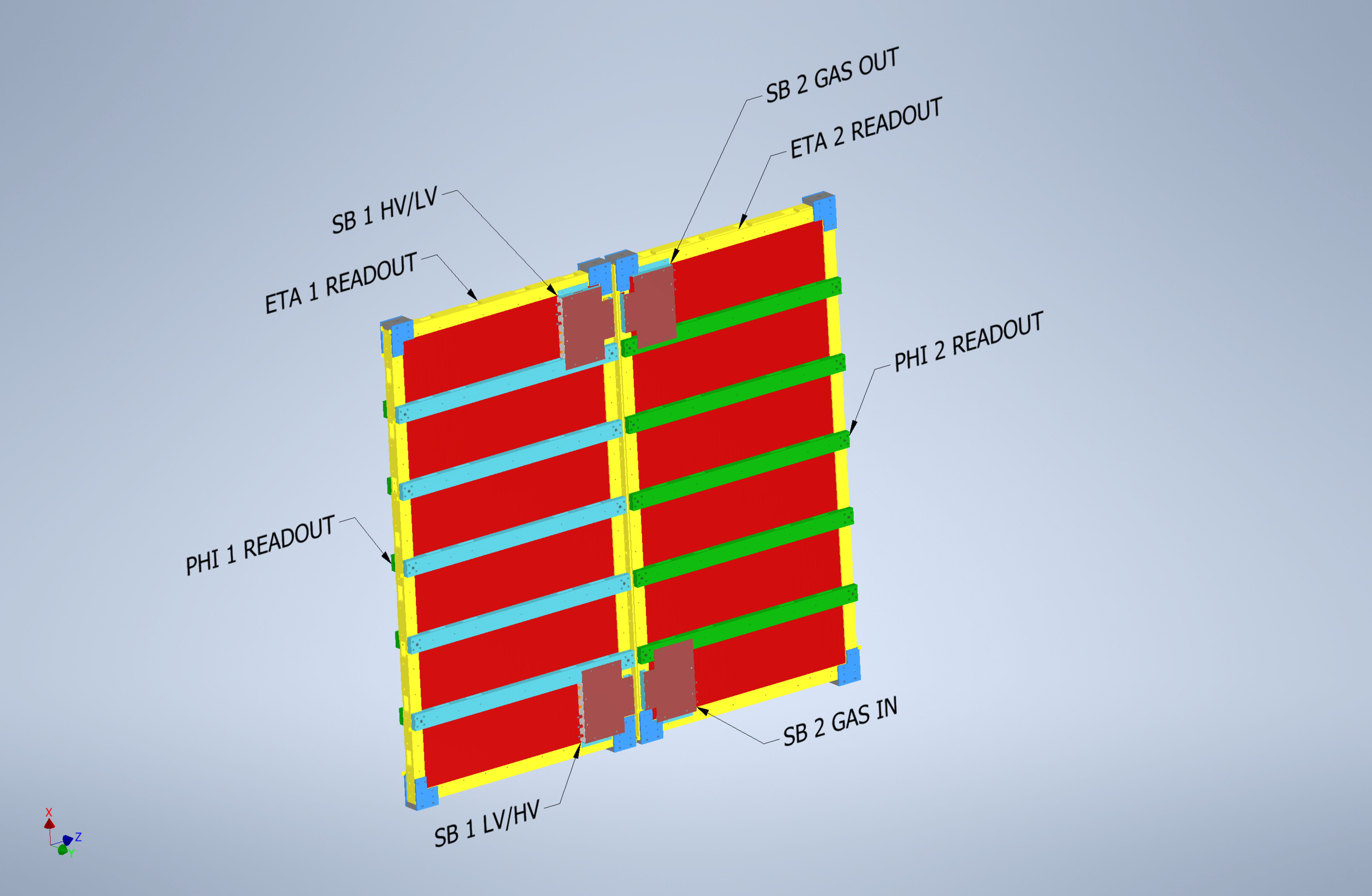}
    \end{center}
    \caption{
        Schematic of a face of \codexbeta, comprising two RPC chambers and their support frames.
        (``PHI'' $\equiv \phi$ and ``ETA'' $\equiv \eta$.)
        Each of the two modules includes a support frame, shown in Fig.~\ref{fig:moduleFrame}.
    }
    \label{fig:RPCTripletSupportBiplab}
\end{figure}

Each module shown in Fig.~\ref{fig:demosketch} comprises a support structure, shown in Fig.~\ref{fig:moduleFrame}, around a stacked triplet of RPC singlets as described in Sec.~\ref{subsec:RPCDesign}, each with an area of $2\times 1\ma$.
The long and short sides of the module are referred to as the $\phi$ and $\eta$ sides, respectively, as in Fig.~\ref{fig:RPCTripletSupportBiplab}.
One $\phi$ side and one $\eta$ side of each module contain readouts for the corresponding perpendicular readout strips of the RPCs.
In order to form a full $2\times 2\ma$ face of the \codexbeta cube, two modules are placed side-by-side along their $\phi$ sides
such that the readouts are on the opposite outer-edges of the module, as in Fig.~\ref{fig:RPCTripletSupportBiplab}.

The modules on each face have a specific orientation depending on which face of \codexbeta they are on (see Fig.~\ref{fig:demosketch} and the beginning of Sec.~\ref{sec:DetectorDesign}):
\begin{itemize}
    \item The modules on the S and G faces will be oriented such that the $\phi$ sides are perpendicular and the $\eta$ sides are parallel to the incoming particles.
    This is due to the expectation that the LLP candidates will mostly originate from the \lhcb IP and travel in the direction perpendicular to the beam-pipe to be incident on \codexbeta.
    The $\phi$ side has finer resolution and therefore should be positioned to obtain the best tracking of these LLP candidates.
    \item The modules on the F, C, and B faces are all oriented such that the $\phi$ sides are parallel to the vertical axis and both the $\phi$ and $\eta$ sides are perpendicular to the incoming particles. The inclusion of the central C face allows for improved tracking of LLP candidates, as it increases the number of detection planes along the expected direction of LLP candidates.
    \item Finally, the modules on the L and R faces are oriented such that the $\phi$ sides are parallel to the vertical axis and the $\eta$ sides are parallel to the incoming particles.
\end{itemize}
Different module locations and orientations require different methods of installation and mechanical support. For example, the modules on the F, C, and B faces form the main rigid structure of \codexbeta, but the modules on the L and R faces are mounted on move-able trolleys that allow access to the interior, as in Fig.~\ref{fig:demosketch}. See Section~\ref{subsec:Mechanical} for further details on the installation procedures.

The cubic geometry of \codexbeta is analogous to the baseline geometry of the full \codexb detector outlined in the \codexb expression of interest~\cite{Aielli:2019ivi}. In this reference, the reasons for such a geometry are outlined in detail, driven primarily by the need for a hermetic detector and the minimal structural component of an RPC module, with full details of the RPC design outlined in Section~\ref{subsec:RPCDesign}.
\subsection{Mechanical Framework}
\label{subsec:MechanicalFramework}
\subsubsection{Introduction}

The installation of the fourteen modules in the D1 barrack requires a frame that is compact and precise and that allows the modules to be transferred directly onto the frame in a simple and efficient manner.
The \atlas experiment's BIS7 module frame served as the basis for the design of the \codexbeta module frame,
and it has been redesigned, reinforced, and equipped with multiple fixation points that allow the \codexbeta modules to be transported, lifted, rotated, and mounted in a number of different ways.

The installation and positioning of all but four of the modules is done with aluminum rails, channels, and tracks. The modules ride either in or on these features using rollers attached directly to the module frames.
The upright modules (those on the F, C, and B faces---those on the L and R faces are a special case, described below) rest on rollers attached to their lower edges that ride inside a channel; these modules are constrained at the top in a similar manner.
The horizontal modules (those on the S and G faces) have rollers attached to the sides of the frames, and these wheels ride on tracks.

To ease the installation process and to allow reasonable access to the electronics located inside \codexbeta, at least one face of the cube must be possible to open.
The four modules on the L and R faces are therefore permanently attached to the same transport carts that are used to transport them, described in Sec.~\ref{subsec:installUpright} and shown in Fig.~\ref{fig:uprightTransportCarts}.
These carts are also used to position the modules to slide them into the rail system and are designed to place the modules at the same height as all the other upright modules.
This allows one or both modules to be rolled away from the structure in order to access the interior without disturbing any other modules.

A more detailed description of the mechanical installation of the modules can be found in Sec.~\ref{subsec:Mechanical}.

\subsubsection{Frame Design}
\label{subsec:frameDesign}
The structural components consist of $100\mm$- and $150\mm$-wide aluminum I-beams, C-channels, and angles that will be bolted together inside the D1 barrack.
The heaviest individual component weighs less than 30\kg, while most weigh less than 15\kg.
Although the profiles are made of US-type extrusions and the dimensions are in inches, the frame is designed to be integrated into the existing array of $150\mm$-wide steel floor-beams, even using those floor beams as an essential part of the structure.
Only structural-type metric fasteners (10, 12, 14\mm) will be used and all the beam connections will make use of $6.4\mm$-thick steel gusset plates as both edge and corner reinforcements for the top plane of the structure.
To properly secure the structure to the floor, either (1, preferred) a series of 14\mm holes will be drilled into the top flanges of the three adjacent floor beams the structure rests upon to enable the structure to be bolted to the floor or (2) clamps will connect the frame to top flanges of these same floor beams.
No welding would be required except during the actual fabrication of the beams and columns.

The frame is designed to support four sets of rails that allow six modules to be positioned upright on the F, C, and B faces and to support tracks that allow four more modules to be installed on both the S and G faces in a horizontal orientation.
These rails are composed of two aluminum profiles welded together to form a ``wineglass'' shape; it is made by attaching a $100\mm$-wide C-channel to the top of a $100\mm$ square wide-flange I-beam with $8\mm$-thick sections across the entire profile.
\subsubsection{Dimensions}
\label{subsec:MechanicalDimensions}
\begin{figure}[tbp]
    \begin{center}
    \includegraphics[width=\textwidth]{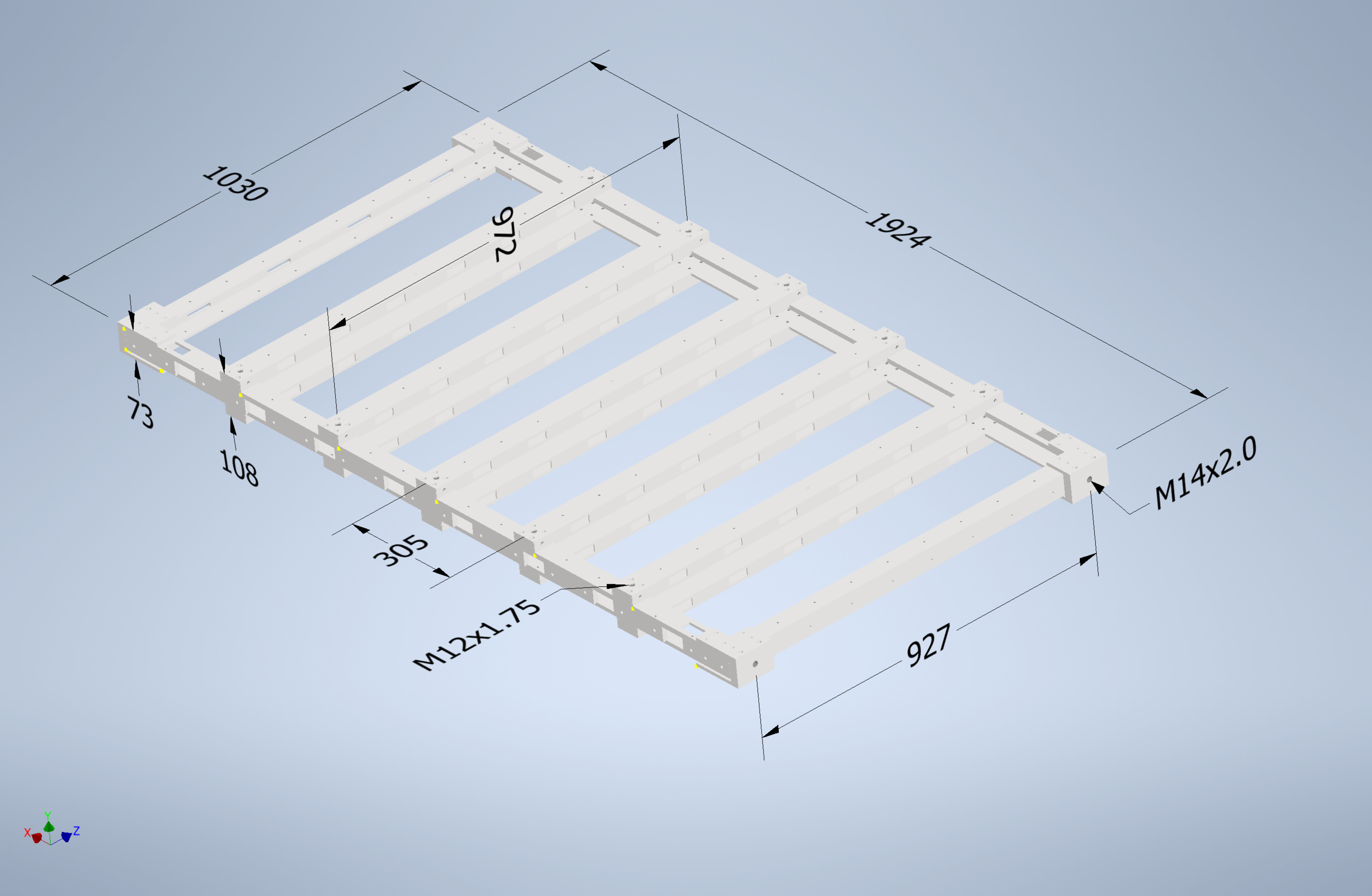}
    \end{center}
    \caption{CX1 module frame for \codexbeta with dimensions in mm. Relevant pick points are also indicated.}
    \label{fig:CX1frame}
\end{figure}

Dimensions of the CX1 frame can be found in Fig.~\ref{fig:CX1frame}, along with the dimensions of the various pick points. 

\begin{figure}[tbp]
    \begin{center}
    \includegraphics[width=\textwidth]{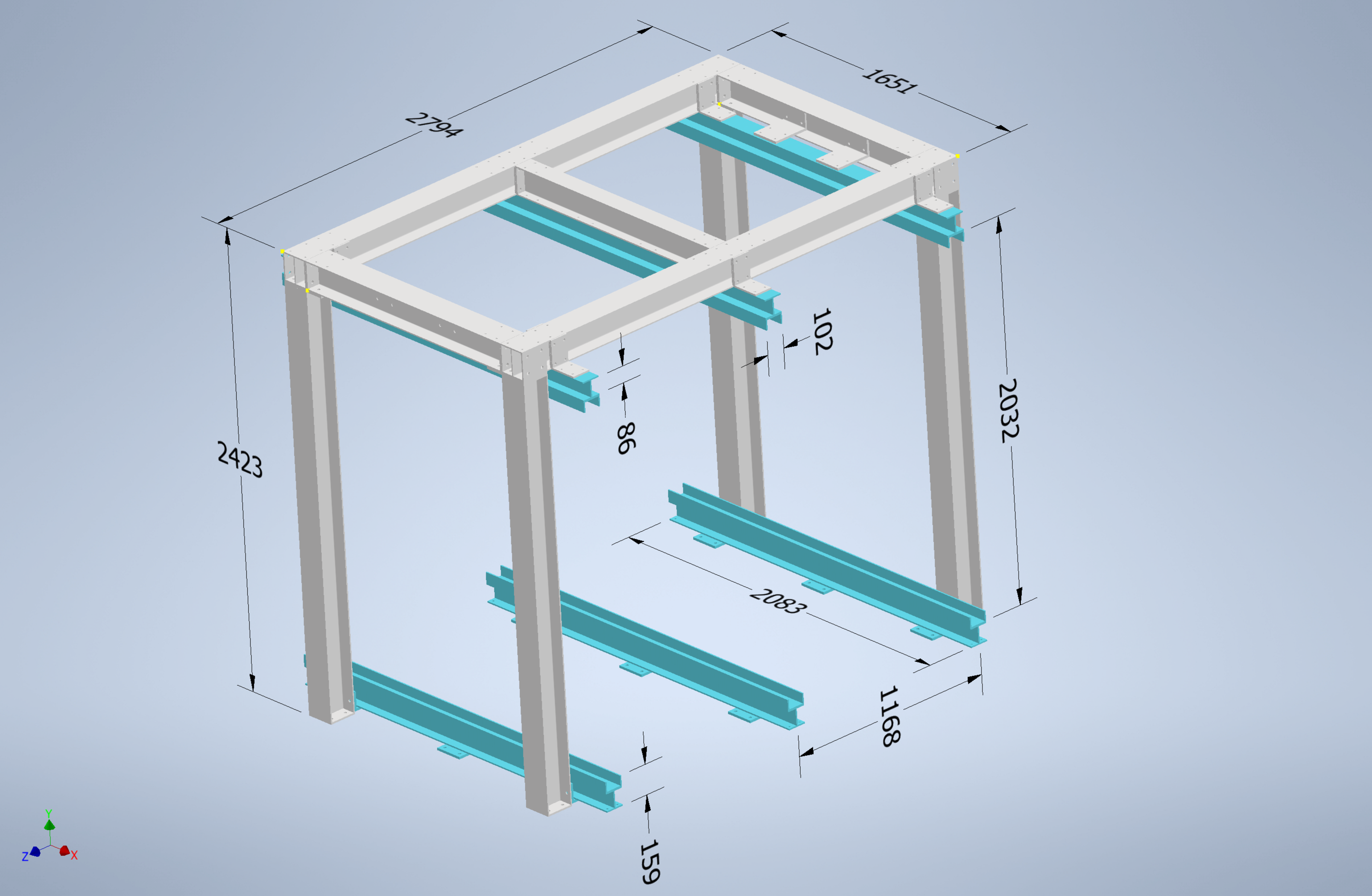}
    \end{center}
    \caption{\codexbeta support structure with dimensions in mm.}
    \label{fig:supportstructure}
\end{figure}

Dimensions of the support structure can be found in Fig.~\ref{fig:supportstructure}. Note the nine plates shown on bottom where the lower wineglass profiles would attach to three of the existing D1 barrack floor beams.

\subsection{DAQ}
\label{subsec:DAQ}
Each module is read out via a low-power GigaBit Transceiver (lpGBT) link which sends data via optical cables to a single TELL40 board located in one of \lhcb's Event Builder nodes. All electronics required at the front end will be mounted on the module itself, minimizing the complexity of cabling inside the D1 barrack. The cables required to go from the barrack to the surface have already been put in place and can be accessed from the barrack. The overall data volume is such that one TELL40 board suffices to process all of it. Once the data is received by the TELL40, it will be treated in exactly the same way as for any other \lhcb subdetector and is made available for decoding and processing in HLT1 and HLT2. 

\begin{figure}[t]
    \centering
    \includegraphics[width=0.9\textwidth]{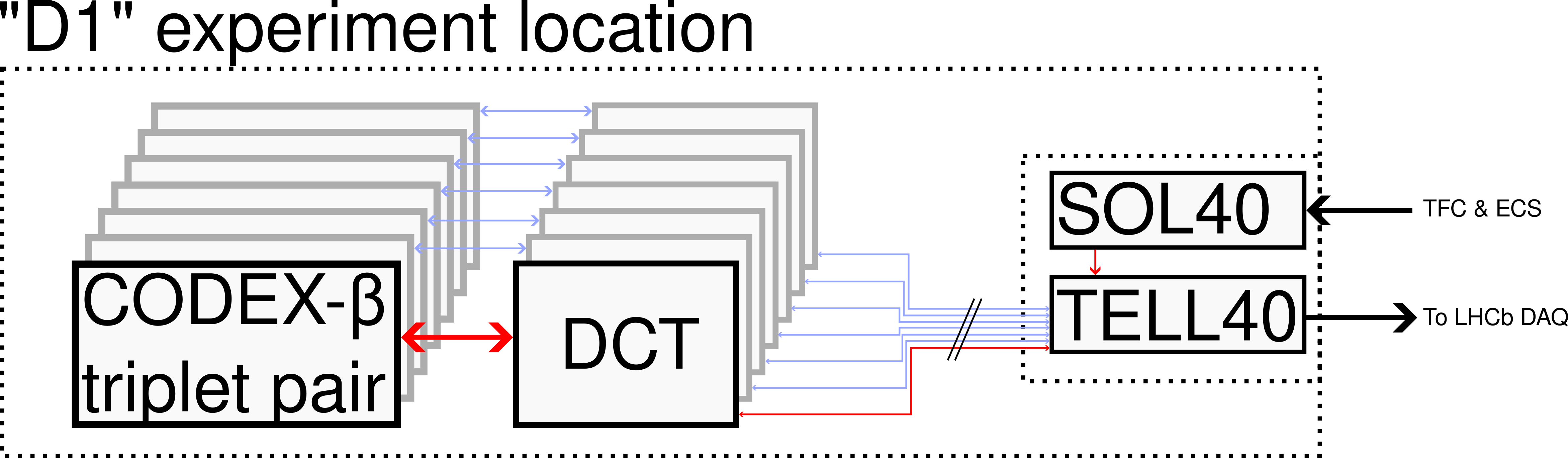}
    \caption{Schematic of the front-end readout of an RPC module.}
    \label{fig:DAQDCT}
\end{figure}

The basic front-end readout is shown in Figure~\ref{fig:DAQDCT}. Each DCT reads out two triplets, so there are
seven DCTs required in total. There are two challenges here compared to the usual LHCb readout schema. First, the use of the lpGBT protocol, which is expected to become available in the second half of 2023, and which the \codexbeta team will help commission (an activity that is useful for LHCb looking towards LS3). The lpGBT protocol used by the DCT operates at $10.24$~Gbs$^{-1}$ while the TELL40 is qualified for $5.12$~Gbs$^{-1}$. However from discussions with firmware specialists, this is not expected to cause major issues. Note that in any case, the actual data volume being transmitted is significantly smaller than the full bandwidth allowed by the protocol.

The second is that the DCT expects to communicate bidirectionally with a FELIX board, sending data and receiving commands and control instructions via the same link. However, within the LHCb DAQ these functions are separated: data is sent up to TELL40 back-end readout boards (one TELL40 in the case of \codexbeta) while a separate SOLL40 board sends commands and control instructions to the front ends. The TELL40 and SOLL40 are both variants of the same basic PCIe40 board, but with different firmware and fulfill different purposes.

In order to reconcile these two architectures another PCIe40 board is required to act as an intermediate layer between the bidirectional DCTs at the front-end and the unidirectional TELL40/SOLL40 boards at the back end. This board will have separate command and data streams looking towards LHCb, and communicate bidirectionally towards the DCT front end. There is a third flavour of the PCIe40 architecture which already mixes command and data which is well suited for this purpose: the MiniDAQ board. We have had preliminary discussions with LHCb readout specialists who believe this approach is generally speaking feasible. We are in the process of working out a detailed design based on this outline which will be submitted to the online and technical teams for review. 

\clearpage

\section{Installation and Commissioning}
\label{sec:Installation}
\subsection{Site Preparation}
\label{subsec:SitePreparation}
The three main challenges confronting site preparation are
\begin{itemize}
    \item removal of the server racks from the D1 barrack and conditioning of the room to host the cube (already complete),
    \item preparation for safe transportation of the RPC modules from the surface to the D1 barrack, and
    \item ensuring that the power supply in the room meets the needs of the demonstrator.
\end{itemize}
Strategies to address these challenges are proposed in the following sections.

\subsubsection{Preparation of the D1 barrack}

\begin{figure}[tbp]
    \begin{center}
        \includegraphics[width=0.7\textwidth, angle=90]{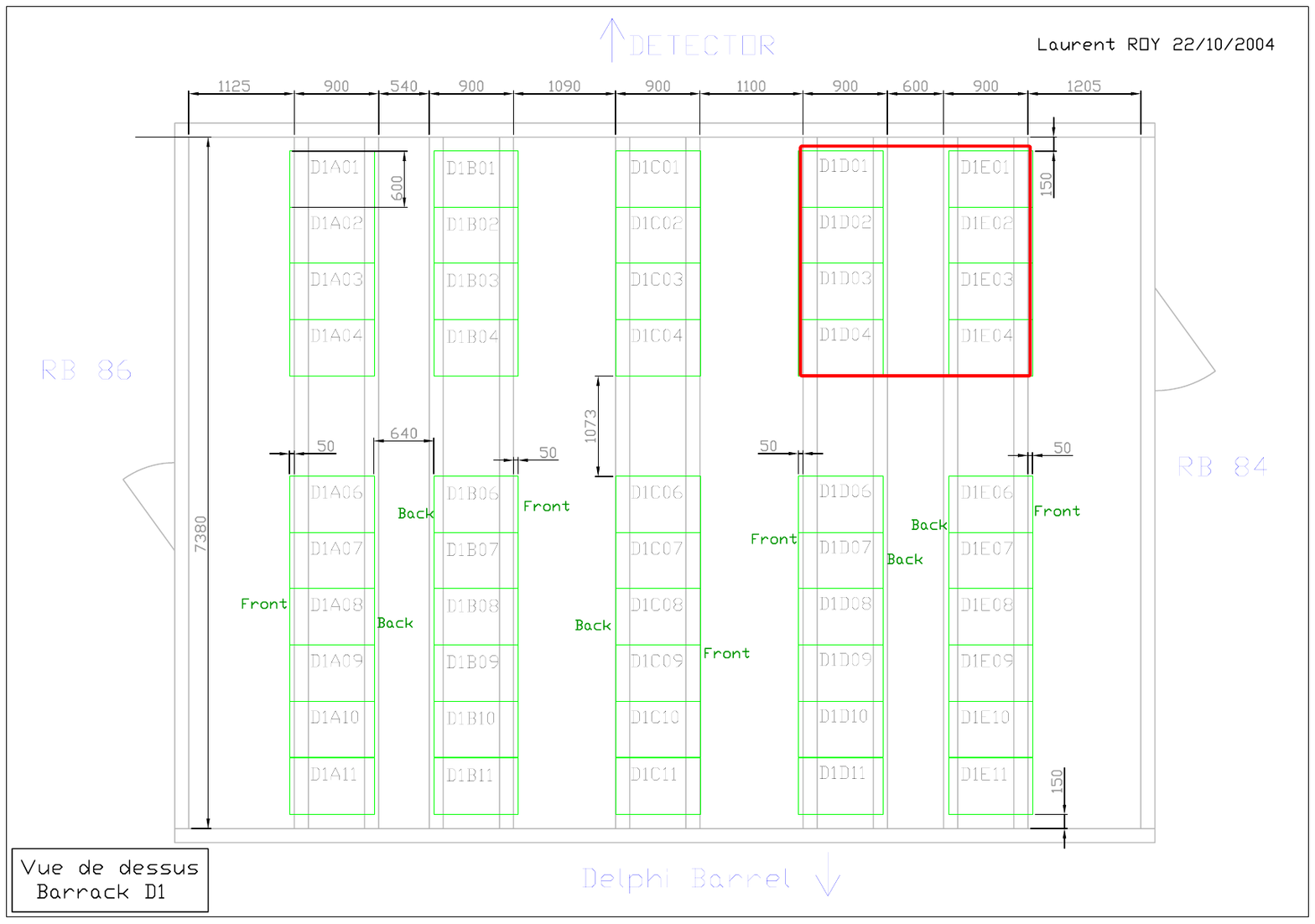}
    \end{center}
    \caption{
        Schematic of the D1 barrack~\cite{D1BARRACK}, where we propose to install \codexbeta.
        The area in red shows the approximate location of \codexbeta.
        Face F of \codexbeta would be along the wall nearest the detector (at the top of the figure), and face L would be along the RB 84 wall (on the right side of the figure).
    }
    \label{fig:D1B}
\end{figure}

\begin{figure}[tbp]
    \begin{center}
        \includegraphics[width=\textwidth]{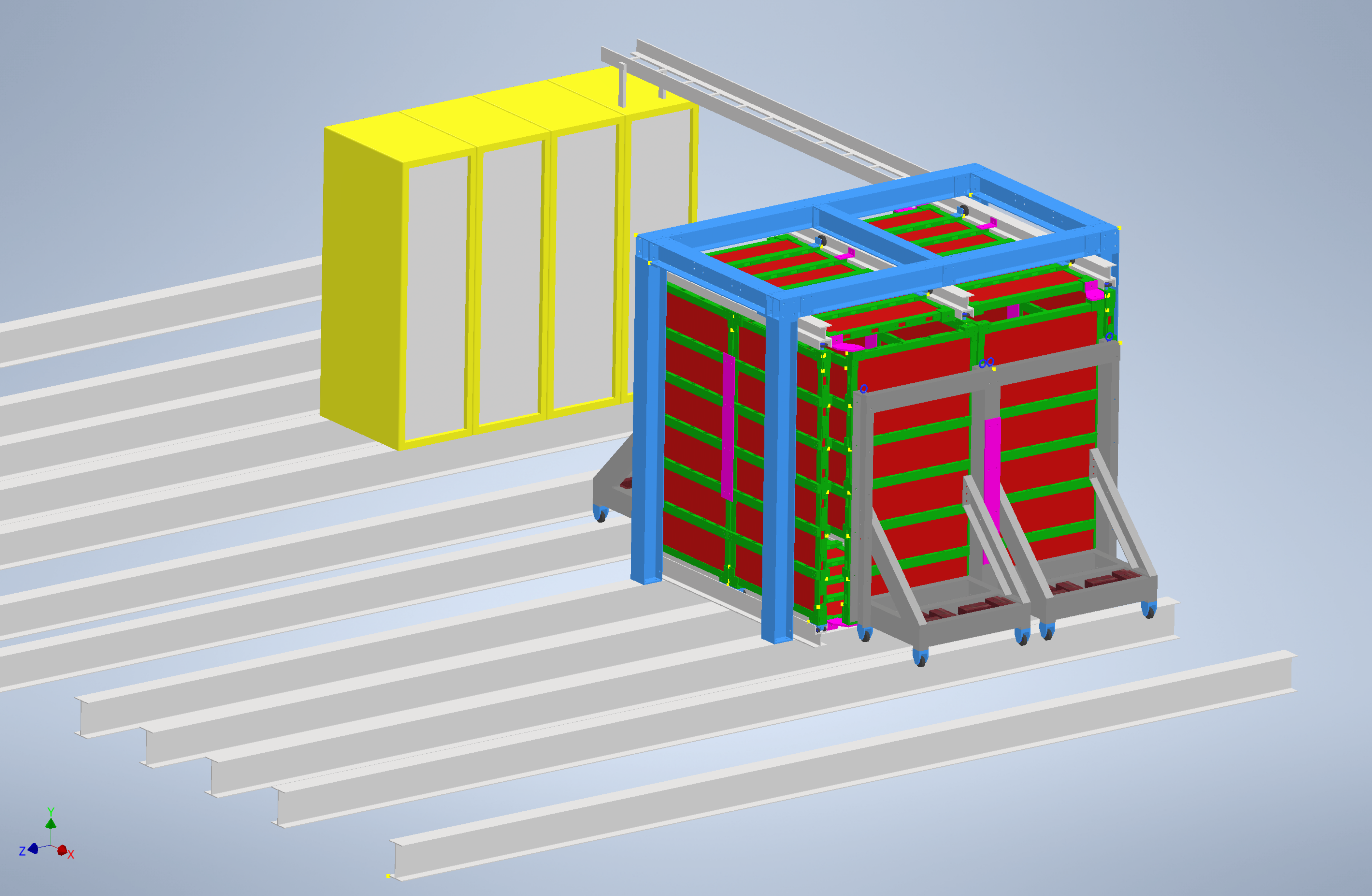}
    \end{center}
    \caption{Diagram of \codexbeta in the D1 barrack with needed services racks and proposed cable tray. The floor panels of the D1 barrack are not shown, but the structural support beams are.}
    \label{fig:CXRracks_cube}
\end{figure}

As described in Sec.~\ref{sec:detectordesign}, \codexbeta will consist of a cube, 2\m on each side.
Because of the size and weight of the detector support structure, it cannot be moved without disassembly. 
Consequently, the very first task towards the installation of \codexbeta is the preparation of the D1 barrack,
shown in Figure~\ref{fig:D1B},
for the in-situ assembly of the demonstrator.
The most natural location for the demonstrator would be where the D1D01 to D1D04 and D1E01 to D1E04 racks were located, next to one of the doors. Figure~\ref{fig:CXRracks_cube} shows \codexbeta situated in the D1 barrack.
Face F of \codexbeta should be approximately 0.5\m from the wall nearest the \lhcb detector to maximize its physics potential while still allowing access to the Face F modules.
We propose placing \codexbeta atop the second, third, and fourth beams from the RB84 wall, thus allowing plenty of space for installation and access on faces B, L, and R.

After the initial preparation is complete, some additional work will be required to prepare three of the existing 150\mm-wide floor beams by drilling a number of 14\mm holes in the top flanges.
This will allow the entire cube structure to be anchored directly to the floor and is the preferred method of attachment, rather than clamping. Note that the currently proposed bottom of the support structure does not have cross-beams that fix the columns in place. Consequently, there may be some lateral stress on the columns which would be mitigated with bolting rather than clamping. However, if the CERN safety teams believe clamping of the support structure will be sufficient, then this is also a possibility.
There may be a need to modify---or even replace---the floor tiles that would be underneath the cube,
to facilitate cable or gas line routing or for structural reasons.
The full weight of the demonstrator frame would be entirely on the floor beams and not on any of the floor tiles; see Sec.~\ref{subsec:frameDesign}.
Temporary floor plates may be required when rolling the chambers in, however, both to reinforce relevant floor tiles and to provide for smooth movement of the components once inside the barrack.

\subsubsection{Access to the D1 barrack and module transportation}

\begin{figure}[tbp]
    \begin{center}
        \includegraphics[width=0.8\textwidth]{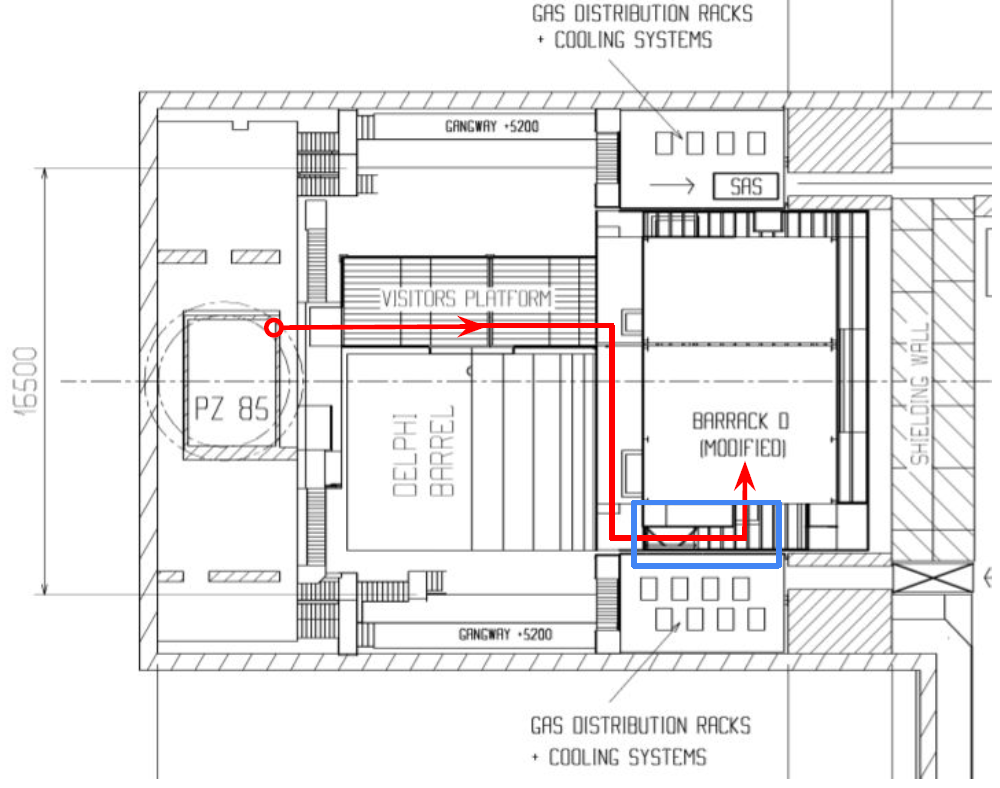}
    \end{center}
    \caption{
        Layout of the UX85 safe side (UXA)~\cite{EDMSUX}. Proposed transportation path of the modules is shown in {\color{red}red}, where the narrow staircase and gangway are highlighted with a {\color{blue} blue} rectangle.
        This {\color{blue} blue} highlighted region is pictured in Fig.~\ref{fig:entryD1B}
    }
    \label{fig:pathD1B}
\end{figure}

A module has the shape of a rectangular parallelepiped of $2 \times 1 \times 0.05\mv$ with an approximate weight of 140\kg.
The module-transportation carts may weigh about the same as a module.
Given their size and weight, each module can be transported from the surface to the underground cavern using the personnel elevator located at the PZ85 shaft.
They will remain on their carts until they are inside the barrack and can be transferred to the support structure;
see Fig.~\ref{fig:pathD1B}.

\begin{figure}[tbp]
    \begin{center}
        \includegraphics[width=1.\textwidth]{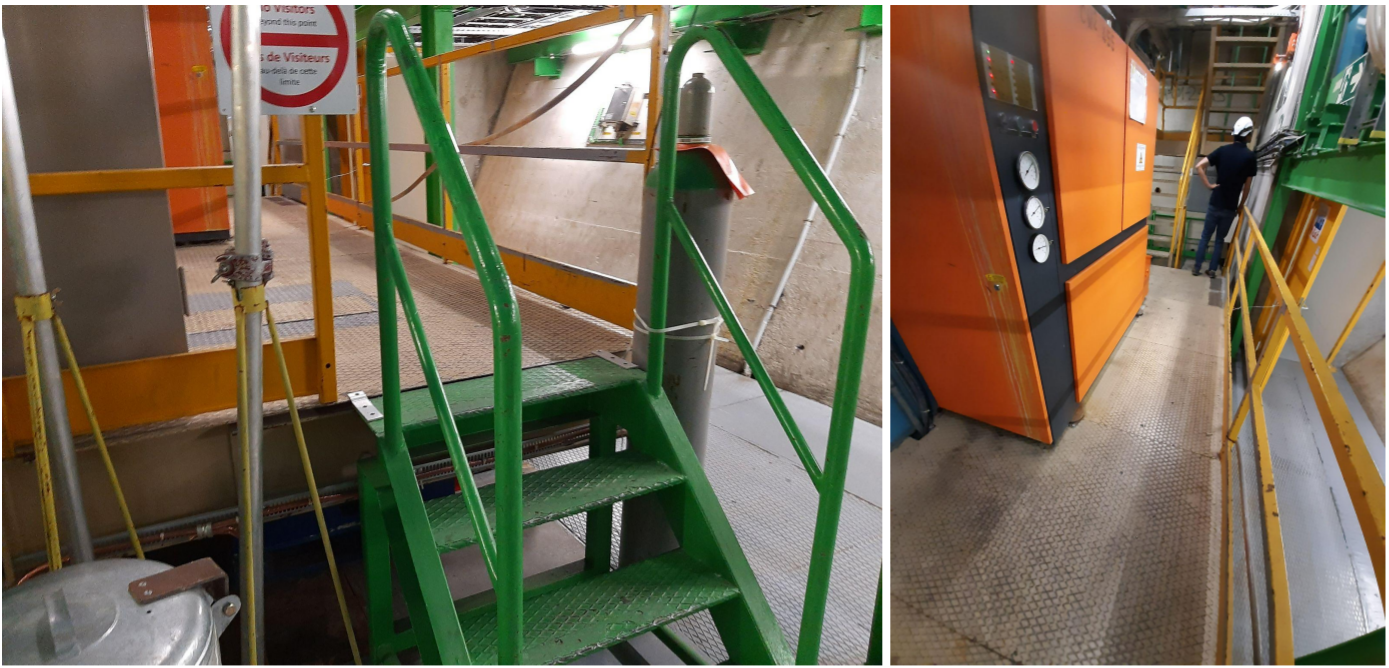}
    \end{center}
    \caption{
        Pictures of the entrance to the D1 barracks, showing the staircase and gangway, and corresponding to the highlighted area with a blue rectangle in Figure~\ref{fig:pathD1B}.
    }
    \label{fig:entryD1B}
\end{figure}

Maneuvering the modules into the UXA-C entrance of the D1 barrack will be challenging because of the narrow gangway (94.5 cm) around an air-conditioning unit, as well as the narrow stairway (80 cm) leading up to it (see Figs.~\ref{fig:pathD1B}~and~\ref{fig:entryD1B}).
Both types of cart (see Figs.~\ref{fig:uprightTransportCarts}~and~\ref{fig:tiltTableTransportCarts}) are designed to fit through the 83\cm-wide doorway.

\subsubsection{Power supply in the D1 barrack}
Remote monitoring and control of the power system is necessary, as well as integration in to a SCADA system (such as WinCC). A standard way to power and control the detector is to adopt a modular integrated system combining HV, LV, and monitoring. Taking as example the CAEN EASY products, the full system would require the following rack-mounted devices adjacent to the demonstrator (vertical cooling is needed):
\begin{itemize}
    \item CAEN low cost mainframe SY5527LC (4U wide) containing one branch controller
    \item 48V DC (1000W) primary power generator with 2 channels. 
    \item 2 EASY A3001 crates (21 slots)
    \item passive power distributors
\end{itemize}
A simple alternative configuration makes use of suitable modules installable directly in the mainframe (in this case aSY4527LC 8U wide is recommended).
Most of the power consumption of each chamber is absorbed by the LVDS driver (5W) and by the DCT (10W). The rest of the power (amplifiers, discriminators, and HV) is contained in 2W per chamber.

\clearpage
\subsection{Mechanical}
\label{subsec:Mechanical}
There are three unique configurations that are used to secure the modules in their proper position, and each makes use of a different set of fixation points located on the module frame.
In the nominal procedure, six of the ten modules oriented in the upright position (those on the F, C, and B faces) will roll into position after being placed in the appropriate tracks with a transport cart that allows for the load to be properly constrained as it is smoothly transferred to the rails;
there are four additional upright modules comprising the L and R faces of the cube that will each remain on their own carts for the duration of the experiment.
(Alternative procedures for installing the upright modules, in addition to the nominal, are discussed below.)
The four remaining modules that are positioned horizontally make up the S and G faces of the cube and must slide into the frame from the L or R face.

The various configurations needed for the different modules require a specific installation technique to position them either flat or upright within the cube.

\subsubsection{Upright modules}
\label{subsec:installUpright}
\begin{figure}[tbp]
    \begin{center}
    \includegraphics[height=0.25\paperheight]{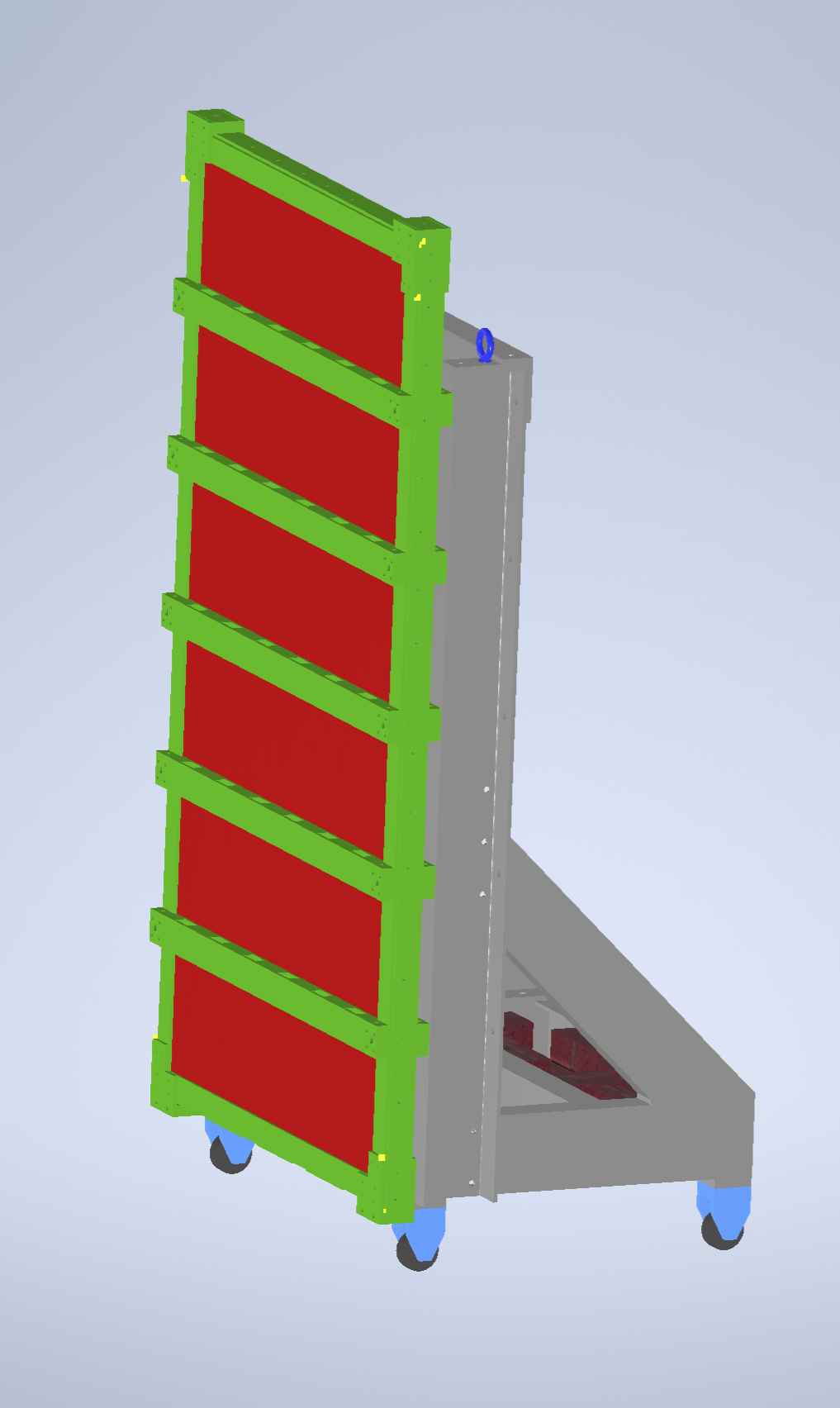}
    \includegraphics[height=0.25\paperheight]{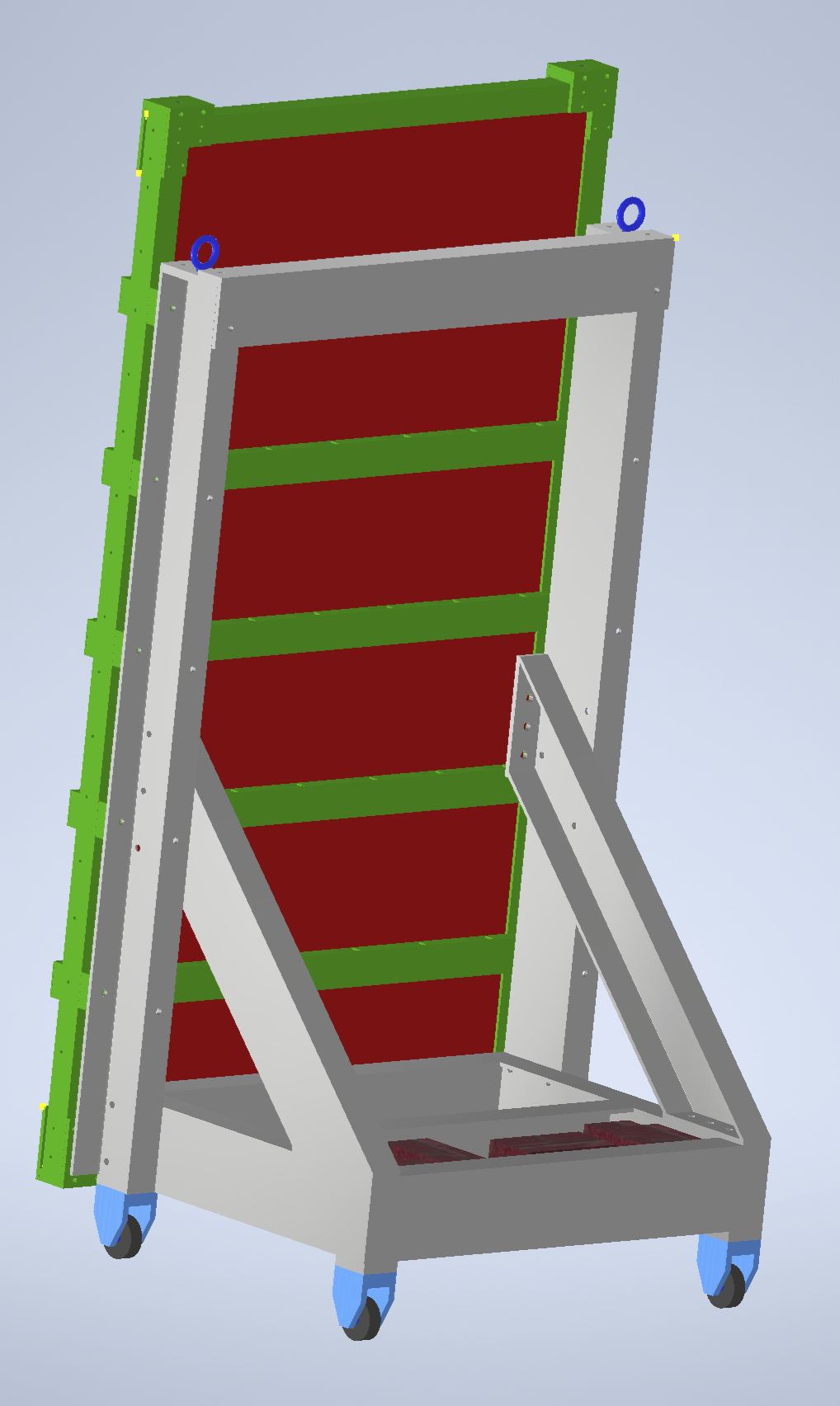}
    \includegraphics[height=0.25\paperheight]{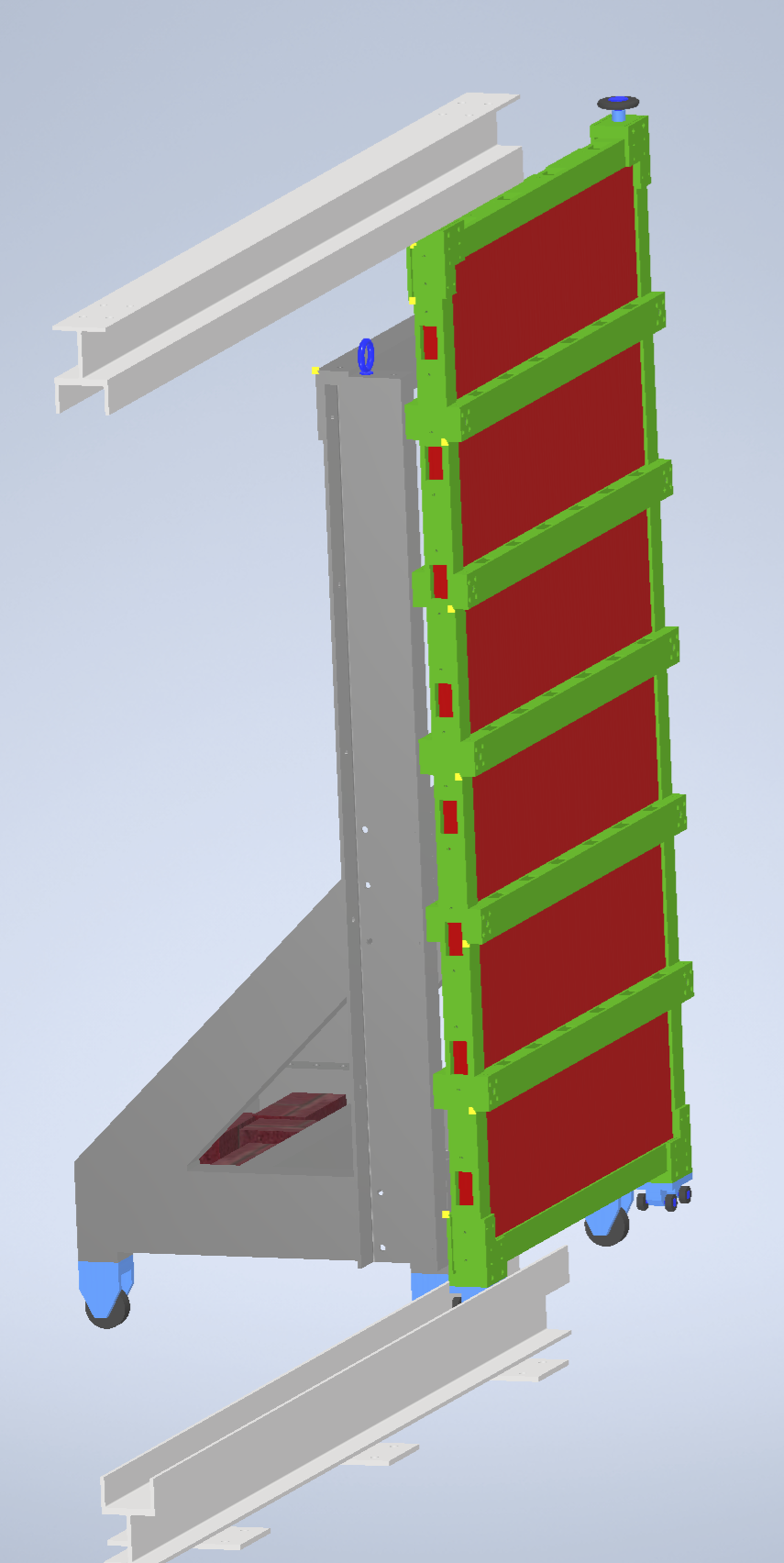}
    \includegraphics[height=0.25\paperheight]{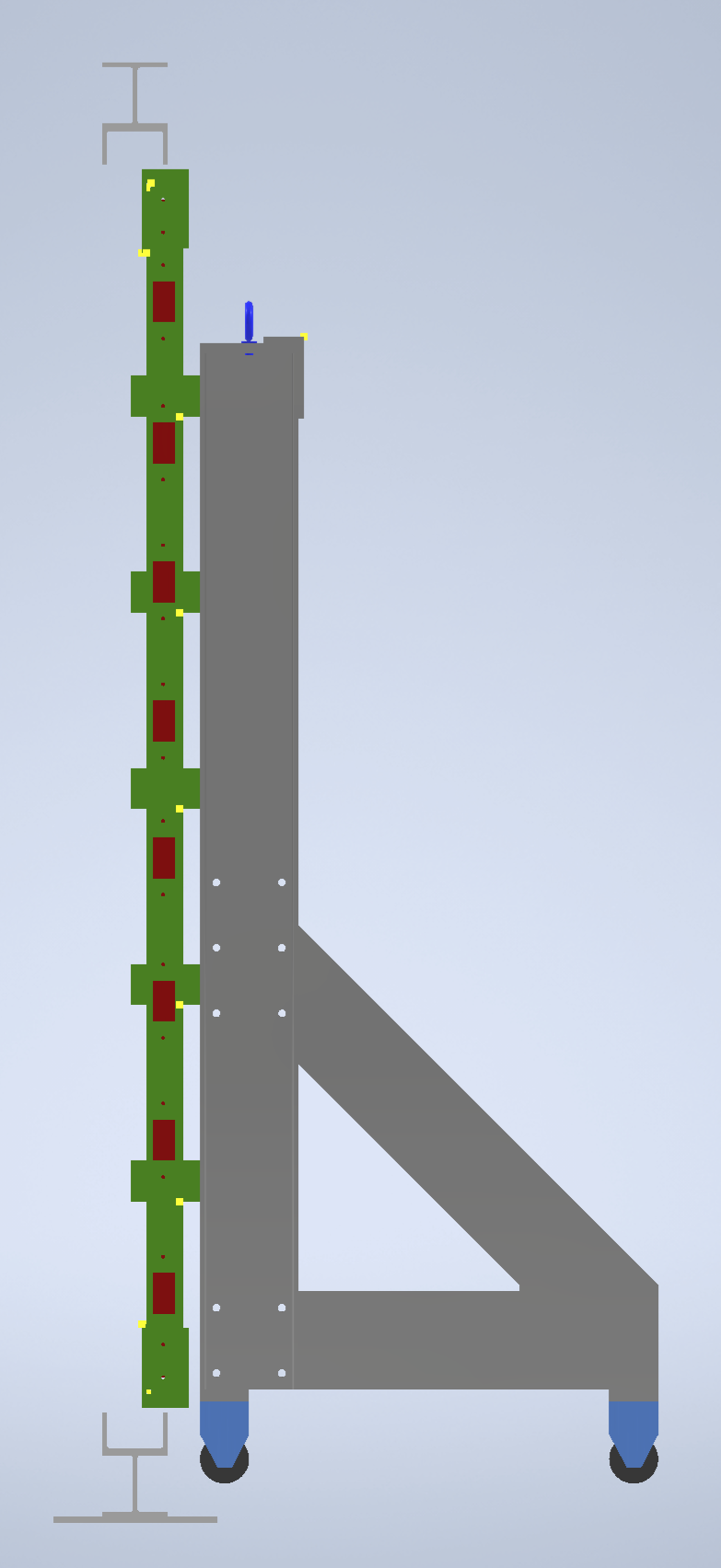}
    \end{center}
    \caption{Perspective diagrams of upright transport carts (with modules attached) used to move F-, C-, B-, L–, and R-face modules into the D1 barrack. The third and fourth demonstrate the positioning w.r.t. the rails for the railcar modules (see text) and for face L/R, respectively.}
    \label{fig:uprightTransportCarts}
\end{figure}

These modules (comprising the F, C, B, L, and R faces) are brought downstairs while mounted to the installation carts and rolled into the D1 barrack with the module in the upright position at all times (see Fig.~\ref{fig:uprightTransportCarts}).
Nominally, these installation carts allow for the module to be attached to the cart asymmetrically (outside the wheelbase of the cart) by placing counterweights in the cart, though alternative designs without counterweights are being considered (see below).

\begin{figure}[tbp]
    \begin{center}
    \includegraphics[width=0.7\textwidth]{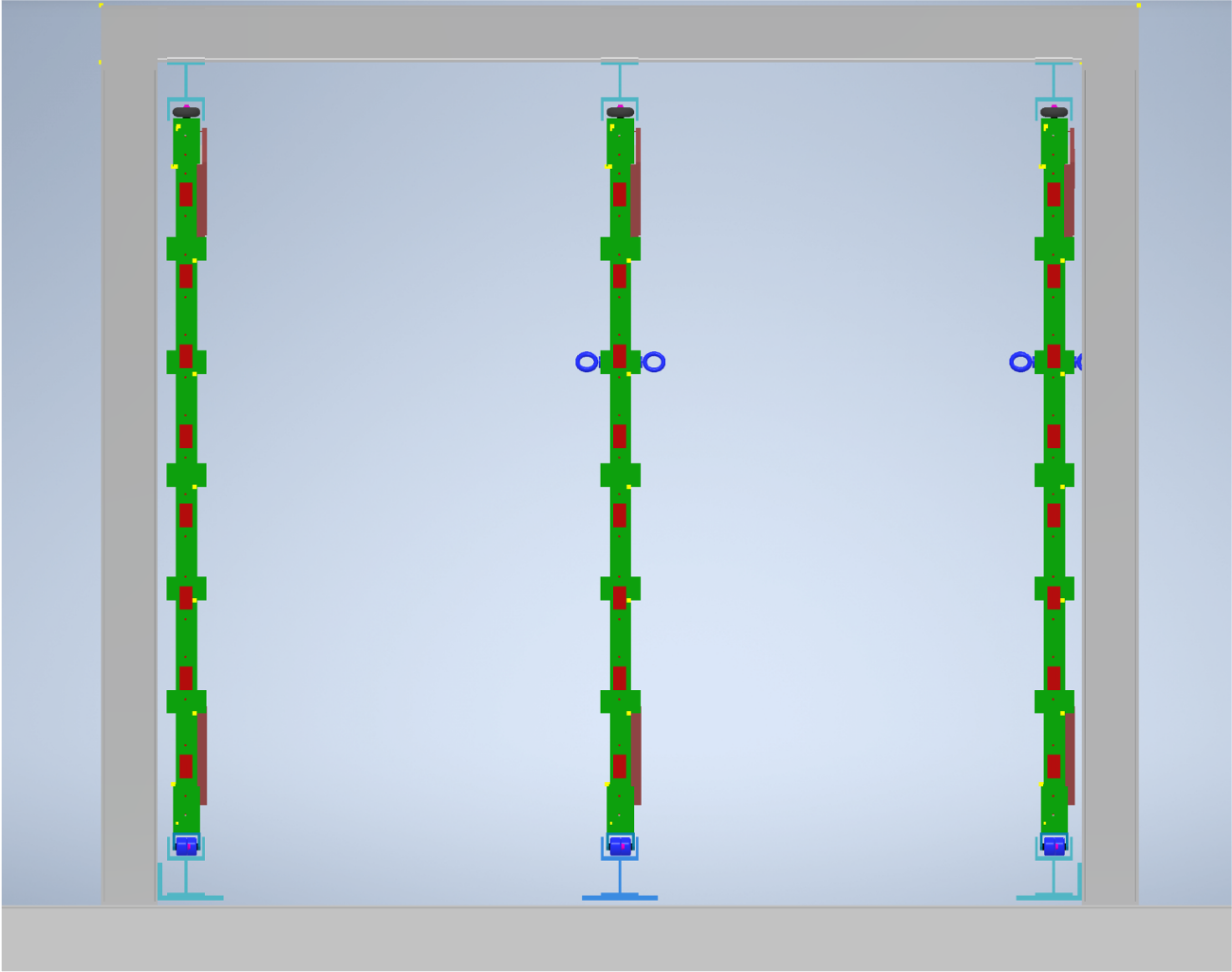}
    \end{center}
    \caption{Profile diagram of the F-, C-, and B-face modules (with some eye-bolts used for transport still attached) after installation in the structural support frame.}
    \label{fig:FCBprofile}
\end{figure}

Six ``railcar'' modules comprise the F, C, and B faces. These modules have rollers mounted to their $\eta$ sides that ride inside the aluminum ``rails'' described in Sec.~\ref{subsec:frameDesign}; see Fig.~\ref{fig:FCBprofile}.
The rollers are attached to each end of the module, which is mounted to the cart, and the module is then guided into the appropriate rail ends while still attached to the cart.
In this manner the module is fully constrained at all times while the load is transferred to the rails.
Once the module is completely inside the rails, the bolts attaching the module to the cart can be removed and the cart rolled away.

The other four upright modules are for faces L and R.
They are each permanently attached to their own installation carts, and are never placed in the rails.
This arrangement simplifies the frame structure, simplifies the installation of the other modules, and allows for easier access to the interior, since the carts may be rolled apart.
The modules mounted in the 3 sequential faces (F, C, and B) cannot be moved at all, once both are in place on a given face.

\begin{figure}[tbp]
    \begin{center}
    \includegraphics[width=0.45\textwidth]{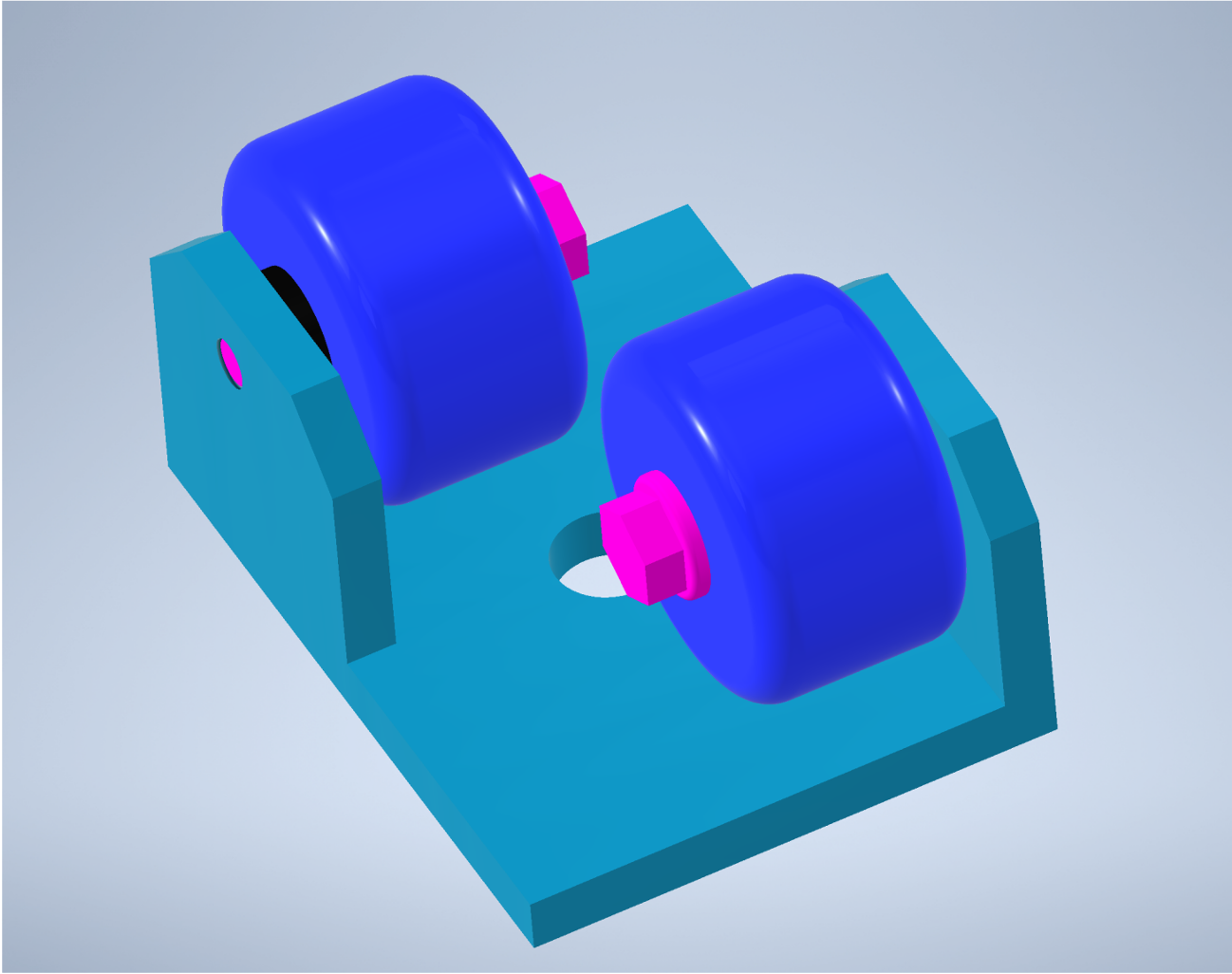}
    \includegraphics[width=0.45\textwidth]{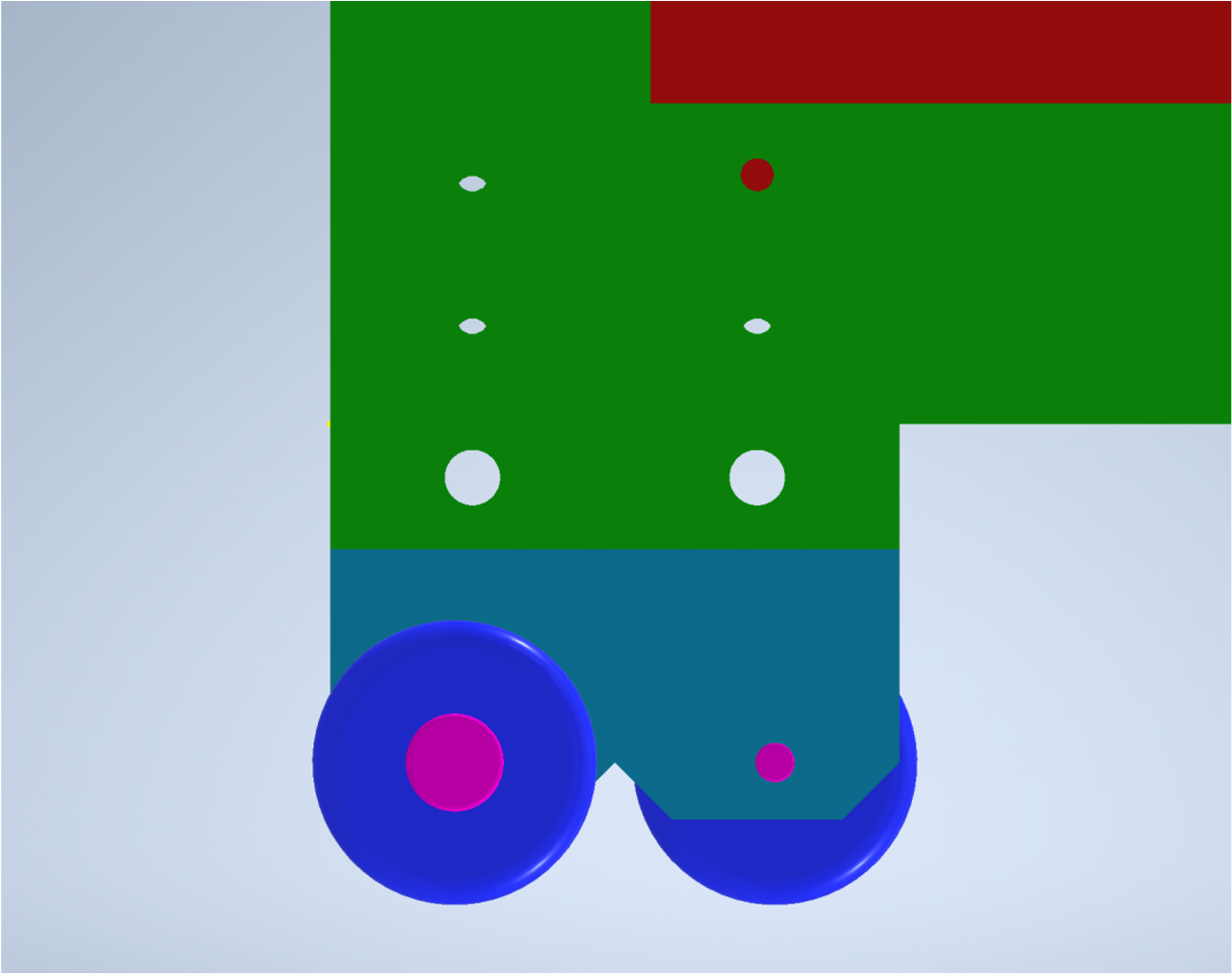} \\
    \includegraphics[width=0.45\textwidth]{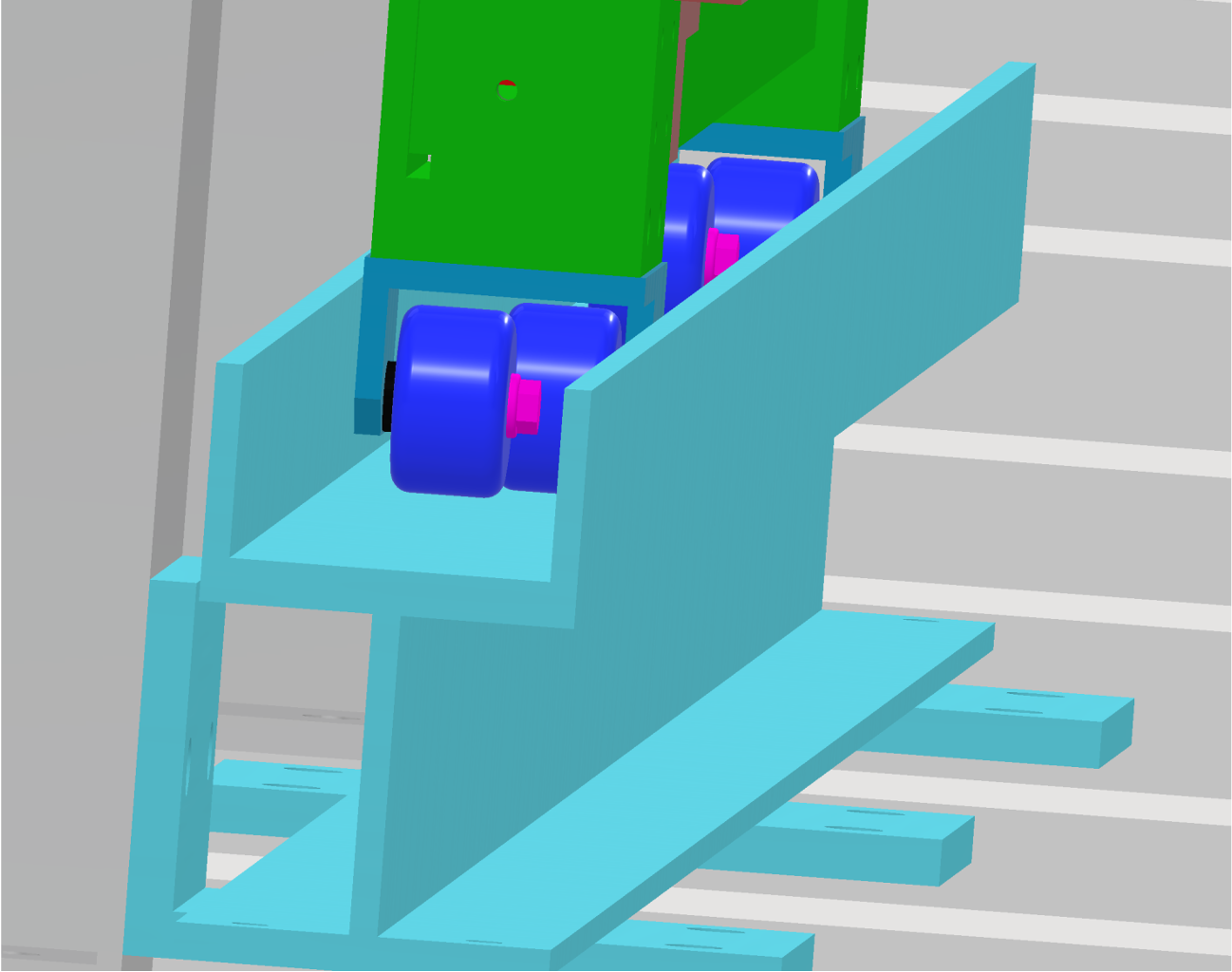}
    \includegraphics[width=0.45\textwidth]{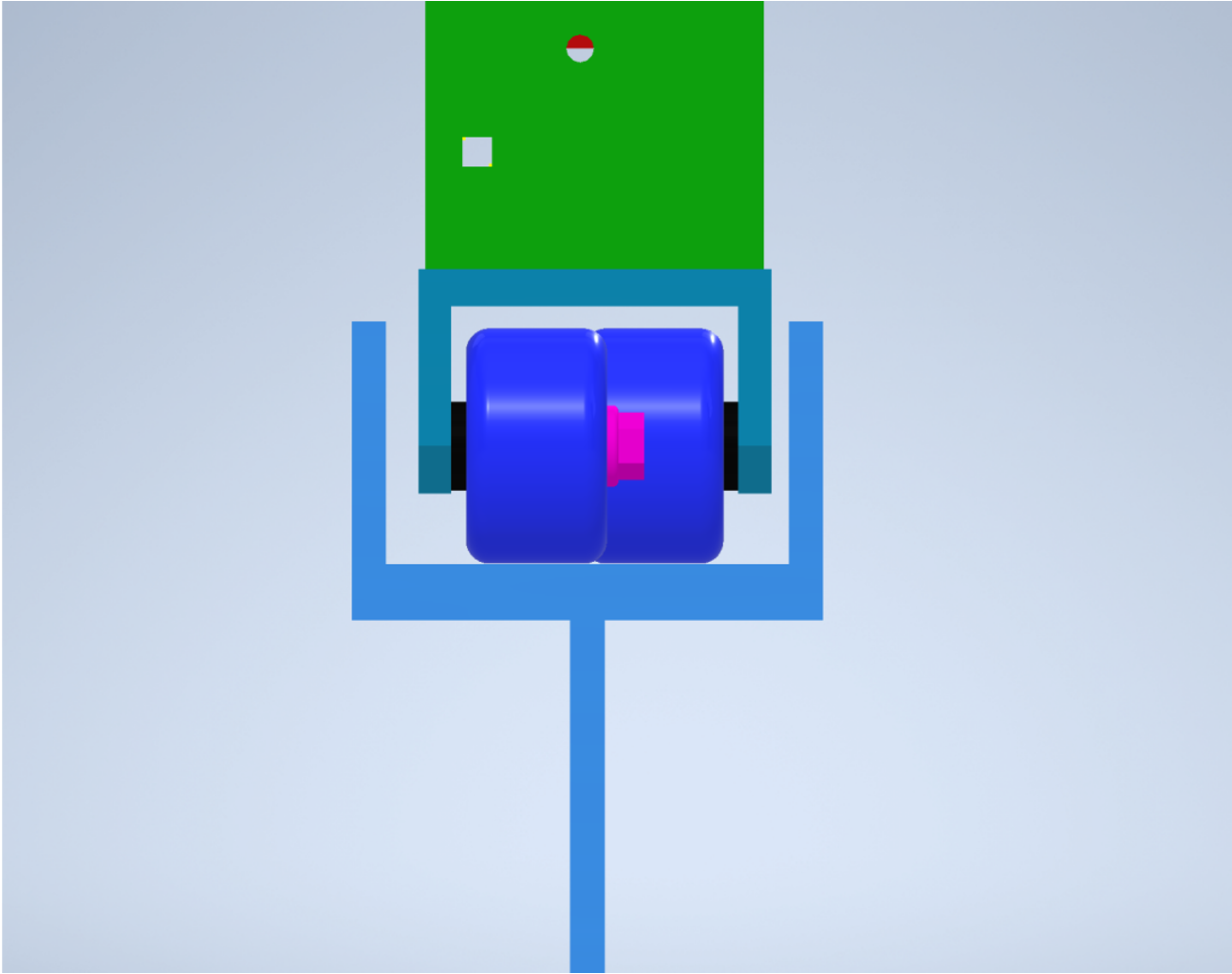}
    \end{center}
    \caption{Perspective diagrams of wheels that could be used to allow the F-, C-, and B-face modules to roll inside the lower aluminum channels; see text.}
    \label{fig:wheels}
\end{figure}

\begin{figure}[tbp]
    \begin{center}
    \includegraphics[width=0.45\textwidth]{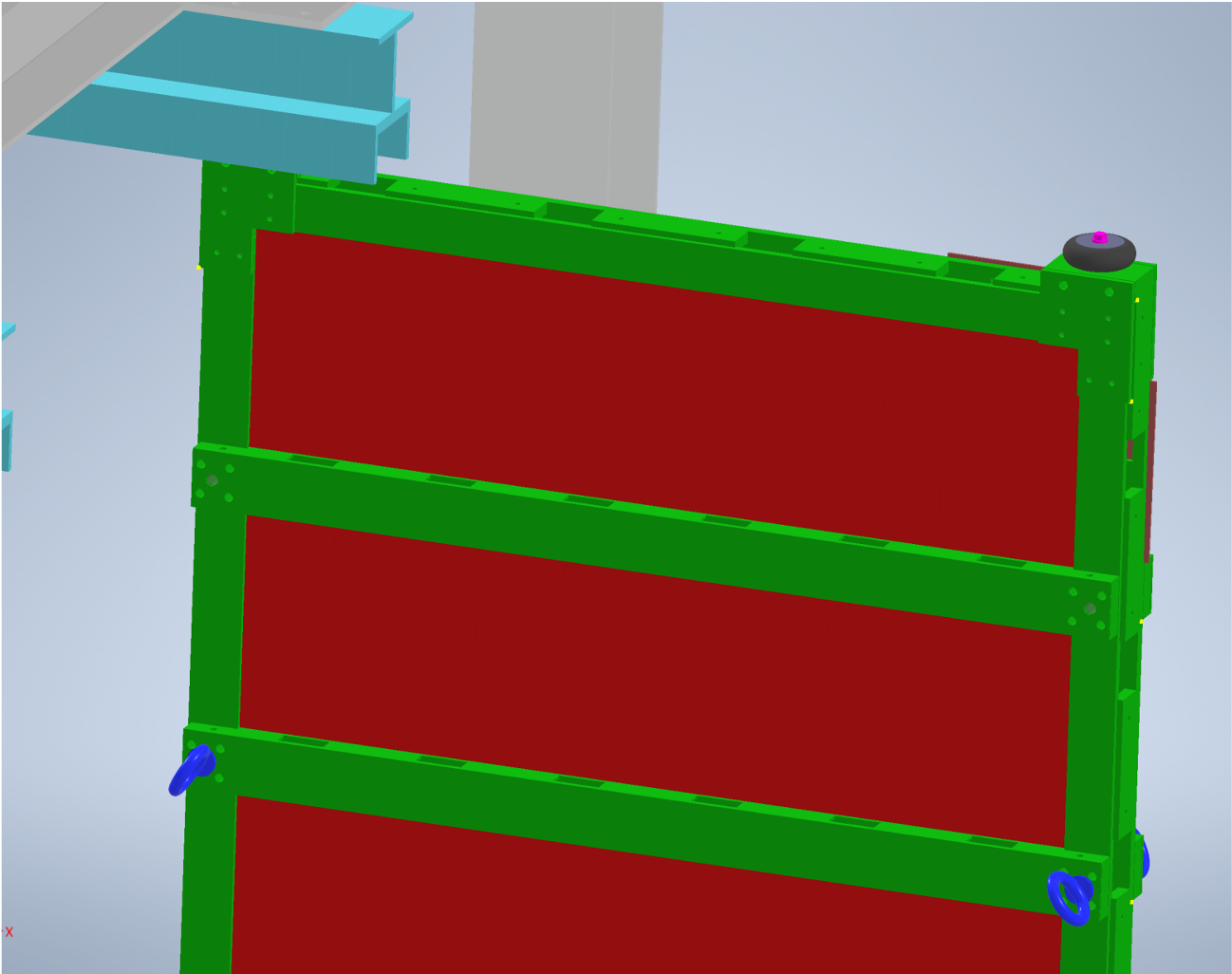}
    \includegraphics[width=0.45\textwidth]{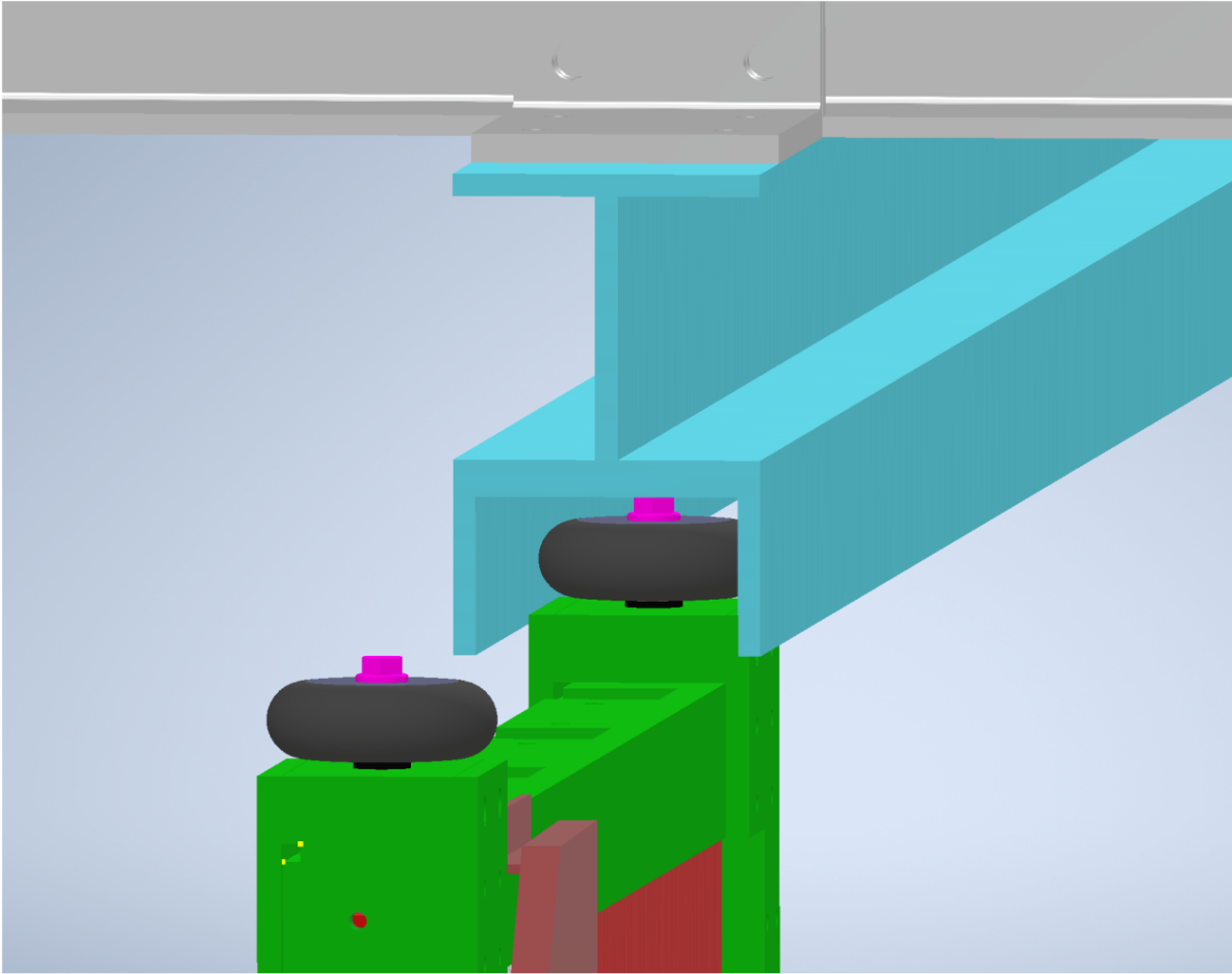}
    \end{center}
    \caption{Perspective diagrams of wheels that could be used to constrain the F-, C-, and B-face modules inside the upper aluminum channels; see text.}
    \label{fig:bumpers}
\end{figure}

\begin{figure}[tbp]
    \begin{center}
    \includegraphics[height=0.4\textheight]{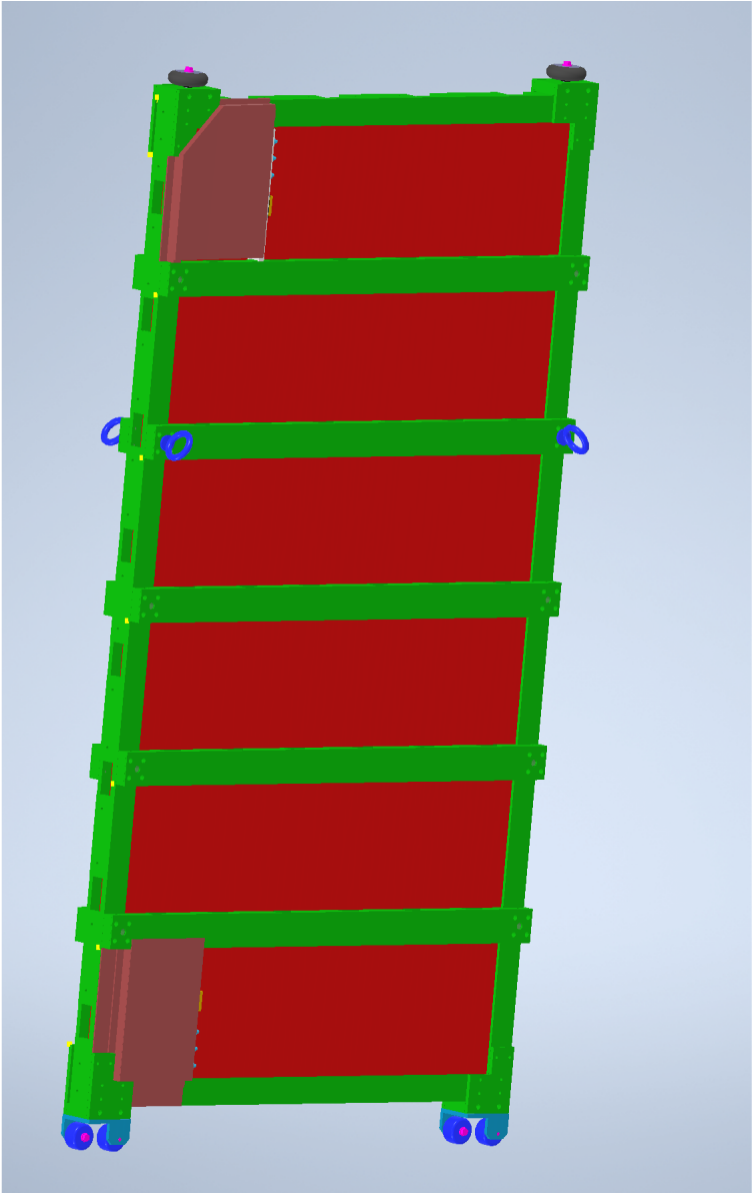}
    \end{center}
    \caption{Perspective diagram of module with upper bumpers and lower rollers attached. Some eye bolts are shown attached in the appropriate positions for hoisting via gantry crane; see text.}
    \label{fig:rollers}
\end{figure}

The ``rollers'' mentioned above may differ for the upper and lower $\eta$ edges of the modules. Two sets of two wheels, shown in Fig.~\ref{fig:wheels}, could be attached to either side of the lower $\eta$ edge (using the appropriate pick points; see Fig.~\ref{fig:CX1frame}) to allow the modules to roll. A set of bumpers, attached using the corresponding holes on the upper $\eta$ edge, would be sufficient to constrain the upper end of the modules inside the upper rails, as shown in Fig.~\ref{fig:bumpers}. Figure~\ref{fig:rollers} shows a module with these components attached.

\begin{figure}[tbp]
    \begin{center}
    \includegraphics[width=0.45\textwidth]{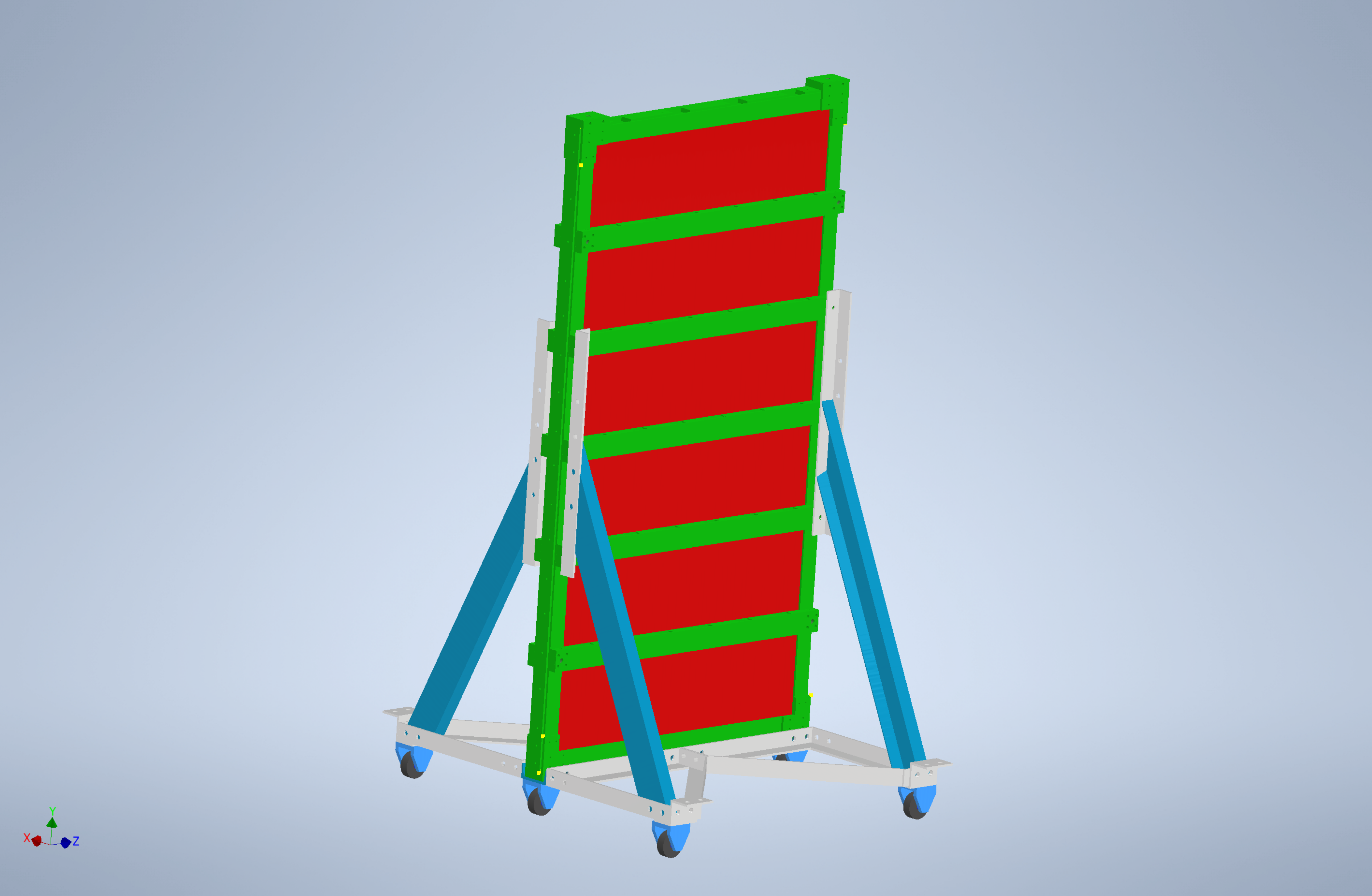}
    \includegraphics[width=0.45\textwidth]{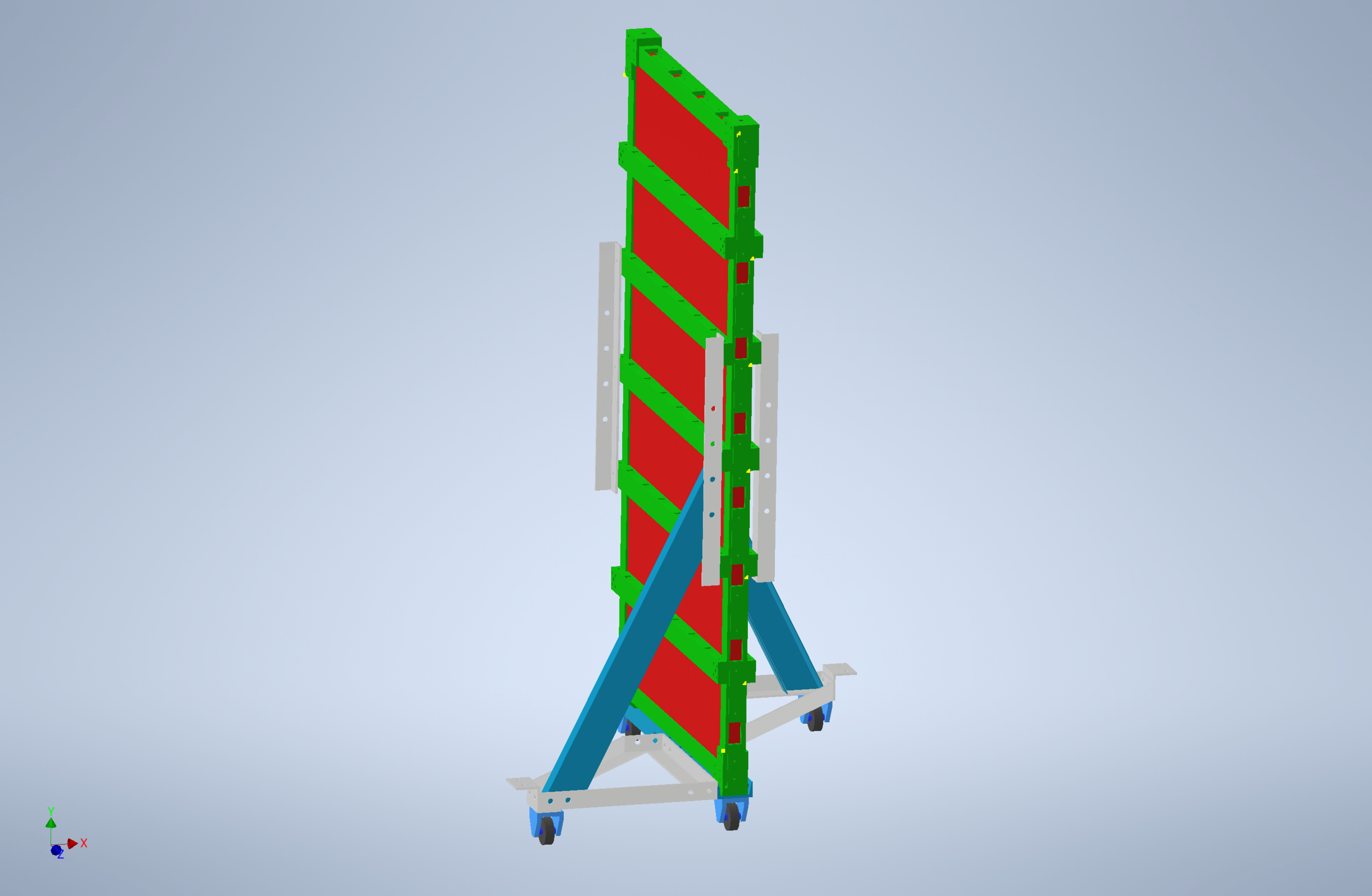}
    \end{center}
    \caption{Perspective diagrams of alternative upright transport carts (with modules attached) that could be used to move F-, C-, B-, L-, and R-face modules into the D1 barrack. The right image shows two of the wheels and their supports removed, narrowing the profile and allowing it to fit through the D1 barrack door.}
    \label{fig:uprightTransportCartsAlternate}
\end{figure}

One potential alternative transport cart design is shown in Fig.~\ref{fig:uprightTransportCartsAlternate}.
These carts do not use counterweights and would not be able to position the modules directly in the frame; were this design adopted, a rigging procedure would need to be developed to locate the upright modules in the support structure, and the support structure may need to be modified to accommodate the modules on the L and R faces and allow them to be rolled apart for access to the interior of the cube.

\begin{figure}[tbp]
    \begin{center}
    \includegraphics[width=0.45\textwidth]{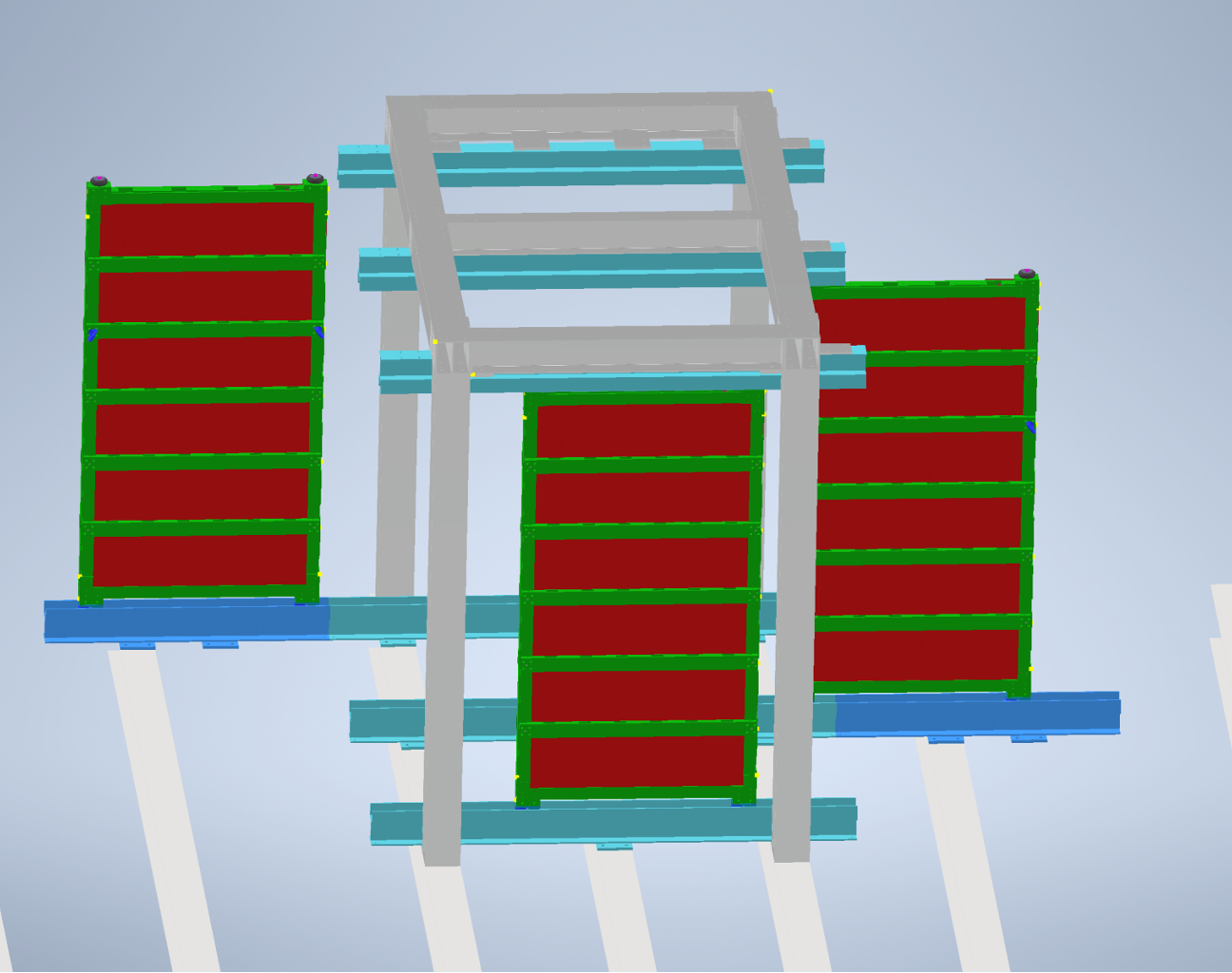}
    \includegraphics[width=0.45\textwidth]{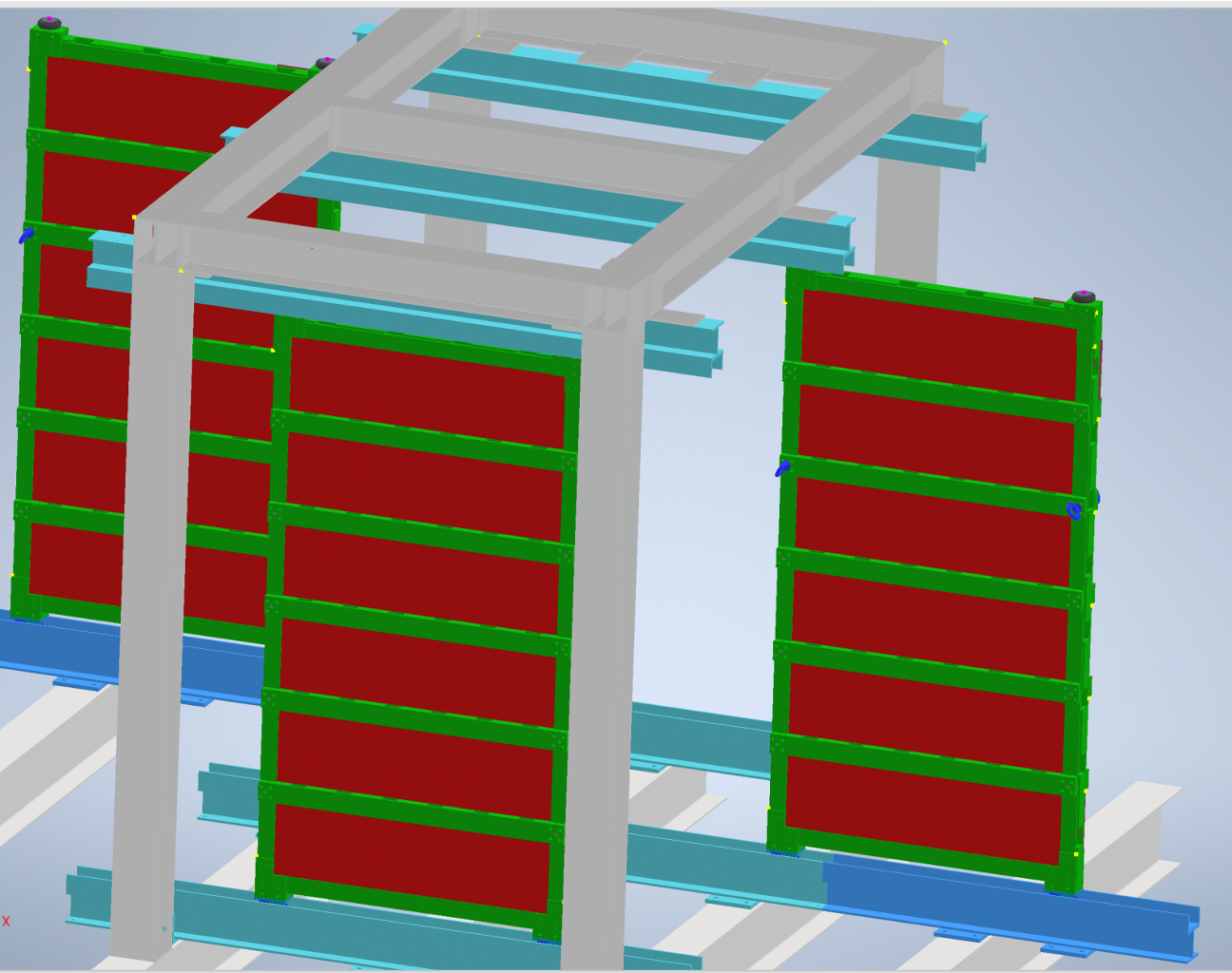}
    \end{center}
    \caption{Potential locations of a guidance rail, shown on the floor in dark blue, that could be used to facilitate an alternative cart design for the upright modules; see text.}
    \label{fig:guidancerails}
\end{figure}

To enable installation with the alternative transport carts, one possibility would be to hoist the modules into the support structure using a portable crane. A system of extendable guidance rails could be designed to facilitate this procedure and to allow access to the interior of the cube. Figure~\ref{fig:guidancerails} shows one such design. This guidance rail would essentially be a shorter, portable version of the same rails that support the F-, C-, and B-face modules inside the cube; it could be placed on either side and affixed to the structural support rails, and a portable gantry crane could be used to lift the modules into the rails and help guide them into position. The modules could be attached to such a crane using eye bolts affixed to the appropriate pick points; see Fig.~\ref{fig:rollers}.

The width of the module and required width of this alternative cart-design's supports prevent entry through the D1 barrack door; for this reason, a modular design is proposed where two of the wheels may be removed, as shown in Fig.~\ref{fig:uprightTransportCartsAlternate}, narrowing the profile for the length of time required to roll through the D1 barrack door.

Despite these additional complications, this alternative design has the merits of simplicity and stability, especially in that it does not require any counterweights. Input from CERN rigging experts would be invaluable to finalizing transport cart designs.

\subsubsection{Flat modules}
There are two types of modules that are configured in the flat position---one type for the G face and one for the S face.
The only difference is the relative height of the rollers (and the clips used to secure them) projecting out from the sides of the modules; these rollers are used to allow the module to run on tracks contained within the rail profiles.
For the G face, the wheels are mounted below the modules, which allows them to roll directly on the floor.
For the S face, the wheels are above the modules (to bring them to the correct elevation within the frame) and cannot be rolled directly on the floor.
Each wheel clip is secured to the module with a single 12\mm bolt attached to the edge of the module frame.

\begin{figure}[tbp]
    \begin{center}
    \includegraphics[width=0.45\textwidth]{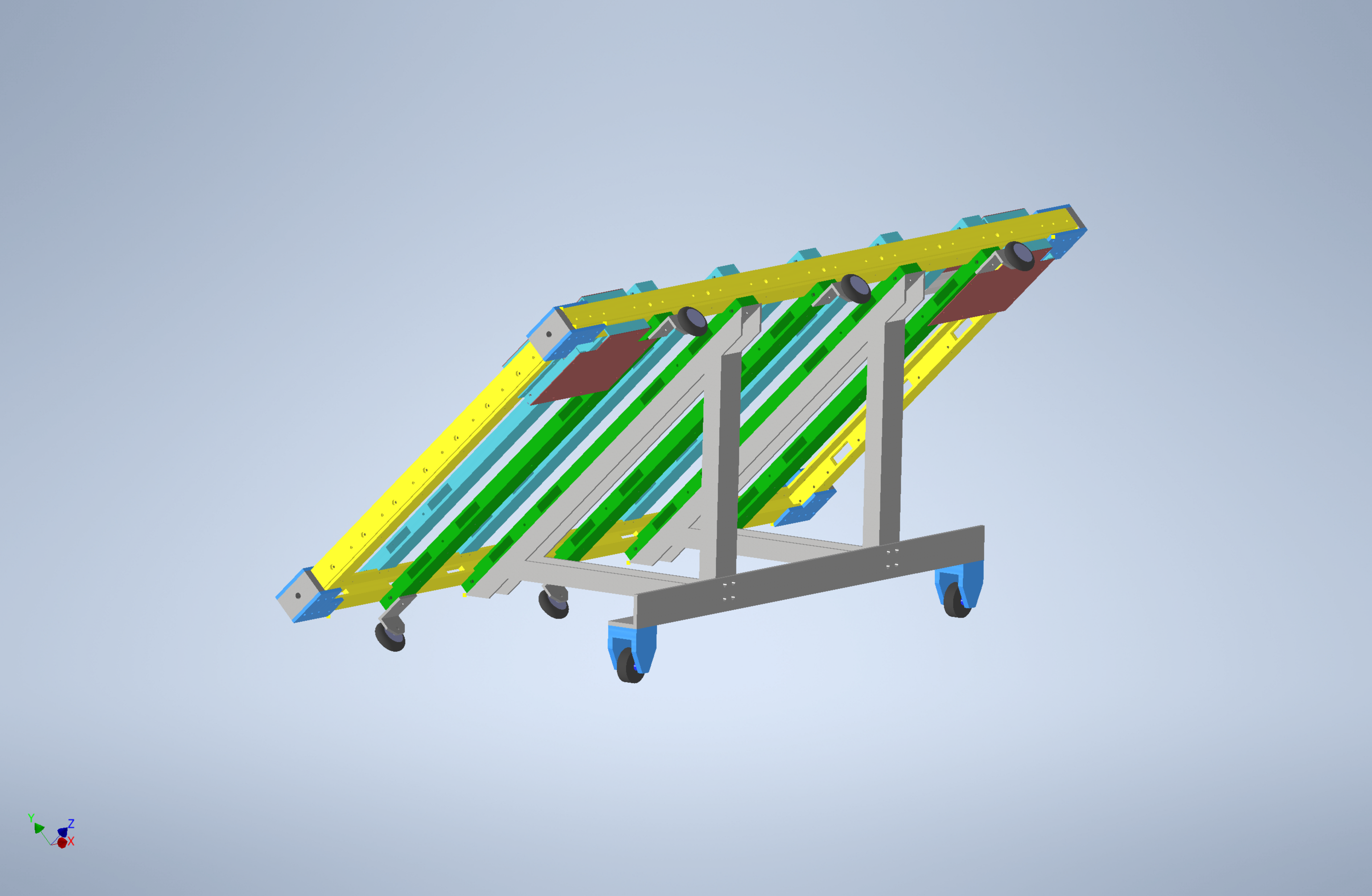}
    \includegraphics[width=0.45\textwidth]{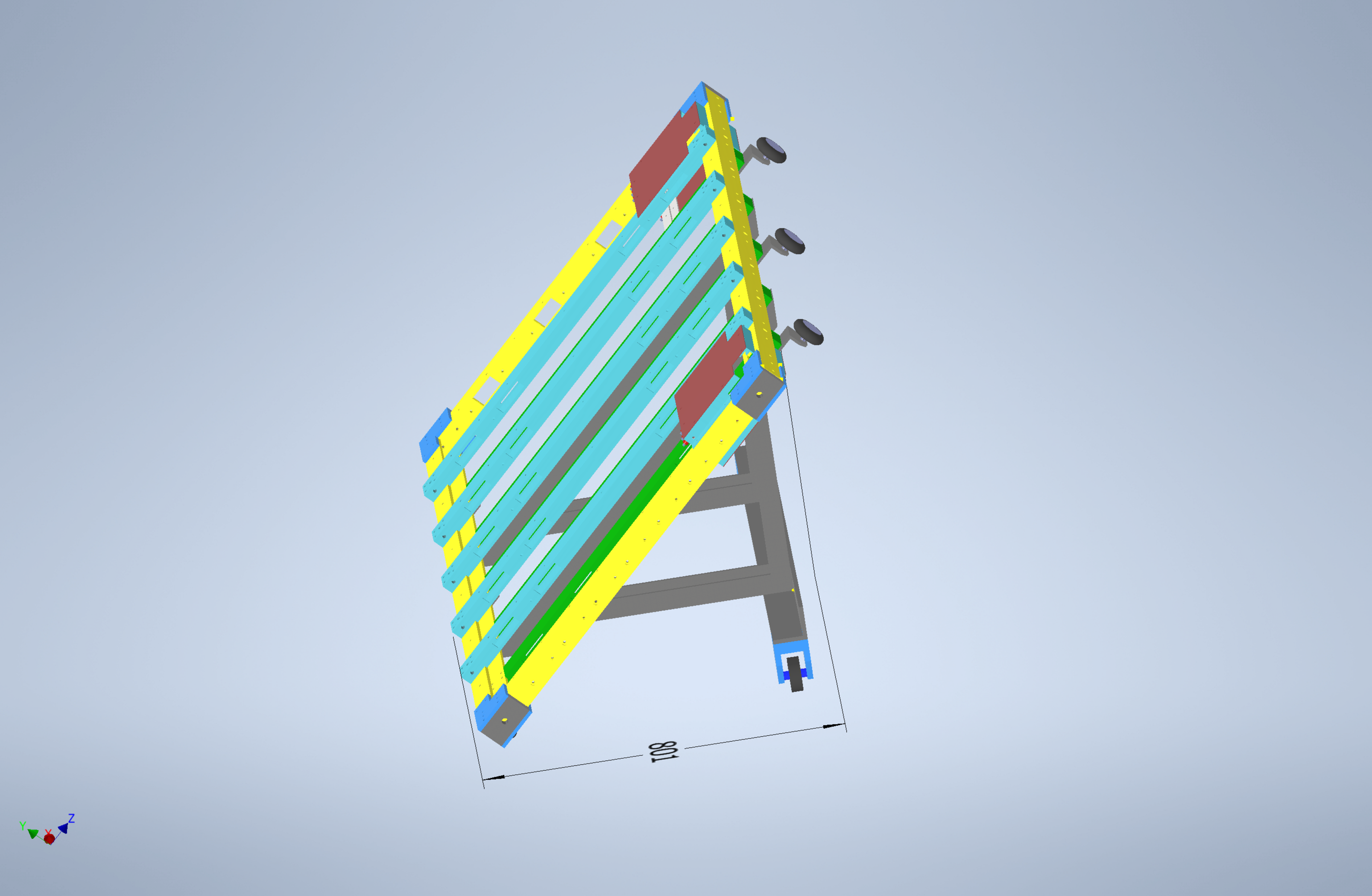} \\
    \includegraphics[width=0.45\textwidth]{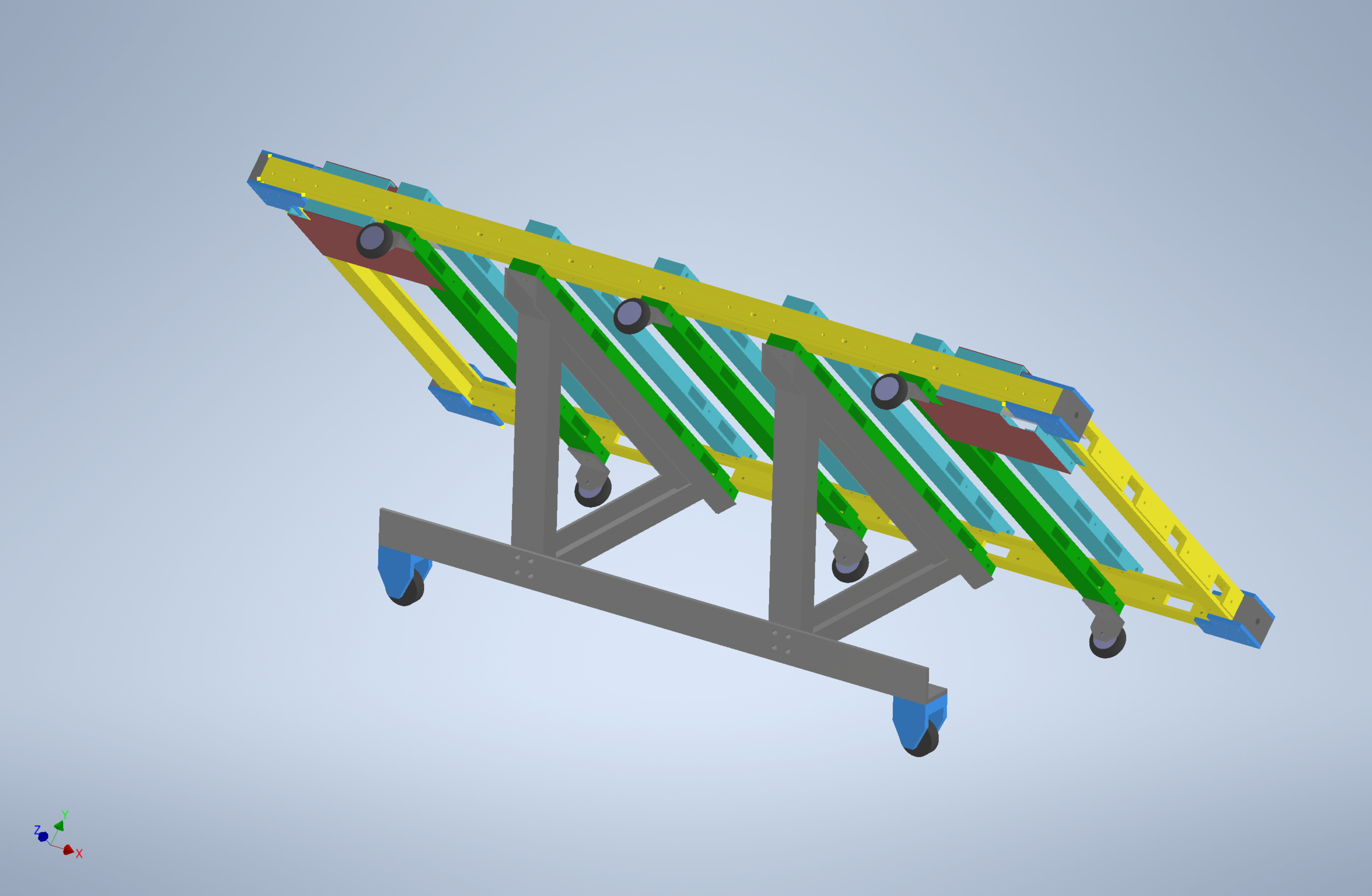}
    \end{center}
    \caption{Perspective diagrams of tilt-table transport carts (with modules attached) used to move S- and G-face modules into the D1 barrack.}
    \label{fig:tiltTableTransportCarts}
\end{figure}

These modules require the use of a different type of installation cart, since it is advantageous to bring them into the D1 barrack in the horizontal position to avoid the need to rotate them down from the upright position once inside.
Because the module is about 110\cm wide and the doorway is only 83\cm, the module must be at an angle that allows for the many tight turns required to reach the final location.
For this reason, the use of a specialized “tilt-table” cart is proposed (see Fig.~\ref{fig:tiltTableTransportCarts}).
In this design, the module would be loaded and held by the cart at a static 45-degree angle, reducing its horizontal dimension to 801\mm and thus allowing it to fit through the door. The cart would be supported on one side by two wheels attached directly to it, while its other side would be supported by three wheels mounted to the $\phi$ edge of the module itself. Care would need to be taken to design safe mounting and dismounting procedures for the modules to this cart if this design were adopted.

For the tilt-table cart, the module is secured to the cart (and rotated within it) by attaching a pair of adjustable struts to 14\mm bolts located at the corners of the module.
These struts are then used to rotate the module from flat to about 60 degrees up from horizontal.
The other two corner 14\mm threads are then used to mount pins that become an axis of rotation for the module.
This configuration allows for a total cart width of about 70\cm for transportation.

Once inside the D1 barrack, the module is then rotated flat to better position it for installation.
After the module is flat, and the wheels positioned on a short ramp, the struts and pins can be removed from both ends of the module.
This then frees the cart so it can be rolled off to the side.
While the G-face modules can roll directly off the ramp onto the tracks within the support frame, the S-face modules will need to be elevated---in the flat position---up the 2.2\m required to reach the tracks they ride on.

The use of structural ``scissors''-type lifting fixtures might be considered both for this 2.2\m lift as well as for lifting the modules the 1\m required just to get them on the same platform the room is situated on.
Each module weighs about 140\kg, and each transport cart may weigh about as much, for a combined load of almost 300\kg.
Since the frame can be lifted at either end by a crane, however, it might be possible to at least get the modules up on the platform with a portable gantry crane, if one is available.
In principle, the modules could be crane-lifted with their carts as a package.
This procedure is still not well understood, and would benefit from suggestions by CERN rigging experts familiar with the location of the D1 barrack.

\subsubsection{Installation sequence}
The design of the cube necessitates an order of installation that does not impede (or prevent) the positioning of successive modules.
For this reason, the following order is proposed:
\begin{enumerate}
    \item S-face modules. Getting them up the 2.2\m and inserting them properly will be less challenging without any of the sequential (F, C, and B face) rail-modules in place.
    \item Sequential (F, C, and B face) rail-modules. (The order is yet-to-be determined.)
    \item G-face modules. Once they are in place, access to the interior becomes limited. 
    \item L- and R-face modules. (The order is yet-to-be determined.) They become the access points for the interior of the cube.
\end{enumerate}

\subsubsection{Alignment}

The RPCs' spatial resolution can reach $\approx 1\,\rm{mm}$, thus a mechanical alignment precision of $<10\,\rm{mm}$ is desirable. A more precise global determination of alignment parameters can be achieved online through track fitting algorithms.

The geometry and alignment of the singlets within each RPC triplet is surveyed as part of the production process. External tests of the functionality will be carried out via geometry measurements and beam tests.

Since the entire detector structure is not located in an area that normally requires precision survey points, the primary focus will be on aligning the RPCs with respect to each other. 
There are a number of external bolt locations on the CX1 module frame that can be referenced to identify the perimeter of the active areas inside the module frame.
Those same bolt locations can also be used to secure the CX1 frame to both the support structure frame and adjacent CX1 frames via precision brackets.
In this manner, the physical position the active element for each RPC triplet is already defined prior to installation, and as long as all the brackets are in place, the modules should be aligned to within 5\mm for adjacent modules. A very precise alignment monitoring system, such as the RasNiK~\cite{vanderGraaf:2021wxg} system developed for the ATLAS muon modules or the LHCb BCAM system, is not necessary for \codexbeta.

The global position and orientation of the detector does not require such a high precision because of its large displacement from the LHCb interaction point. Its coordinates will be determined by measuring the positions of the frames' external bolt locations with respect to known reference points in the D1 barrack using common indoor distance measurement techniques like laser-based instruments. These measurements do not require any contributions from the CERN surveyor team. The stability of the detector on the false floor of the D1 barracks is ensured by placing it on the floor support beams, which are strong enough to sustain the much heavier computing racks that used to be installed in this place. 

The alignment and calibration parameters will be determined independently with cosmic muon data and measurements of decays of \KL mesons produced by neutral particles from the \lhcb interaction point interacting with the shielding.
This will allow cross checks and improvements of the alignment determined from the mechanical setup.

\subsection{Cabling and Power}
\label{subsec:CablingAndPower}
The power and control system is installed on a dedicated rack in the neighbourhood of the detector to avoid excessive length of the low voltage cables. The power system, which will be connected to the power supply in the D1 barrack, is based on the CAEN mainframe SY4527 equipped with a supplementary power unit bringing the total power to 1200 W.
The mainframe hosts the power distributors including the following functions: High voltage, Low voltage, monitoring.

\subsubsection {High voltage}
The RPCs are operated to maximum 6 kV with the standard mixture, the expected current driven by a single gas gap being $< 5 \mu$A. There are, in total, 42 gas gaps. Two suitable boards are CAEN A1580 HP (single unit, 16 channels, 8 kV 20 $\mu$A) or CAEN A1590 HP (double unit, 16 channels, 9 kV 50 $\mu$A). These boards use high density Radiall connectors and need a transformation kit and a patch panel for the final distribution. There are 3 options: 3 boards (48 channels no fan out); 2 boards (1590 type only), 24 channels with a distribution fan out 1 to 2; 1 board (1590 type only), 12 channels with a distribution fan out 1 to 3.
In any case there will be 42 RG58 coaxial HV cables, installed in suitable cable trays, connecting the HV distributors to the detector.

\subsubsection {Low voltage}
The low voltage supply is needed for:
\begin{itemize}
    \item powering the front-end boards and providing the threshold
    \item powering the readout FPGA 
    \item powering the sensors
\end{itemize}
The LV system can be based on six CAEN A2551 (single slot, 8 channels, 8V, 12A) power boards.

\begin{figure}[tbp]
\centering
    \includegraphics[width=0.8\textwidth]{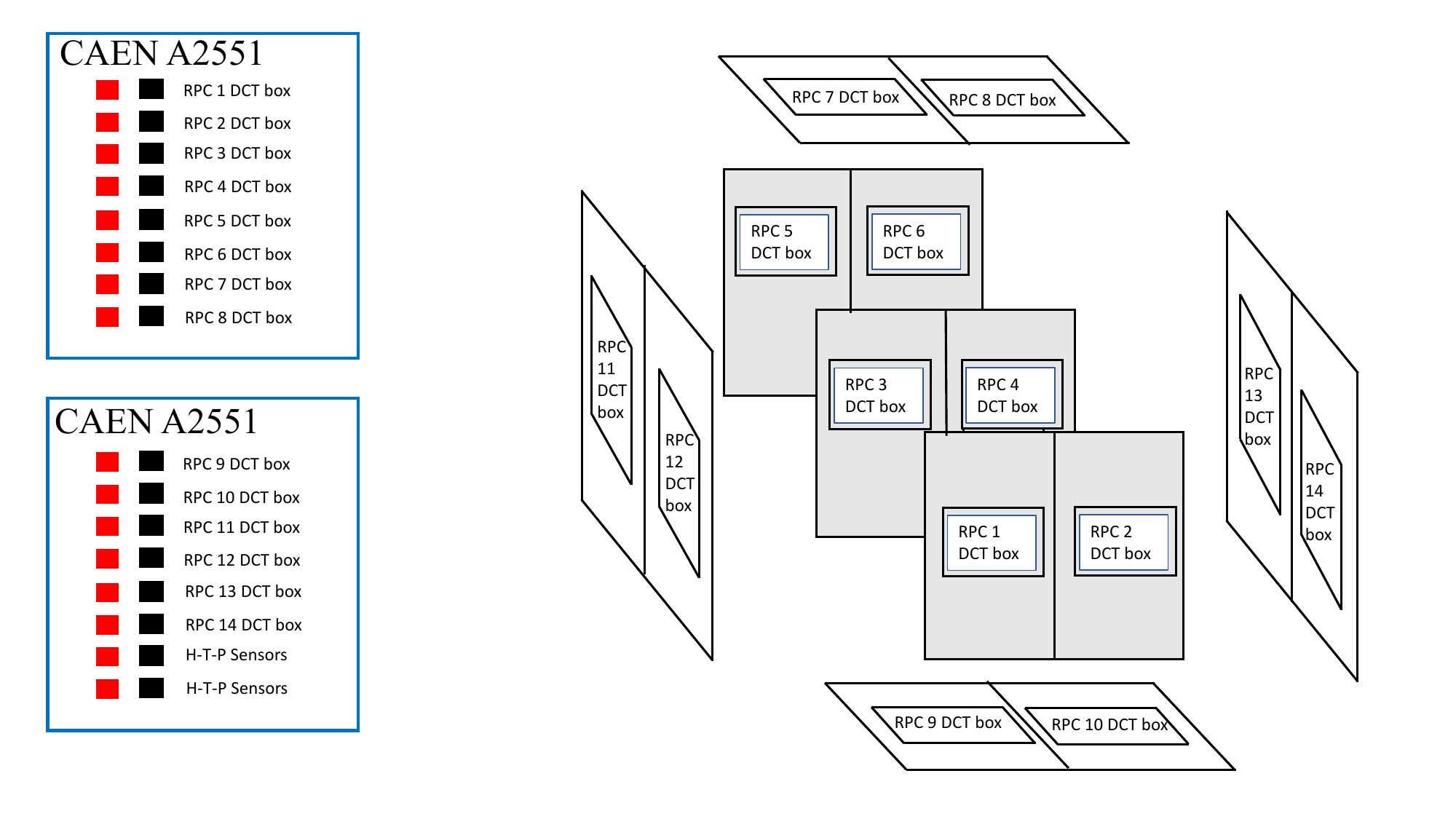}
    \caption{Schematics of the Low Voltage distribution for the DCT boxes, from the LV modules to the detector. Each DCT box has its own dedicated LV channel. The last 2 channels are dedicated to the H-T-P sensors distributed along the detector.}
    \label{fig:DAQpwdistri}
\end{figure}

As shown in Fig.~\ref{fig:DAQpwdistri}, fourteen channels of two of the boards are needed for the Data Collection and Transmission (DCT) boxes, which contain the FPGAs and are typically powered at 5V and absorb 4A. It is advisable to maintain the single FPGA granularity to ease detector management. Given the one-to-one relation between a power channel and a readout box, we foresee 14 point-to-point cables for this function.

An additional channel is reserved for sensor power and the other is left as a spare, so in total, 15 of the 16 available channels across two of the boards are used.

\begin{figure}[tbp]
\centering
    \includegraphics[width=0.8\textwidth]{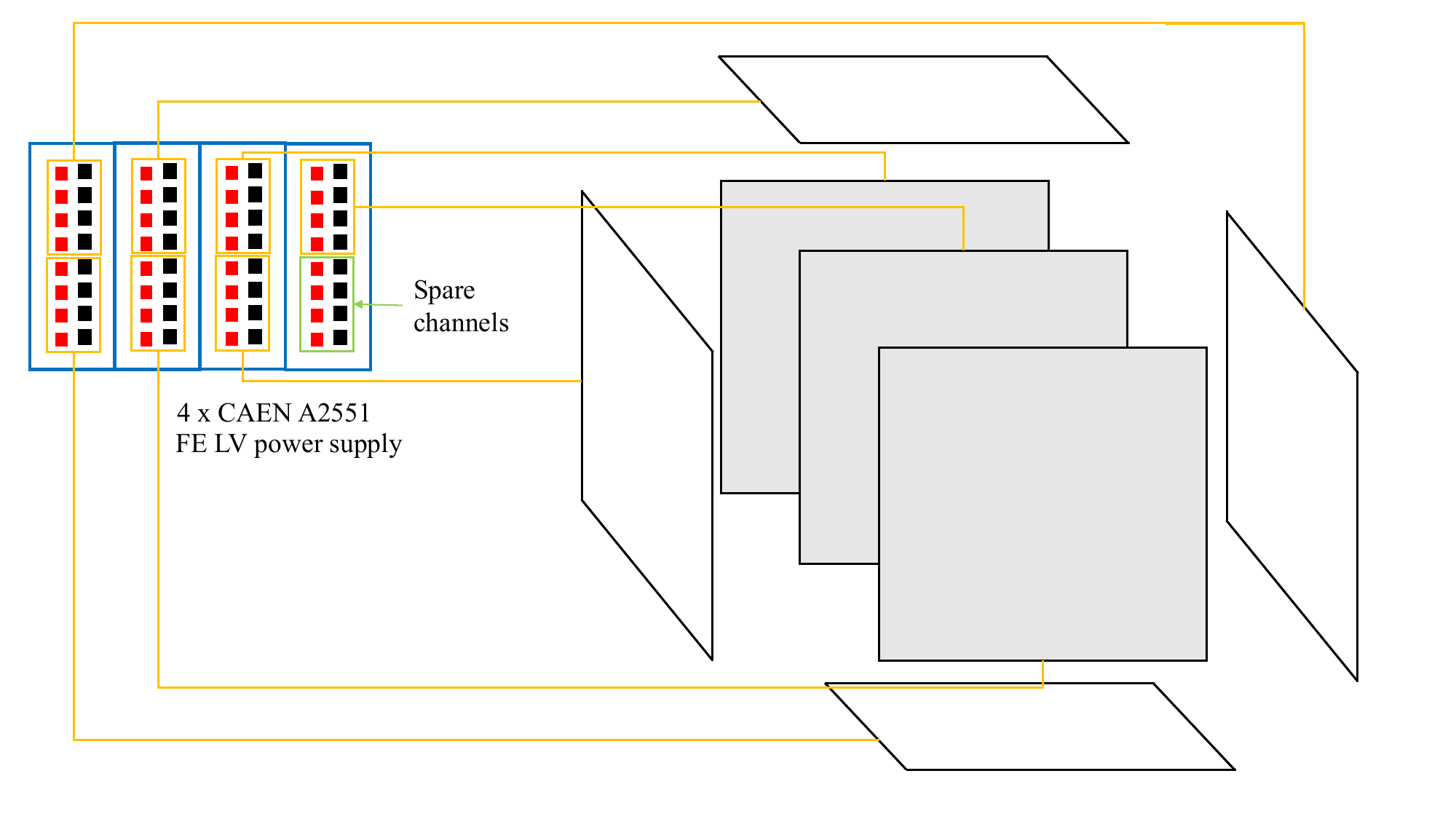}
    \caption{Schematics of the Front-End power distribution along the detector. Each detector face has its own block of LV channels. Each block is composed of 4 channels which provide the various voltages to the Front-End electronics.}
    \label{fig:LVdistri}
\end{figure}
\begin{figure}[tbp]
\centering
    \includegraphics[width=0.8\textwidth]{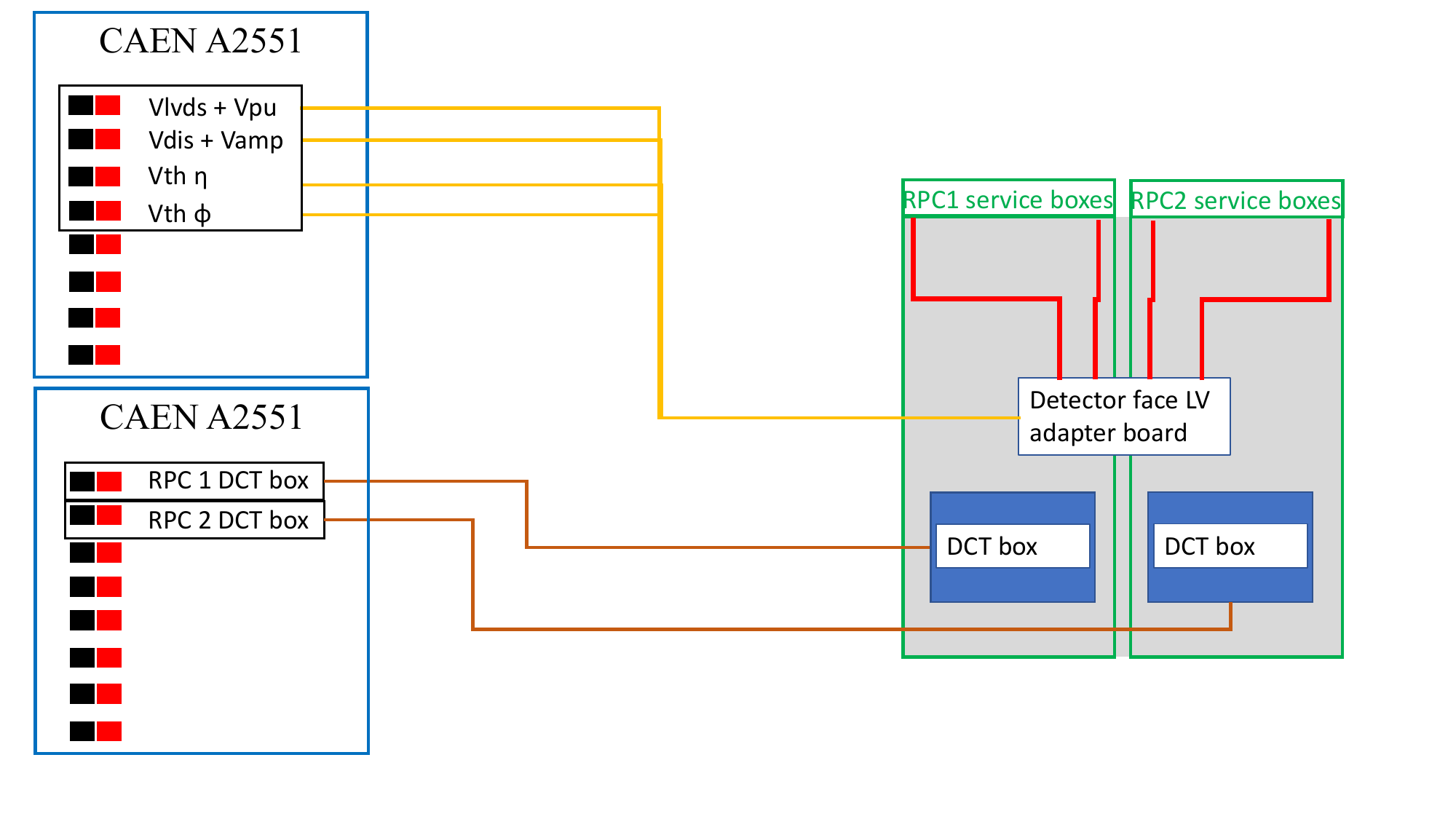}
    \caption{Schematics of the LV distribution for a single detector face. The 4-channel block is used to provide the 4 voltages needed to the FE electronics of a detector face (see Fig.~\ref{fig:LVdistri}). Those 4 voltages are then split to each triplet by means of an adapter board. Each triplet has a dedicated DCT box which is powered by its own LV channel (see Fig.~\ref{fig:DAQpwdistri}).}
    \label{fig:LVdistrising}
\end{figure}

The front-end boards are powered through 4 main supplies: V digital (V$_{d}$), V analog (V$_{a}$), and V threshold (V$_{th}$), where V$_{th}$ is given separately for the two different orientations of the readout strips ($\phi$ and $\eta$). 
To have a reasonable compromise between cost and functional system granularity and to respect the limits of the power boards, we propose to distribute these 4 supplies to every individual face of the cube (comprising 2 modules).
In this case, each 8-channel power board can serve up to two faces of the cube, as shown in Fig.~\ref{fig:LVdistri}, so only four power boards are needed to power all of the front-end boards.
An adapter board would distribute the incoming LV supplies to each module, as shown in Fig.~\ref{fig:LVdistrising} and described in the next section.

\subsubsection{On-chamber power distributor}
A special distributor, or detector-face LC-adapter board, is mounted on each face of the cube and serves the following functions:
\begin{itemize}
    \item receive the 4 primary LV cables and generate, by means of linear and low-dropout (LDO) voltage regulators, the needed voltages for the FE boards of each of the 2 modules (V$_{amp}$, V$_{discr}$, V$_{pull}$, V$_{lvds}$, V$_{th_\eta}$, V$_{th_\phi}$), as shown in Fig.~\ref{fig:LVdistrising};
    \item distribute to each of the two modules the voltages by means of 2 cables per module, one for the $\eta$ and one for the $\phi$ strips;
    \item distribute the necessary supply to the local sensors (gas flow, T, RH);
    \item collect monitoring information coming from the 3 gas gap current readouts, the 3 sensors, and the main LV supply currents (4 currents x 6 panels) per module, in total 30 values per module. Two 64-wires flat cables deliver directly to the ADC boards placed in the power rack. We are assuming that the monitoring is performed through the CAEN A3801 boards (128 channels each).
\end{itemize}

\subsection{LV patch panel and cables}
The front-end boards and sensors power supplies outputs are interfaced to the main cables by a patch panel mounted on the power rack. 
In output of the LV patch panel there are 7 multi-pin connectors, one for each face of the cube, each carrying the 5 different voltages, 4 for the main supplies of the FE boards and one for the sensors.

The cables interconnecting the power rack to the cubes, through appropriate cable trays, are listed below:
\begin{itemize}
    \item 42 GR58 HV cables (suitable for at least 8 kV)
    \item 7 multi-wire LV cables. Minimum 10 wires, suitable for carrying 6-8 A per couple
    \item 14 LV cables for powering the 14 readout boxes
    \item 14 flat cables, 64 wires each, to carry the analog monitoring information to the ADCs
\end{itemize}
\subsection{Gas}
\label{subsec:Gas}
The gas mixture required for the \codexbeta demonstrator is the same typically used on other RPCs made of bakelite, including those in use by both \cms and \atlas. It is composed of $94.7\%$ R-134a (1,1,1,2 tetrafluoro-ethane), $5\%$ isobutane, and $0.3\%$ SF6 (sulfur hexafluoride), with enough water vapor added to bring the relative humidity of the mixture to about $30$--$40\%$. The mixture is not flammable. The full volume of the detector array is about $80\L$, and assuming one volume change every $4$ hours, $20\L/\mathrm{hour}$ of this mixture will be needed for normal operation. Nitrogen, possibly humidified, will also be needed as an inert purge gas.

Since none of the working gases are presently in use at LHCb, several new high-pressure cylinder stations will need to be installed in the upstairs gas handling/storage area. To avoid creating redundancies in the LHCb supply area, CODEX-$\beta$ would like to make occasional use of the nitrogen source that already exists by tapping into the upstairs piping at an appropriate location. During the leak check process a maximal nitrogen rate of $20~\mathrm{L/hour}$ will be required, but it is unlikely that nitrogen will be required afterwards.

\begin{figure}[tbp]
\centering
    \includegraphics[width=0.65\textwidth]{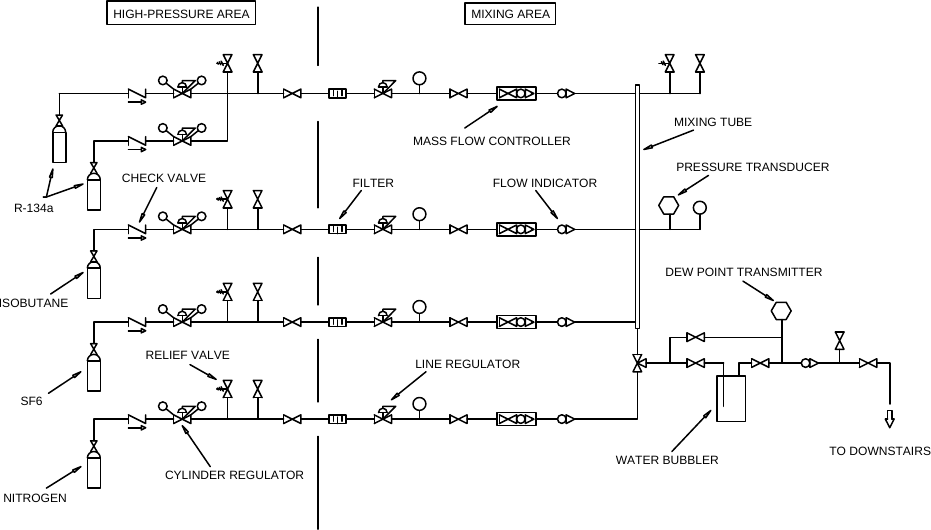}\\
    \includegraphics[width=0.65\textwidth]{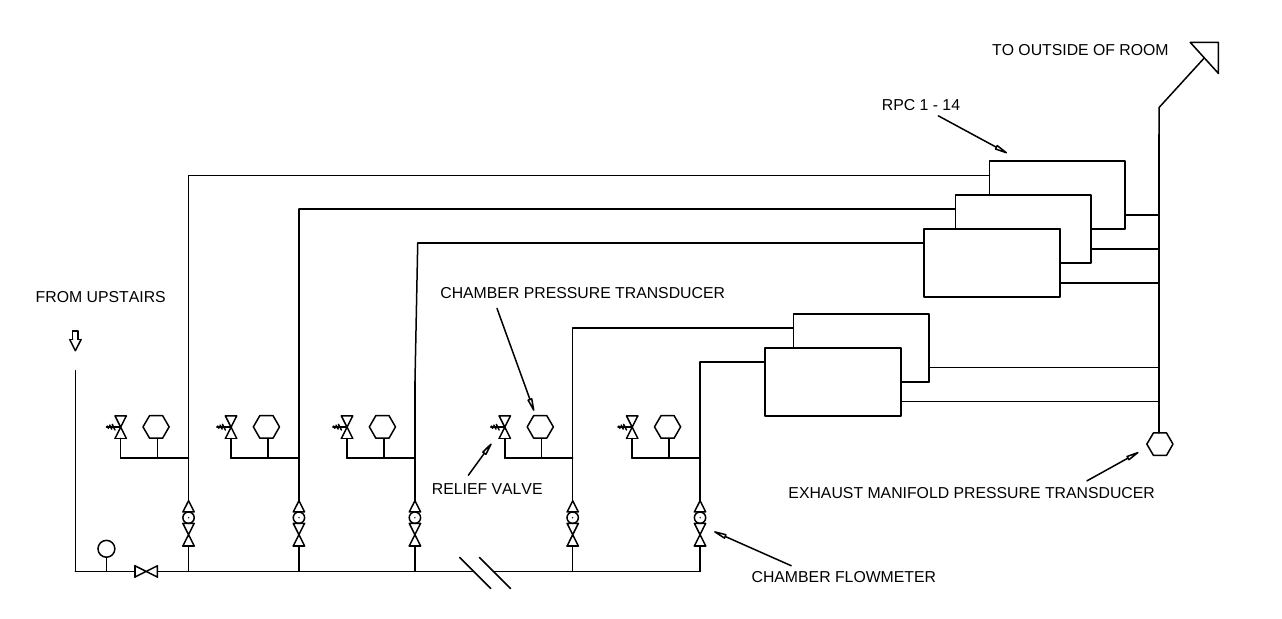}\\
    \includegraphics[width=0.65\textwidth]{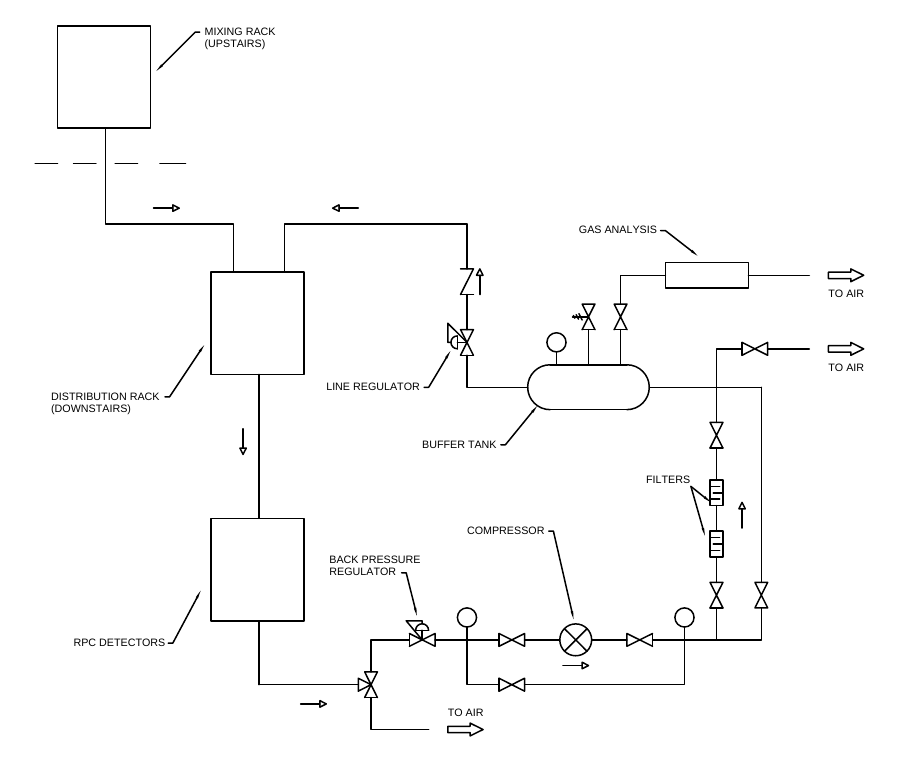}
    \caption{Schematics of the (top) upstairs, (middle) downstairs gas systems. More detail about the proposed recirculation system is given in the bottom figure. NB: Humidifier to be located {\it downstairs}, not upstairs as currently shown, to avoid possible condensation.}
    \label{fig:CDXgas}
\end{figure}

The upstairs part of the system, shown in Fig.~\ref{fig:CDXgas}, consists of both the mixer and the gas storage area. A single line will run the mixture to the downstairs distribution panels and then to each individual detector. A humidifier will be installed in the near proximity of the detector apparatus. The flow rate through the detector is relatively low, but the high global warming potential (GWP) of the mixture (approximately $1400$: $.94.7 \times 1400$ from the R-134a, and $.003 \times 23,000$ from the SF6) 
is sufficient to merit the installation of a recirculation system~\cite{Guida:2017pik} to minimise emissions. This system will be connected to a single return line, to return the residual gas from the recirculation system to the surface.

All of the gases used for normal operation require special consideration. While all three gases come as pressurized liquids, only the isobutane cylinder needs to be kept at room temperature because the vapor pressure is so low; both the R-134a and SF6 have vapor pressures high enough so that any online cylinders could be kept in an unheated area. A leak detector will be installed in the mixing rack. This will be a level-3 leak detector which will be installed and coordinated through the relevant CERN channels. Note, this leak detector is for the mixing rack and will shut down the gas supply from the surface. It does not necessarily shut down power underground, and should not affect D2-D3. Because of the relatively low flow rate, only a single cylinder of isobutane would need to be connected to the system at any one time. Most likely, only one small SF6 cylinder is needed and will last the entire run.

Assuming a nominal year's operation of 300 days, and recirculation efficiency of 60\% (a conservative estimate; much higher efficiency could be achieved), the following consumptions could be expected: R134a 300 kg/year, iC4H10 8 kg/year, and SF6 1.2 kg/year. The total annual emissions correspond to 450 tCO2eq.

The details of the gas system infrastructure are being developed in conjunction with the CERN gas group, who will work alongside the \codexbeta group to manage installation.

\subsection{Chamber Certification}
\label{subsec:QA}
The RPC chamber commissioning is structured in several steps that lead to the validation of the fully assembled detector.
The first step consists of the validation of the singlet as an individual detector. Basic structural controls are performed on the singlet:
\begin{enumerate}
    \item Gas tightness: The gas gap is checked for eventual gas leakage.
    \item High Voltage: The HV distribution is checked and the voltamperometric curve of the detector is performed in order to verify the functionality of the detector and its stability in terms of current driven by the gas gap.
    \item Low Voltage: The LV distribution is checked and proper current absorption is verified in order to ensure the functionality of the front-end electronics.
    \item Cosmic-ray test: The singlet performance and overall behaviour is tested by means of cosmic rays and by using an external tracking system as reference.
\end{enumerate}

The validation of the singlets allows the construction of the complete chamber, composed of three independent singlets (forming a triplet), along with the entire on-chamber services.
The fully assembled triplet is also subject to the above validation tests in order to be certified.
The last step of commissioning is performed using cosmic rays, and all the detector aspects in terms of performance are checked.
The cosmic-ray commissioning steps are reported in Fig.~\ref{fig:CRcomm}.

\begin{table}[tbp]
\centering
    \includegraphics[scale=0.7]{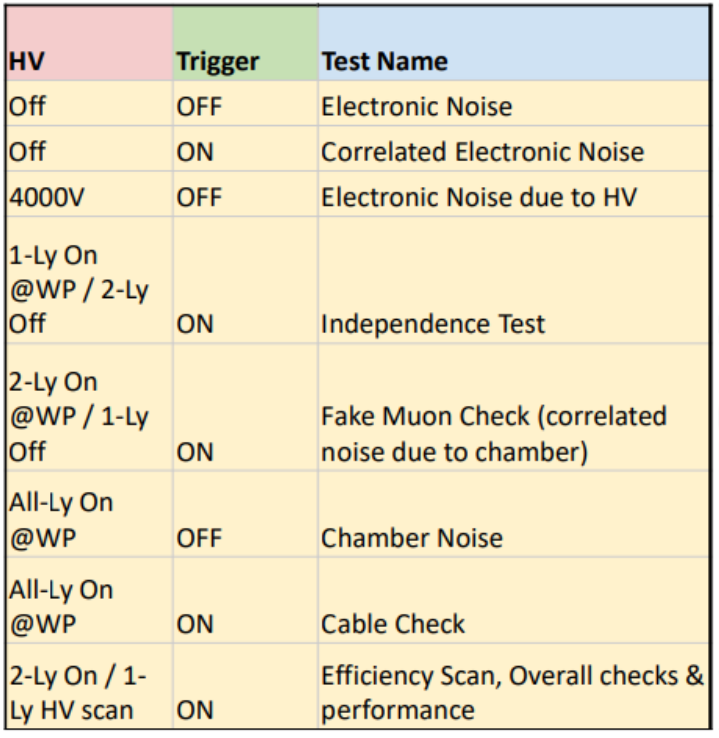}
    \caption{Table of the various chamber configurations during the cosmic rays commissioning steps.}
    \label{fig:CRcomm}
\end{table}  

The channels noise map allows to verify the electronic noise of the chamber and permits the masking of the noisy channels which would generate fake events.
The diagonal maps correspond to the correlation plots between the same readout view of different singlets.
These plots verify that the signal cabling is performing correctly.
The independent tests aim to verify the absence of any fake trigger produced by the chamber itself.
The completion of the controls previously described allows the study of the performance of the detector in terms of efficiency (per layers and overall), time-over-threshold distributions, cluster size, and time-of-flight.

\section{Conclusion}
The design for a new demonstrator apparatus, the \codexbeta experiment, has been presented. The components of the detector are under construction as an R\&D project of the LHCb experiment. It is intended that the demonstrator experiment will be integrated with the LHCb DAQ and collect data during 2025. 

The demonstrator project is an essential step in the journey towards a proposed future installation of the larger \codexb experiment during HL-LHC operation. The demonstrator will verify estimates of background levels in the underground area near LHCb. It will demonstrate the potential to integrate the future experiment with the LHCb DAQ and provide the CODEX-b collaboration with operational expertise of an RPC-based tracking detector. Construction of the demonstrator is a valuable opportunity to prototype elements of the support structure that will be needed for the future \codexb experiment.

\clearpage

\bibliographystyle{jhep}
\bibliography{main}

\section*{Acknowledgments}
We are grateful for the support we have received from the LHCb collaboration. 
We thank the technical and administrative staffs at CERN and at other CODEX-b institutes
for their contributions to the success of the CODEX-b and \codexbeta effort.
In addition, we gratefully acknowledge the computing centers and personnel of the
Worldwide LHC Computing Grid and other centres for delivering so effectively the computing infrastructure essential to our analyses.

Individuals have received support from the Science and Technology Facilities Council (UK) [grant reference ST/W004305/1]
and National Research, Development and Innovation Office (NKFIH) research grant (Hungary) [contract number TKP2021-NKTA-64].
The work of IGFAE members is supported by the Spanish Research State Agency under project PID2022-139514NA-C33;
by Xunta de Galicia (Centro singular de investigación de Galicia accreditation 2019-2022), by European Union ERDF;
and by the Spanish Ministry of Universities under the NextGenerationEU program.
The work of LBNL staff/researchers is supported by the Office of High Energy Physics of the U.S. Department of Energy under contract DE-AC02-05CH11231. The work of the University of Cincinnati personnel is in part supported by the United States National Science Foundation under grant NSF-PHY-220976.
The work of the DESY personnel is supported by DESY (Hamburg, Germany), a member of the Helmholtz Association HGF, and by the Deutsche Forschungsgemeinschaft
(DFG, German Research Foundation) under Germany’s Excellence Strategy -- EXC 2121 ``Quantum Universe'' -- 390833306.
The work of CVS is supported by Agencia Estatal de Investigación (Spain) through the Ramón y Cajal program RYC2023-043804-I.

\end{document}